\documentclass[12pt]{article}
\usepackage{amsmath,amssymb,epsfig,amsfonts}
\usepackage{graphicx,subfigure}
\usepackage[usenames, dvipsnames]{color}
\usepackage{multirow}
\usepackage[backref]{hyperref}
\usepackage[nosort]{cite}
\usepackage{lscape}
\usepackage{rotating}
\usepackage[normalem]{ulem}

\usepackage{extarrows}

\usepackage{verbatim} 

\makeatletter

\newcommand{\be}{\begin{equation}}
\newcommand{\ee}{\end{equation}}
\newcommand{\ba}{\begin{aligned}}
\newcommand{\ea}{\end{aligned}}

\def\half{{\frac{1}{2}}}

\def\unit{{1\kern-.65ex {\rm l}}}
\def\1{{1\kern-.65ex {\rm l}}}

\newcommand{\bfive}{\mathbf{5}}
\newcommand{\obfive}{\bar{\mathbf{5}}}
\newcommand{\bten}{\mathbf{10}}

\newcount\hour \newcount\minute
\hour=\time \divide \hour by 60
\minute=\time
\def\now{%
\ifnum \hour<13
  \ifnum \hour=0 \advance \hour by 12 \number\hour:\else \number\hour:\fi%
     \ifnum \minute<10 0\fi%
     \number\minute%
\ A.M.%
\else \advance \hour by -12 \number\hour:%
  \ifnum \minute<10 0\fi%
  \number\minute%
  \ P.M.%
\fi%
}

\makeatother

\begin{document}

\baselineskip=18pt  
\numberwithin{equation}{section}  
\allowdisplaybreaks  


%
%


\thispagestyle{empty}

\vspace*{-2cm} 
\begin{flushright}
{\tt KCL-MTH-15-06} 
\end{flushright}

\vspace*{0.8cm} 
\begin{center}
{\Huge Froggatt-Nielsen meets Mordell-Weil:}\\
\bigskip\smallskip

{\Large A Phenomenological Survey of Global F-theory GUTs with $U(1)$s}\\

 \vspace*{1.5cm}
{Sven Krippendorf$\, ^\diamond$, Sakura Sch\"afer-Nameki$\, ^\star$ and Jin-Mann Wong$\, ^\star$ }\\

\vspace*{.8cm} 
 {\it $^\diamond$ Rudolf Peierls Centre for Theoretical Physics, University of Oxford\\
1 Keble Road, Oxford, OX1 3NP, England }\\
 {\tt {{sven.krippendorf@physics.ox.ac.uk}}}\\

 \vspace*{.5cm} 
 {\it $^\star$ Department of Mathematics, King's College, London \\
 The Strand, London WC2R 2LS, England }\\
 {\tt {gmail:$\,$  sakura.schafer.nameki, jinmannwong}}\\

\vspace*{0.8cm}
\end{center}
\vspace*{.1cm}
\noindent
In F-theory, $U(1)$ gauge symmetries  are encoded in rational sections, which generate the Mordell-Weil group of the elliptic fibration of the compactification space. 
Recently the possible $U(1)$ charges for global $SU(5)$ F-theory GUTs with smooth rational sections were classified 
\cite{Lawrie:2015hia}. In this paper we utilize this classification to probe global F-theory models for their phenomenological viability. 
After imposing an exotic-free MSSM spectrum, anomaly cancellation 
(related to hypercharge flux GUT breaking in the presence of $U(1)$ gauge symmetries),  
absence of dimension four and five proton decay operators and other R-parity violating couplings, and the presence of at least the third generation top Yukawa coupling, we generate the remaining quark and lepton Yukawa textures by a Froggatt-Nielsen mechanism. 
In this process we require that the dangerous couplings are forbidden at leading order, and when re-generated by singlet vevs, lie within the experimental bounds. We scan over all possible configurations, and show that only a small class of $U(1)$ charge assignments and matter distributions satisfy all the requirements. 
The solutions give rise to the exact MSSM spectrum with
realistic quark and lepton Yukawa textures, which are consistent with the CKM and PMNS mixing matrices.
We also discuss the geometric realization of these models, and provide pointers to the class of elliptic fibrations with good phenomenological properties.

\newpage
\setcounter{tocdepth}{2}
\tableofcontents


\section{Introduction and Overview}

{Remarkable progress in the construction of global F-theory compactifications in recent years has resulted in both conceptual and technical advances. 
{After the initial surge  in particle physics explorations of local F-theory Grand Unified Theories (GUTs),  the study of phenomenological implications was somewhat 
side-stepped in recent advances in global model building. }

{Global models have} in particular seen much progress {in} view of a comprehensive understanding of F-theory vacua -- both in terms of the base as well as the fiber geometry.} In view of this, it is timely to conduct a survey of 4d  F-theory vacua and their phenomenological viability. The goal of this paper is to provide such an analysis, by imposing the most stringent phenomenological requirements upon the F-theory compactifications with additional $U(1)$ symmetries and their  4d effective theories, in particular an exotic-free {Minimal Supersymmetric Standard Model (MSSM)} spectrum, absence of dangerous couplings, such as proton decay operators, as well as consistent flavor physics generated by a Froggatt-Nielsen mechanism. 

Central to both guaranteeing the absence of dangerous couplings {and the applicability of a Froggatt-Nielsen mechanism} is the presence abelian gauge symmetries. 
One of the string theoretic inputs in our analysis is the classification of $U(1)$ charges in $SU(5)$ F-theory GUTs, which was recently performed in \cite{Lawrie:2015hia}. This classification result utilizes general insights from codimension two fibers in \cite{Hayashi:2014kca}, which realize the matter fields,  and consistency of rational sections, which give rise to $U(1)$ gauge potentials. The one assumption in this classification is that the rational sections are smooth. 
The resulting analysis does not provide a constructive way of obtaining the elliptic fibrations, but gives a classification of all consistent fibers with rational sections, 
which in turn determines the set of matter $U(1)$ charges. It reproduces all charges known to exist in explicit geometric constructions based on hypersurfaces and complete intersections \cite{Grimm:2010ez, Braun:2011zm, Mayrhofer:2012zy,
Morrison:2012ei, Borchmann:2013jwa, Cvetic:2013nia, Borchmann:2013hta, Cvetic:2013jta, Cvetic:2013qsa, Braun:2013yti, Braun:2013nqa, Kuntzler:2014ila, Lawrie:2014uya, Braun:2014qka, Klevers:2014bqa,  Morrison:2014era, Baume:2015wia, CKPT}, but {the set of possible charges from this classification} is strictly larger than the ones arising from known geometries. {This \"ueber-set obtained in \cite{Lawrie:2015hia}  contains all charges that can potentially arise in global F-theory compactifications, under the assumption of smooth rational sections, and will be referred to  as {\it F-theoretic $U(1)$ charges}. }

{A second constraining factor in F-theory GUT model building is the requirement of cancellation of anomalies that arise in the context of GUT breaking via hypercharge flux \cite{Buican:2006sn, Donagi:2008kj, Beasley:2008kw}, which to date is the only known mechanism to break the GUT group in F-theory without immediately introducing exotics, such as is the case for Wilson line breaking \cite{Donagi:2008kj, Marsano:2012yc}.  In the presence of additional $U(1)$ symmetries, hypercharge flux induces a chiral spectrum, which can be anomalous. The  MSSM-$U(1)$ mixed anomalies were determined in \cite{Dudas:2010zb, Marsano:2010sq, Dolan:2011iu,  Palti:2012dd} and form a stringent constraint on the matter spectra and associated $U(1)$ charges. It is worth noting,  that none of the models with charges in known geometric constructions  solve these anomaly constraints without introducing exotics or dangerous proton decay operators\footnote{We will determine a new class of elliptic fibrations, which do in fact solve the anomalies and suppress the couplings of dangerous operators. This will be discussed in section \ref{sec:TwoU1Geo}. However these models are not amenable for an FN-type generation of flavor textures.}. However in the F-theory charge set obtained in \cite{Lawrie:2015hia} we do find solutions, including models with realistic flavor physics. One of the goals of this paper is to identify these phenomenologically sound models, provide the corresponding charge patterns as well as fiber types, and thereby give guidance towards their geometric construction.  }

Before diving into a summary of the results of our analysis, we begin with a brief overview of F-theory phenomenology, in particular in view of flavor physics, which will play a key role in our analysis. 
The most promising particle physics results were thus far obtained in local F-theory GUTs and their associated spectral cover models, i.e. 4d supersymmetric GUTs obtained from compactifications of the 7-brane effective theory on a four-cycle, that is embedded in a Calabi-Yau four-fold. Proton decay was studied in the context of local spectral cover models in \cite{Marsano:2009gv, Marsano:2009wr, Dudas:2010zb, Hayashi:2010zp, Dolan:2011iu, Dolan:2011aq} 
The anomalies of \cite{Dudas:2010zb, Marsano:2010sq, Dolan:2011iu,  Palti:2012dd} in conjunction with constraints on proton decay operators were surveyed in \cite{Dolan:2011iu, Dolan:2011aq} and in particular it was shown that in local spectral cover models, the anomalies were in conflict with $U(1)$ symmetries required for suppression of dimension five proton decay operators. The only way to consistently combine these two effects was to allow for exotics.

Flavor in local F-theory models has a long history starting with the initial exciting insight that the top Yukawa coupling is generated at order one at a local $E_6$ enhancement point \cite{Donagi:2008ca, Beasley:2008dc, Beasley:2008kw, Hayashi:2009ge} and further more refined developments regarding corrections to the leading order Yukawa  matrices \cite{Heckman:2008qa, Font:2008id, Tatar:2009jk, Bouchard:2009bu, Marchesano:2009rz, Hayashi:2009bt, Conlon:2009qq, Cecotti:2009zf,King:2010mq, Leontaris:2010zd, Camara:2011nj, Ludeling:2011en, Aparicio:2011jx, Font:2012wq, Font:2013ida, Marchesano:2015dfa}, see \cite{Heckman:2010bq, Maharana:2012tu} for reviews of various particle physics implications of F-theory models. 
Local flavor models  have undergone various stages of accurateness. The present status is that  world-volume gauge fluxes do not lead to any corrections at all, but non-commutative fluxes in combination with non-perturbative effects can potentially give rise to suitable corrections. 
Froggatt-Nielsen models in local F-theory models were studied comprehensively in \cite{Dudas:2009hu}, however it was shown that unless one imposes by hand an R-parity, the local models universally suffer from regeneration of dangerous couplings. 
Clearly, global constraints, such as the type of $U(1)$ charges, fluxes and most likely the base geometry provide an additional set of constraints. The local models, by now are understood to be incomplete in that they do not seem to give rise to all possible $U(1)$ symmetries that can be constructed globally -- this holds true for the geometrically realized charges, and even more so for the charge classification in \cite{Lawrie:2015hia}. This leads then to the question whether global models can more successfully implement these phenomenological constraints, and whether there are any distinct features in such models. 

Phenomenological studies of global models have been rather scarce. The toric top-models were shown not to give rise to appealing flavor models and a stable proton \cite{Krippendorf:2014xba}. {As an alternative to GUTs, recent work has considered direct construction of the MSSM in F-theory \cite{Lin:2014qga, Grassi:2014zxa, Cvetic:2015txa}, which however requires further careful analysis of the phenomenology. }
In this paper we will assess the question of phenomenological implications of the $U(1)$ symmetries in F-theory based on the 
\"uber-set obtained in \cite{Lawrie:2015hia},  in conjunction with consistency requirements such as anomalies, and provide some insights into how to construct the relevant geometries.

\subsection*{Overview of Results and Search Strategy}

To give the reader an overview of the results, we now summarize our framework and constraints, and provide pointers to where these are found in the main text of the paper. The setups we consider are 
 $SU(5)$ GUTs with hypercharge flux GUT breaking in F-theory compactification on an elliptically fibered Calabi-Yau four-fold.
In addition, the following consistency requirements are imposed:
\begin{enumerate}
\item[1.] Exact MSSM spectrum and absence of anomalies (A1.)$-$(A5.) listed in (\ref{A1MSSM})$-$(\ref{A5Higgs}).
\item[2.] $U(1)$ charges within the classification of \cite{Lawrie:2015hia} as summarized in (\ref{FCharges}).
\item[3.] $U(1)$ symmetries forbid all couplings (C1.)$-$(C7.) listed in (\ref{C1})$-$(\ref{C7}).
\item[4.] $U(1)$ symmetries compatible with one generation top Yukawa coupling.
\item[5.] Froggatt-Nielsen (FN) mechanism to generate remaining Yukawa textures for both quarks and leptons, by giving vevs to $U(1)$-charged GUT singlets without getting in conflict with the constraints (C1.)$-$(C7.).
\end{enumerate}
A more detailed exposition of these conditions can be found in section \ref{sec:Const}. The survey is organized by number of $U(1)$ symmetries, number of ${\bf 10}$ and ${\bf \bar{5}}$ representations, $\mathcal{N}_{\bf 10}$ and $\mathcal{N}_{\bf \bar{5}}$, respectively.
{The models with a single $U(1)$ generically do not allow for very interesting flavor physics, without further input, such as non-perturbative effects,  going beyond an FN-type mechanism. }For $\mathcal{N}_{\bf 10}=1$ there is exactly one solution, which satisfies all anomaly and (C1.)--(C7.) constraints, given by I.1.4.a in table \ref{tab:I.1.4}. All other models for any $\mathcal{N}_{\bf \bar{5}}$ regenerate dangerous couplings at the same order as Yukawa couplings, or include exotics (for high enough number of matter multiplets).

Models with two additional $U(1)$ symmetries allow for a more interesting solution space. We find a large set of solutions to the constraints, and 
and focus on two subclasses: either the models  satisfy conditions 1.$-$5., or they satisfy 1.$-$4., but have a geometric realization. 
The models satisfying 1.$-$5., which will be referred to as {\it F-theoretic FN-models},  are discussed in section \ref{sec:flavor}, and their spectra are summarized in  tables \ref{tab:E8} and  \ref{tab:NewTexture4}. These models generate known Yukawa textures for the quarks, and furthermore provide realistic lepton and neutrino sectors. The matter charges of these solutions are within the set of  F-theory $U(1)$ charges,  however we do not yet know of an explicit construction.  Nevertheless, to guide such geometric endeavours, we summarize  the fiber types of these models in section \ref{sec:FibsFN}.

The second subclass of two $U(1)$ models satisfy 1.$-$4., but not $5.$, i.e. do not allow for a realistic FN-mechanism. However, 
they have the advantage that we can construct the corresponding geometries:
\begin{enumerate}
\item[$\widetilde{5}.$] Geometric construction in terms explicit realization of the elliptic fiber. 
\end{enumerate}
The existence of such global solutions to the anomalies and constraints on couplings is in stark contrast to local models, where there are no solutions satisfying all the conditions 1.$-$4. (with 2. modified to mean local spectral cover $U(1)$s). This class of global models are discussed in section \ref{sec:TwoU1sGeo} and their geometric realization is given in section \ref{sec:TwoU1Geo}.

%
%
%
%
%
%
%
%


\section{Constraints}
\label{sec:Const}
This section provides an overview of all the constraints, and outlines the scope and strategy of our search.
The setup in the following will be $SU(5)$ supersymmetric GUTs, with additional $U(1)$ symmetries with a realization in F-theory compactifications on Calabi-Yau four-folds. 

The first type of conditions arise from basic consistency of the 4d effective theories, namely an {\it exotic-free} MSSM spectrum and superpotential couplings, as well as absence of dangerous couplings that render the models inconsistent, which arise for instance through proton decay and R-parity violation. Throughout this paper we will impose that suppressions of couplings will be administered through additional $U(1)$ symmetries, which will be one of the F-theoretic inputs into the models. 
Additional phenomenological requirements arise from flavor constraints. There is somewhat more flexibility in how the flavor hierarchies are engineered, and we will do a systematic analysis including flavor considerations using Froggatt-Nielsen type models in section \ref{sec:flavor}.

The second type of constraints are specific to the class of theories, namely GUTs with a UV completion within F-theory. 
Here, one class of constraints arise from the GUT breaking, which in F-theory can be realized in terms of hypercharge flux breaking, i.e. non-trivial flux in the direction of the $U(1)_Y$ \cite{Donagi:2008kj,Beasley:2008kw}. In addition to imposing geometric conditions on the class of this background flux\footnote{The requirement is that it is topologically trivial as a two-form in the Calabi-Yau, but non-trivial on the 4-cycle that realizes the GUT theory.
Examples of geometries realizing such classes are known 
see e.g. \cite{Marsano:2009ym, Blumenhagen:2009yv}. However constructions of the hypercharge flux in terms of an M-theory $G_4$ flux is thus far been elusive, although recent progress was made in \cite{Braun:2014pva} for the $U(1)$-restricted Weierstrass model of \cite{Grimm:2010ez}. Extending this work to models with rational sections would be of vital importance. }, if the model has in addition abelian symmetries,  the mixed MSSM-$U(1)$ anomalies need to be cancelled \cite{Dudas:2010zb, Marsano:2010sq, Dolan:2011iu,  Palti:2012dd}.The second class of F-theoretic constraints is the type of $U(1)$ symmetries. In the recent work \cite{Lawrie:2015hia}, constraints on these have been determined. The combination of F-theoretic $U(1)$ charges and the hypercharge flux induced anomalies result in additional constraints on the possible $U(1)$ charges and distributions of the matter fields. 
In the following we will discuss both classes of constraints  in detail.


\subsection{MSSM Spectrum and Anomalies}

We consider $N=1$ supersymmetric GUTs  with $SU(5)$ gauge group and matter in the ${\bf 10}$ and ${\bf \bar{5}}$ representation.
The Higgs doublets of the MSSM arise from fundamental and anti-fundamental representations of the $SU(5)$. 
In F-theory the GUT multiplets are geometrically localized on complex curves, so-called matter curves inside a 4-cycle $S_{\text{GUT}}$, which is wrapped by 7-branes in F-theory. The low energy theory on the 7-brane realizes the gauge degrees of freedom.  Chirality is induced by $G_4$-flux, and will be labeled by $M_a$ and $M_i$ for ${\bf 10}$ and ${\bf \bar{5}}$ matter. GUT breaking is achieved by non-trivial flux in the $U(1)_Y$ direction, $\langle F_Y \rangle$. This lifts both the XY bosons of the gauge group $SU(5)$, as well as ensures that the Higgs triplets are massive. 
The restrictions of the hypercharge flux on the ${\bf 10}$ and ${\bf \bar{5}}$ matter curves will be referred to in terms of integers $N_a$ and $N_i$, respectively. 

In summary the matter content of the $SU(5)$ GUT, with $M$ chiral generations  and restriction of hypercharge flux $N$ is parameterized as follows:
\be 
{\renewcommand{\arraystretch}{1.5}
\begin{array}{c|c|c|c}
\text{$SU(5)$ representation} & \text{MSSM representation} & \text{Particle} & \text{Chirality} \\ \hline\hline
 & ({\bf 3}, {\bf 2})_{1/6} & Q & M_a \\
{\bf 10}_a & (\bar{\bf 3}, {\bf 1})_{-2/3} & \bar{u} & M_a - N_a \\
 & ({\bf 1}, {\bf 1})_{1} & \bar{e} & M_a + N_a \\\hline
 \multirow{2}{*}{$\obfive_i$} & (\bar{\bf 3}, {\bf 1})_{1/3} & \bar{d} & M_i \\
 & (\bar{\bf 1}, {\bf 2})_{-1/2}  & L & M_i + N_i \\
\end{array}
}
\ee
The integers $M$ and $N$ have to satisfy basic requirements of realizing the exact MSSM spectrum. In this paper we will in particular impose 
that the spectra are free from exotics. In addition to placing constraints on the values of $M$ and $N$, the absence of exotics places a bound on the number $\mathcal{N}$ of distinctly charged $\bten$ and $\obfive$,
\be \ba 
\mathcal{N}_{\bten} &\leq 3 \\
\mathcal{N}_{\obfive} &\leq 8 \,.
\ea \ee
To derive these bounds, note that if we were to consider more than three $\bten$s then some of these must have $M_a = 0$ as there are only three generations of left-handed quarks. Allowing a non-zero restriction of hypercharge flux over these allows the presence of either a right-handed quark or lepton with the wrong chirality for the MSSM spectrum, which results in the presence of exotics. Likewise, the maximum number of $\obfive$s is given by the sum of three generations of left-handed leptons and right-handed quarks, in addition to $H_u$ and $H_d$.

In addition to the GUT gauge symmetry, we require additional abelian gauge factors, $U(1)_{\alpha}$, $\alpha= 1, \ldots, A$, under which the $SU(5)$ representations  ${\bf 10}_a$ and $\bar{\bf 5}_i$ {carry} charges $q^{\alpha}_a$ and $q^{\alpha}_i$, respectively.  
The type of $U(1)$ charges are determined in terms of the F-theory geometry and will be the subject of section \ref{sec:FU1}.
The combined system of $F_Y$ hypercharge flux breaking and additional $U(1)$ symmetries implies that there can potentially be mixed MSSM-$U(1)$ anomalies. 

Anomaly cancellation and the requirement of three generations imply the following set of constraints on the chiralities $M$, hypercharge flux restriction $N$ and charges $q^\alpha$ -- all sums $\sum_{a=1}^{\mathcal{N}_{\bf {10}}}$ are over ${\bf 10}$ representations, $\sum_{i=1}^{\mathcal{N}_{\bf \bar{5}}}$ over ${\bar{\bf 5}}$s, with $\mathcal{N}_{\bf R}$ corresponding to the number of matter multiplets in the representation ${\bf R}$ with distinct $U(1)$ charge:
\begin{itemize}
\item[(A1.)] MSSM anomalies
\be \label{A1MSSM}
\ba
\sum_i M_i &= \sum_a M_a \,.
\ea
\ee
\item[(A2.)]  Mixed $U(1)_Y$-MSSM anomalies \cite{Dudas:2010zb, Marsano:2010sq, Dolan:2011iu}
\be
\sum_i q_i^\alpha N_i +  \sum_a q_a^\alpha N_a =0 \,,\qquad \alpha = 1, \ldots, A\,.
\ee
\item[(A3.)] Mixed  $U(1)_Y$-$U(1)_\alpha$-$U(1)_\beta$ anomalies \cite{Palti:2012dd}
\be
3 \sum_a q_a^\alpha q_a^\beta N_a +  \sum_i q_i^\alpha q_i^\beta N_i  =0\,,\qquad \alpha, \beta = 1, \ldots, A\,.
\ee 
\item[(A4.)] Three generations of quarks and leptons:
\be
\sum_a M_a = \sum_i M_i =3  \,.
\ee
\item[(A5.)] Absence of exotics:
\be
\sum_a N_a =\sum_i N_i =0 \,.
\ee
\item[(A5.)] One pair of Higgs doublets:
\be\label{A5Higgs}
\sum_i |M_i +  N_i| =5 \,.
\ee
\end{itemize}
The set of constraints (A1.)$-$(A5.) will be strictly imposed on any model, as a minimal requirement for realistic 4d physics. 
Note that we have not as yet imposed any Yukawa couplings -- which Yukawas will be required to be compatible with the $U(1)$ charges will be discussed in section \ref{sec:Yuks}. 
We now turn to additional conditions on the $U(1)$ charges, on top of the anomaly constraints, which will ensure absence of 
dangerous couplings, such as proton decay.


\subsection{Proton Decay, $\mu$-Term and R-parity Violation}

Rapid proton decay and R-parity violation (RPV) can cause supersymmetric  GUTs to become phenomenologically unfit. 
In this paper we will utilize $U(1)$ symmetries to forbid these couplings. The $U(1)$ symmetries are broken, at a higher scale
and for some of these couplings we will require that they are not regenerated, e.g. by giving vevs to $U(1)$-charged singlets.

\subsubsection{Summary of Dangerous Couplings}

Let us first summarize the various problematic couplings
and then discuss the bounds  on their suppression -- $i, j,\cdots$ and $a, b, \cdots$ label matter representations, whereas $I, J, \cdots= 1, 2, 3$ and $A, B, \cdots =1, 2, 3$ label 
generations:
\begin{itemize}
\item[(C1.)] $\mu$-term: 
\be\label{C1}
\mu {\bf 5}_{H_u} \bar{\bf 5}_{H_d} \ee
\item[(C2.)] Dimension five proton decay: 
\be
 \delta^{(5)}_{abci} {\bf 10}_a {\bf 10}_b  {\bf 10}_c {\obfive}_i   
\ee
\item[(C3.)] Bilinear lepton number violating superpotential coupling: 
\be
\beta_i \bar {\bf 5}_i {\bf 5}_{H_u} \supset \beta_I L_I H_u    
\ee
\item[(C4.)] Dimension four proton decay:
\be
\lambda^{(4)}_{ija} \bar{\bf 5}_i   \bar{\bf 5}_j {\bf 10}_a    
\ee
\item[(C5.)] Tri-linear lepton number violating K\"ahler potential couplings:
\be
\kappa_{abi} {\bf 10}_a {\bf 10}_b \bar{\bf 5}^\dagger_i  \supset \kappa_{ABI} Q_A \bar{u}_B L_I^\dagger
\ee
\item[(C6.)] Dimension five lepton violating superpotential coupling:
\be
\gamma_i  \bar{\bf 5}_i   \bar{\bf 5}_{H_d} {\bf 5}_{H_u} {\bf 5}_{H_u} \supset  \gamma_I L_I H_d H_u H_u 
\ee
\item[(C7.)] Dimension five lepton violating  K\"ahler potential coupling:
\be\label{C7}
\rho_a \bar{\bf 5}_{H_d} {\bf 5}^\dagger_{H_u}  {\bf 10}_a  \supset  \rho_A {H_d} H_u^\dagger  \bar{e}_A 
\ee
\end{itemize}
We require these to be absent at leading order. Furthermore, if a Yukawa matrix element is generated by a singlet vev, we require that 
these operators do not re-appear with the same singlet suppression. In the case that multiple singlet vevs are required to generate a certain forbidden coupling, 
we study in detail whether the suppression is within the bounds that we summarize below. This occurs frequently in our analysis for dimension four and 
five proton decay operators. 

Note that, if the top and bottom Yukawas are generated for all generations, then 
compatibility of the $U(1)$ symmetries with the Yukawas as well as absence of the $\mu$-term (C1.) implies (C2.) with opposite sign. 
However, this needs to be checked in addition, if not all Yukawas are generated perturbatively, as in most of the following models. 

Imposing one top Yukawa coupling (for at least one generation, see (\ref{Ytop})), as well as the absence of (C5.) implies that 
there cannot be  ${\bf \bar{5}}$ matter on the same curve as $H_u$
i.e. 
\be\label{Conu}
Y^t,\  (C5.) \qquad \Rightarrow \qquad   M_{i} =0 \,, \quad N_i =- 1 \qquad i = H_u  \,.
\ee
Likewise, imposing that the bottom Yukawa couplings are realized (either at leading order or regenerated by singlet vevs, see (\ref{DownTypeYukawa})), as well as the absence of the coupling 
(C4.) implies that there cannot be ${\bf \bar{5}}$ matter on the same curve as ${H_d}$, i.e.
\be\label{Cond}
Y^b,\  \hbox{(C4.)} \qquad \Rightarrow \qquad  M_{i} =0 \,, \quad N_i = 1 \qquad i = H_d \,.
\ee


\subsubsection{$\mu$-Term}

The $\mu$-term  is a supersymmetric Higgsino mass term. Radiative electroweak supersymmetry without much tuning in the MSSM requires $\mu$ to be around $O(100)$GeV. If this coupling is generated at tree-level, this cannot be achieved without a fair amount of fine-tuning and low-energy supersymmetry does not address the hierarchy problem. One way to avoid this problem is to forbid the $\mu$-term at the high scale with a $U(1)$ symmetry -- a so-called PQ $U(1)$ symmetry, i.e. the charge of the $H_u$ and $H_d$ do not add up to zero. The $\mu$-term can then be generated by a coupling to a charged singlet $S$ (or products of singlets) either via the superpotential or the K\"ahler potential. Concretely the $\mu$-term can for instance be generated as follows 
\be
{S^\dagger \over \Lambda}  H_u H_d \,, 
\ee
where the ${\langle F_S \rangle \over \Lambda}$ then generically sets the scale of the $\mu$-term, which is the  Giudice-Masiero mechanism \cite{Giudice:1988yz}. This type of $\mu$-term has wide application in gravity \cite{Kim:1983dt,Giudice:1988yz} but also in gauge mediated supersymmetry breaking models, see for instance also in F-theory \cite{Marsano:2008jq}. 
We shall not discuss the specific mechanisms of supersymmetry breaking in this paper, as these are highly model dependent, thus deviating from the goal that we set out here, to be as comprehensive and general as possible. For our purposes we will impose that the coupling (C1.) is absent at tree-level. 


\subsubsection{ Dimension 4 Proton Decay Operators}

Dimensions four and five proton decay operators are highly constrained in GUT models, and one of the requirements in our search is the compatibility of the models with the bound on the lifetime of the proton given by $\tau_p \geq 10^{34}$ years \cite{pdg}. 
The dimension four proton decay operators originate in the coupling 
\be  
\lambda_{ija}^{(4)}\obfive_{i}\obfive_{j}\mathbf{10}_{a}, \label{eq:dim4pdsu5} 
\ee
where the $\obfive_i$ and $\bten_a$ denote matter representations.
This operator results in the following couplings,
\be 
\lambda^0_{IJA}L_I L_J \bar{e}_A + \lambda^1_{IJA}\bar{d}_I L_J Q_A + \lambda^2_{IJA} \bar{d}_I\bar{d}_J\bar{u}_A \,,\label{eq:dim4pd}
\ee 
where $I,J,A$ label the generation index. Dimension four proton decay occurs via interactions involving products of $\lambda^i$, the main decay channel being $p \rightarrow \pi^0 e^+$ \cite{BaerTata} which involves the product $\lambda^1 \lambda^2$. If both operators with couplings $\lambda^1$ and $\lambda^2$ are present this results in very fast proton decay. Proton lifetime results in the following bounds on the coupling constants {for the lightest generation}\cite{Hinchliffe:1992ad}
\begin{align} 
\sqrt{\lambda^1 \lambda^2} & \leq \left(\frac{M_{SUSY}}{\text{TeV}}\right) 10^{-12} \,.
\end{align}
Here $M_{SUSY}$ is the mass scale of the supersymmetric particles entering the process, and will be taken of $O(1)$ TeV.
Bounds on the other generation couplings for GUT models are discussed in \cite{Smirnov:1995ey} and an up to date summary of all bounds can be found in section 6.5 of \cite{Barbier:2004ez} from indirect searches, and section 7 from direct searches.  In particular for $\lambda^0$ there are bounds which are much weaker $\sim 10^{-5}$, cf. (6.100), and for the other components of $\lambda^1 \lambda^2$ see (6.110) in \cite{Barbier:2004ez}.
These operators violate baryon or lepton number and thereby R-parity. In this analysis we require the $U(1)$ symmetry forbid all operators of this type, and furthermore that singlets do not regenerate them outside of the bounds listed above.


\subsubsection{Dimension 5 Proton Decay Operators}

\begin{figure}
  \centering
  \includegraphics[width= 10cm]{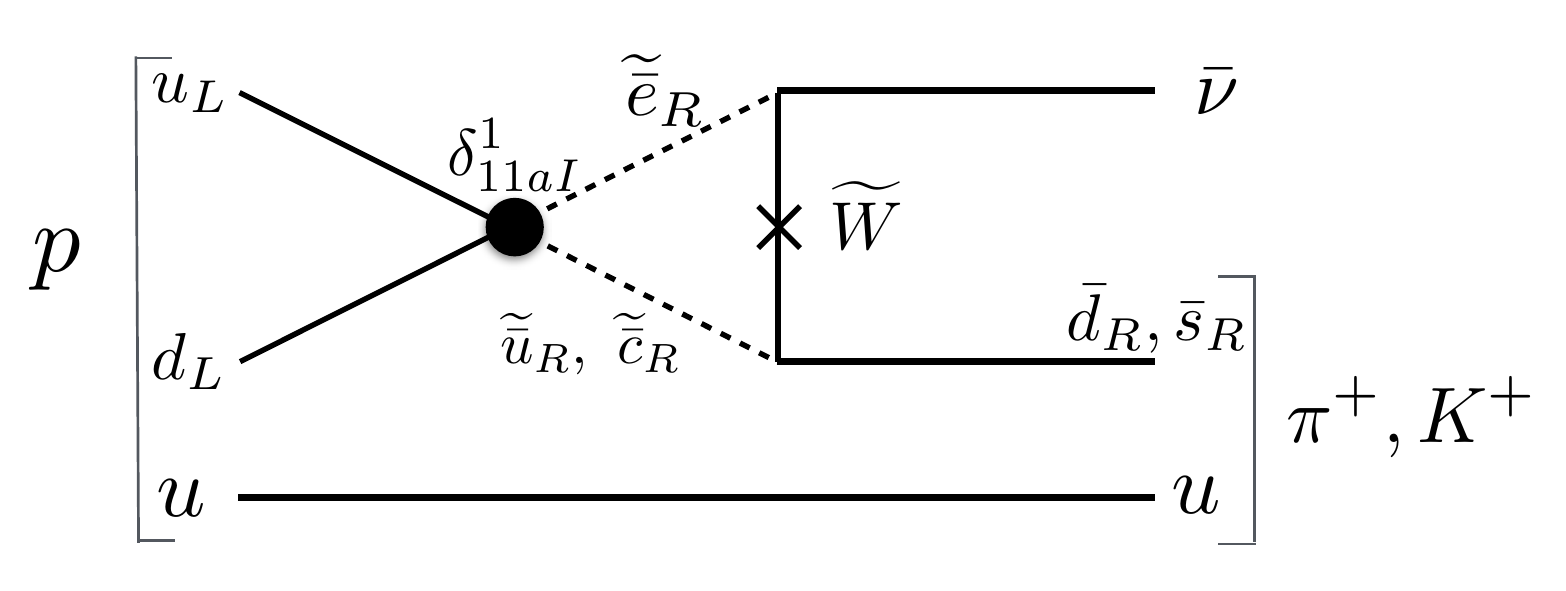}
  \caption{Process giving rise to dimension five proton decay parameterized by $\delta^1_{11aI}$.}
    \label{fig:dim5}
\end{figure}

The main contribution to proton decay from dimension five operators occurs through the coupling
\be
\delta^{(5)}_{abci}\mathbf{10}_a \mathbf{10}_b\mathbf{10}_c \obfive_i \,,\label{eq:dim5pdsu5}
\ee
which gives rise to the operators
\be
\delta^1_{ABCI}Q_AQ_BQ_CL_I + \delta^2_{ABCI}\bar{u}_A \bar{u}_B\bar{e}_C \bar{d}_I + \delta^3_{ABCI}Q_A \bar{u}_B\bar{e}_C L_I. \label{eq:dim5pd}
\ee
The bound on the coupling constant due to the interaction involving $\delta^1$ is \cite{Hinchliffe:1992ad}
\be
\delta^1_{112I} \leq 16\pi^2 \left(\frac{M_{SUSY}}{M_{GUT}^2}\right) \qquad I = 1,2,3\,,
\label{dim5bound}
\ee
where the relevant diagram is shown in figure \ref{fig:dim5}. The mass $M_{SUSY}$ is set by the mass of the sfermions contributing to the loop diagram.
The operators involving other (s)quark generations are suppressed with appropriate {flavor insertions}, i.e. at least the appropriate CKM elements have to be inserted. This ameliorates the bounds, in particular for operators involving the third generation.


\subsubsection{Remaining B/L violating operators}

The remaining couplings are also constrained in particular from limits on flavor changing processes, see \cite{Ibanez:1991pr} and for a review \cite{Barbier:2004ez} . 
The bilinear coupling (C3.) violates lepton number and leads to a mixing between the Higgs and lepton sectors. 
At tree-level we will forbid this coupling, however in section \ref{sec:leptonmixing} use it to generate neutrino masses.
The couplings (C5.), (C6.) and (C7.) violate either lepton or baryon number, and thus contribute to proton decay in combination with the other B/L violating operators.


\subsection{Flavor Constraints and FN-models}
\label{sec:Yuks}
The assignment of the $U(1)$ charges of matter must be such that it allows for a top Yukawa coupling for the third generation, which amounts to requiring at least one charge neutral coupling of the form
\be\label{Ytop}
Y^t_{ab}: \qquad \lambda^t_{ab} {\bf 10}_a  {\bf 10}_b  {\bf 5}_{H_u} \supset Y^u_{AB} Q_A \bar{u}_B H_u \,,
\ee
where $A,B$ label the quark generations. As the mass of the bottom quark is suppressed with respect to the mass of the top, a bottom Yukawa coupling
\be
Y^b_{ai}: \qquad \lambda^b_{ai}  {\bf 10}_a  \bar{\bf 5}_i \bar{\bf 5}_{H_d} \supset Y^d_{AI}Q_A \bar{d}_I H_d + Y^L_{IA} L_I \bar{e}_A H_d \,,
\label{DownTypeYukawa}
\ee
is not necessarily imposed at leading order. Both cases of rank 0 and rank 1 bottom Yukawa matrices at first order are studied. The up-type and down-type Yukawa matrices, $Y^u$ and $Y^d$ are the matrices formed from the couplings $Y^u_{AB}$ and $Y^d_{AI}$, respectively after the distribution of MSSM matter has been assigned to the $\bten$ and $\obfive$ representations.

In the present context we will apply a FN-type mechanism \cite{Froggatt:1978nt} to generate the remaining Yukawa matrix entries, i.e. {the} $U(1)$-charged couplings are generated by giving suitably charged singlets a vev. 
In a generic F-theory models, the couplings between conjugate fields are always geometrically generated, i.e. the singlets required for all possible couplings of the form
\be \ba
& {\bf 1}\, \obfive_i \bfive_j \cr
& {\bf 1} \, \overline{\bten}_m \bten_n\,,
\ea \ee
are always present. Giving these singlets a vacuum expectation value breaks the $U(1)$ symmetry under which the singlet is charged, and generates the remaining Yukawa couplings. Whether or not such a vacuum expectation value can indeed be obtained, is a question of moduli stabilisation, which is beyond the scope of this paper. For a singlet $S$ of charge $q$ a coupling, with charge $n q$, is regenerated with suppression
\be 
s^n = \left(\frac{\langle S \rangle}{M_{GUT}}\right)^n \,.
\ee

Experimentally masses, mixing angles and CP violation are measured at low energies compared to the UV scale at which we are calculating (see PDG flavor reviews \cite{pdg} for the latest experimental summary). To compare UV models of Yukawa couplings to low-energy data, one needs to appropriately renormalize the couplings via the RG equations. This evolution allows for additional effects that can explain the flavor structure at low energies. For example flavor-violating effects from soft supersymmetry breaking can give large contributions to flavor observables, in fact they could even generate the entire flavor structure \cite{Weinberg:1972ws,Lahanas:1982et}. 

However, the question here is different, namely, can the pattern that we can obtain from the additional $U(1)$s account for the entire flavor physics, i.e. with minimal RG evolution effects. This limit can be achieved when large flavor violating effects are absent in the soft-terms and canonical kinetic terms are present. In this context the RG evolution of quark and lepton masses as well as mixing parameters to high energies, e.g. the GUT scale at around $10^{16}$ GeV, has been performed (see for instance \cite{Ross:2007az}). Roughly speaking one observes with the above assumptions that the quark mixing parameters and masses do not run very much. To first approximation, we hence aim at obtaining the following mass ratios and mixing angles in the quark and lepton sector \cite{Ramond:1993kv,Nir:1995bu}:
\be
\ba\label{MassRatios} 
m_t:m_c:m_u &\sim 1:\epsilon^4:\epsilon^8\cr 
m_b:m_s:m_d &\sim 1:\epsilon^2:\epsilon^4\cr 
m_\tau:m_\mu:m_e &\sim 1:\epsilon^2:\epsilon^{4,5}  \,,
\ea
\ee
 and quark mixing angles 
\be
 \theta_{12} \sim \epsilon\,, \quad \theta_{23}\sim \epsilon^2\,, \quad \theta_{31} \sim \epsilon^3\,,
\ee
where $\epsilon \approx 0.22$ is the Wolfenstein parameter \cite{Wolfenstein:1983yz}.
We do not determine the ratio 
$ \frac{m_b}{m_t}=\epsilon^x \tan^{-1}{\beta}$
as this is part of a full-fledged supersymmetry breaking model, which is not part of our analysis. Furthermore, we do not discuss CP violation here as the $U(1)$ symmetries used for obtaining the hierarchical scaling do not constrain the complex phases of the singlet insertions. 

This experimentally constrained structure of masses and flavor mixings does not fix the entire structure in the Yukawa matrices. 
There are various popular models for this such as \cite{Ramond:1993kv,Ibanez:1994ig, Dreiner:2003hw}. More systematically, by focusing on generating all hierarchies with one singlet, one can classify all viable textures for the quark masses \cite{Dreiner:2003hw}. In the present context, the only model in this classification, which is consistent with $SU(5)$, is the following hierarchical scaling of the Yukawa matrices first obtained by Haba in \cite{Haba:1998wf} 
\be
\label{HabaText}
Y_{\rm Haba}^u\sim\left(
\begin{array}{c c c}
\epsilon^8 & \epsilon^6 &\epsilon^4\\
\epsilon^6 & \epsilon^4 &\epsilon^2\\
\epsilon^4 & \epsilon^2 & 1\\
\end{array}
\right)\ , \qquad 
 Y_{\rm Haba}^d\sim\left(
\begin{array}{c c c}
\epsilon^4 & \epsilon^4 &\epsilon^4\\
\epsilon^2 & \epsilon^2 &\epsilon^2\\
1 & 1 & 1\\
\end{array}
\right) \,.
\ee
Another texture which will be shown to be consistent with the F-theoretic setup was already obtained by Babu, Enkhbat  and Gogoladze (BaEnGo) in \cite{Babu:2003zz}
and is given by 
\be\label{BaEnGoText}
Y_{\rm BaEnGo}^u \sim\left(\begin{array}{ccc}
\epsilon^8 & \epsilon^6 & \epsilon^4\\
\epsilon^6 & \epsilon^4 & \epsilon^2\\
\epsilon^4 & \epsilon^2 & 1
\end{array}\right) \,,\qquad 
Y_{\rm BaEnGo}^d \sim\left(\begin{array}{ccc}
\epsilon^5 & \epsilon^4 & \epsilon^4\\
\epsilon^3 & \epsilon^2 & \epsilon^2\\
\epsilon & 1 & 1
\end{array}\right)\,.
\ee
The $U(1)$ symmetries and associated singlet vevs  generate these hierarchies in the couplings, but do not  predict the exact values for the masses. These are obtained by ${O}(1)$ numbers in front of each coupling, whose string theoretic origin can for instance be non-canonical contributions to the kinetic terms. The couplings do not only depend on the singlets but also on uncharged complex structure moduli. 
In practice we will determine $O(1)$ numbers which generate the experimentally favored values, in particular for the lepton sectors, which will be discussed in section \ref{sec:leptonmixing}.

A detailed analysis of the flavor constraints in the context of Froggatt-Nielsen models will be done in section \ref{sec:flavor}. We find F-theoretic models consistent with the above two hierarchies as in \cite{Haba:1998wf} and \cite{Babu:2003zz}. In appendix \ref{app:OtherTextures} we consider other known textures \cite{Ross:2007az, Dudas:2009hu} and show that it is not possible to find F-theoretic charges, which solve the anomaly cancellation conditions and generate the required quark Yukawa matrices.


\subsection{F-theory $U(1)$s and the Mordell-Weil group}
\label{sec:FU1}

The key input from F-theory -- apart from the anomalies -- is the set of possible $U(1)$ charges for matter fields as determined in \cite{Lawrie:2015hia}.
Much recent progress in F-theory model building has resulted in constructions of examples of global elliptic Calabi-Yau four-folds, which realize both, GUT gauge groups in terms of singularities in the elliptic fiber, as well as additional abelian gauge symmetries (see introduction for a list of references). Abelian gauge symmetries are constructed in terms of so-called rational sections, which are maps from the base of the fibration back to the fiber \cite{Morrison:1996pp}. None of the explicit algebraic realizations, however, resulted in a complete classification of the possible $U(1)$ symmetries. 
The collection of rational sections form a finitely generated abelian group (under the elliptic curve group law), called the Mordell-Weil group, which is isomorphic to $\mathbb{Z}^r \oplus \Gamma$, where $\Gamma$ is the torsional part, which will not play a role in the current discussion. If the Mordell-Weil group  has rank $r$, then the resulting compactification will have $r$ additional $U(1)$ symmetries. Realizing elliptic fibrations with multiple matter curves of distinct $U(1)$ charges is technically a highly challenging enterprise. Therefore an alternative approach that would yield the charges, without necessarily constructing the corresponding geometries is highly desirable. 

\subsubsection{Models with one $U(1)$}

Such a full classification of possible $U(1)$ symmetries for $SU(5)$ was obtained in  \cite{Lawrie:2015hia}\footnote{There is an assumption, that the section is a smooth divisor in the resolved Calabi-Yau four-fold. A discussion of this particular point and potential extensions beyond that can be found in \cite{LSW}.}. 
There the starting point is not a concrete realization of the elliptic fiber, but  a more abstract approach pursued by studying the constraints on the possible $U(1)$ charges in terms of general consistency requirements between the rational sections and  codimension two fibers from the classification in \cite{Hayashi:2014kca}. This approach has the great advantage of giving rise to a super-set of $U(1)$ charges, that can be realized in F-theory, without however providing a direct geometric construction. We take this set of charges as an input for our {analysis and} show  that certain charges in this set are phenomenologically favored, as they satisfy all constraints and provide realistic flavor physics. In this way, we provide a   pointer towards which geometric constructions can yield globally consistent compactifications. 
We shall give some details on geometries of this type later in the paper in section \ref{sec:GeoRealisation}. 

\begin{figure}
  \centering
  \includegraphics[width= 7cm]{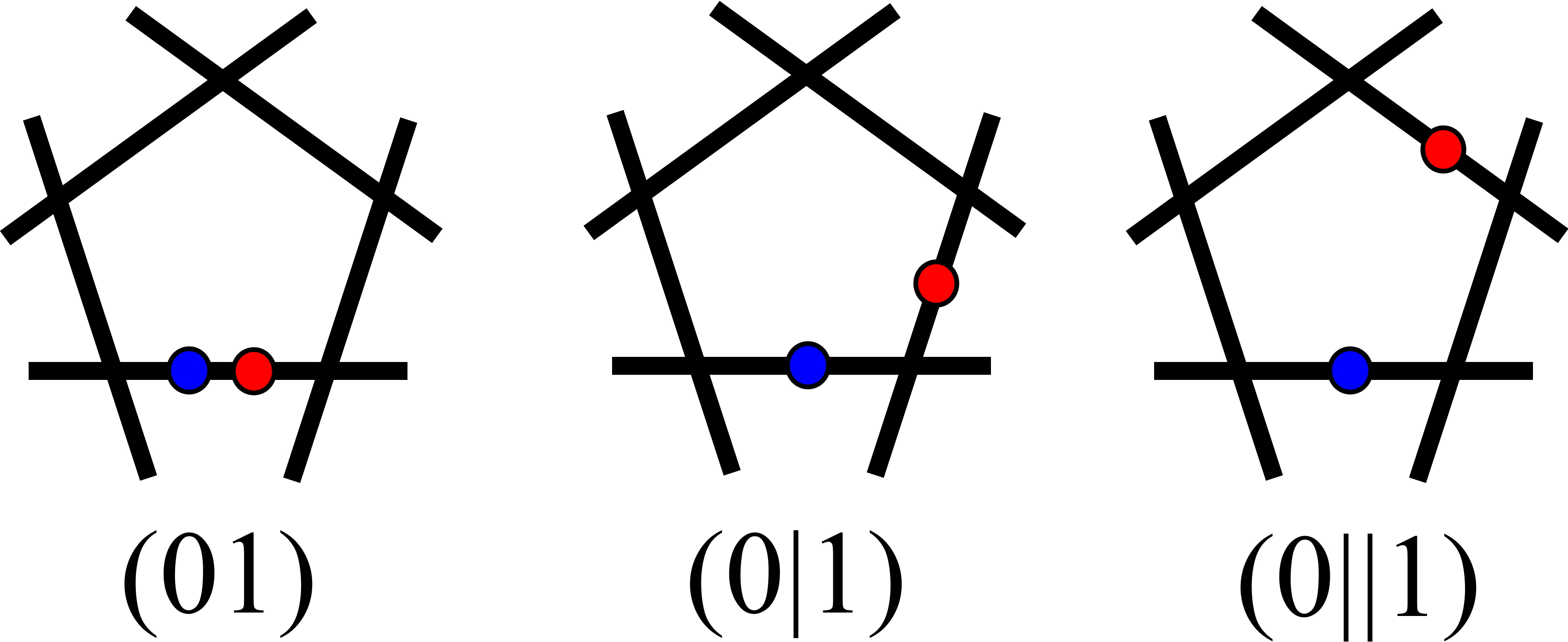}
  \caption{The configurations of fibers for an $SU(5)$ model with one $U(1)$ symmetry. The $I_5$ fiber, realized by the five black lines (corresponding to $\mathbb{P}^1$s in the fiber) gives rise to the $SU(5)$ gauge bosons, and the sections, shown as colored dots corresponding to the zero-section (blue) and the additional section (yellow), give rise to the additional abelian gauge factor.  }
    \label{fig:I5Fib}
\end{figure}

The input from the classification result in \cite{Lawrie:2015hia} for F-theory compactifications to 4d with $r$ $U(1)$ symmetries, and matter in the ${\bf 10}$  and ${\bar{\bf 5}}$ of $SU(5)$ is the set of possible charges. 
For a single $U(1)$, are three types of distinct distributions of sections in the codimension one fiber\footnote{Recall that sections can be thought of as marked points on the elliptic curve. A model that realizes an $SU(5)$ gauge theory has special, so-called singular $I_5$ fibers above a codimension one locus in the base of the fibrations. Geometrically these are a ring of five rational curves, i.e. two-sphere, which intersect in the affine $SU(5)$ Dynkin diagram -- as shown in  figure \ref{fig:I5Fib}. To describe a model with a single $U(1)$ there is a zero-section (origin of the elliptic curve) and the additional rational section, which generates the Mordell-Weil group. The codimension one fibers with rational sections are thus decorations  of the affine $SU(5)$ Dynkin diagram with two marked points, modulo trivial relabeling. These are shown in figure \ref{fig:I5Fib}.}
 -- for the reader interested solely in the model building aspects, it is sufficient to understand that there are three set of charges, labeled by 
\be\ba\label{I5fibs}
&\ I_5^{(01)} \,, \qquad   I_5^{(0|1)} \,, \qquad   I_5^{(0||1)} \,.
\ea
\ee
and are shown in figure \ref{fig:I5Fib}.
For a given codimension one distribution of sections labeled by $I_5$ with the superindex indicating the separation between the zero-section (0) and the extra section (1), it was found in \cite{Lawrie:2015hia} that, for smooth rational sections, the possible charges for $\bf 10$ and $\bf \bar{5}$ matter that can arise in $SU(5)$ F-theory models are as follows\footnote{For some studies it will be useful to rescale the models in $I_5^{(01)}$ by 5, so that a uniform treatment is possible, i.e. that the unit charges is ``normalized" to $5$.}:
\be\label{FCharges}
\ba
 I_5^{(01)} :\ & \left\{ 
 	\ba
 	q_{\bf 10}  \in &\left\{-3,-2,- 1, 0, +1,+2,+3 \right\}  \cr
  	q_{\bf \bar{5} }\in &\left\{-3,-2, - 1, 0, +1,+2, +3 \right\}
	\ea\right.\cr 
 I_5^{(0|1)} :\ & \left\{ 
 	\ba
 	q_{\bf 10}  \in &\left\{-12,-7,-2,+3,+8,+13 \right\}  \cr
  	q_{\bf \bar{5} }\in &\left\{-14, -9,-4,+1, + 6,+ 11 \right\}
	\ea\right.	\cr 
 I_5^{(0||1)} :\ & \left\{ 
 	\ba
 	 q_{\bf 10}  \in &\left\{-9,-4,+1, + 6,+11 \right\}   \cr
  	q_{\bf \bar{5} }\in &\left\{- 13,- 8, -3, +2, + 7, + 12 \right\}\,.
	\ea\right.		 
\ea
\ee
A natural question is then to determine, whether there are integral solutions for the $M$s and $N$s, such that the resulting charge assignments solve anomaly conditions (A1.)$-$(A5.) and do not give rise to proton decay. 

Finally we should comment on the $U(1)$ charges of GUT singlets, which will play a role later on in the Froggatt-Nielsen inspired
flavor construction. In F-theory $SU(5)$ GUTs, the singlets arise at the intersection of any two ${\bf \bar{5}}$ curves (as well as two 
${\bf 10}$s). Hence, we can read off the singlet charges from the difference of charges 
\be
q_{{\bf 1}_{ij}}=q_{\bf \bar{5}_i} - q_{\bf \bar{5}_j}\,,\qquad i\not= j \,,
\ee
 for each of the three codimension one models. 

Throughout the main text we will consider only these F-theory charges (\ref{FCharges}). In certain cases it is possible to use 
methods from solutions of Diophantine equations to solve the anomalies in general and we will provide these in appendix \ref{app:Mordell}.


\subsubsection{Models with two $U(1)$s}

As we will see in section \ref{sec:OneU1}, there are only very few viable solutions with one $U(1)$ symmetry. 
 To construct models with two additional $U(1)$s we need two additional rational sections, $\sigma_1$ and $\sigma_2$ in the elliptic fibration, in addition to the zero-section, $\sigma_0$. There are nine possible codimension one fiber types in this case, up to a reordering of the simple roots and exchanging the two rational sections. These are given by,
\be 
I_5^{(012)}, \quad I_5^{(01|2)}, \quad I_5^{(01||2)}, \quad I_5^{(0|12)}, \quad I_5^{(0|1|2)}, \quad I_5^{(0|1||2)}, \quad I_5^{(0|1|||2)}, \quad I_5^{(0||12)}, \quad I_5^{(0||1|2)} \,,
\label{Codim12U1sFake}
\ee
where $i = 0,1,2$ denotes the position of the section $\sigma_i$. For each codimension one fiber there is a collection of codimension two fibers, and thus charge-sets.

\begin{figure}
  \centering
  \includegraphics[width= 7cm]{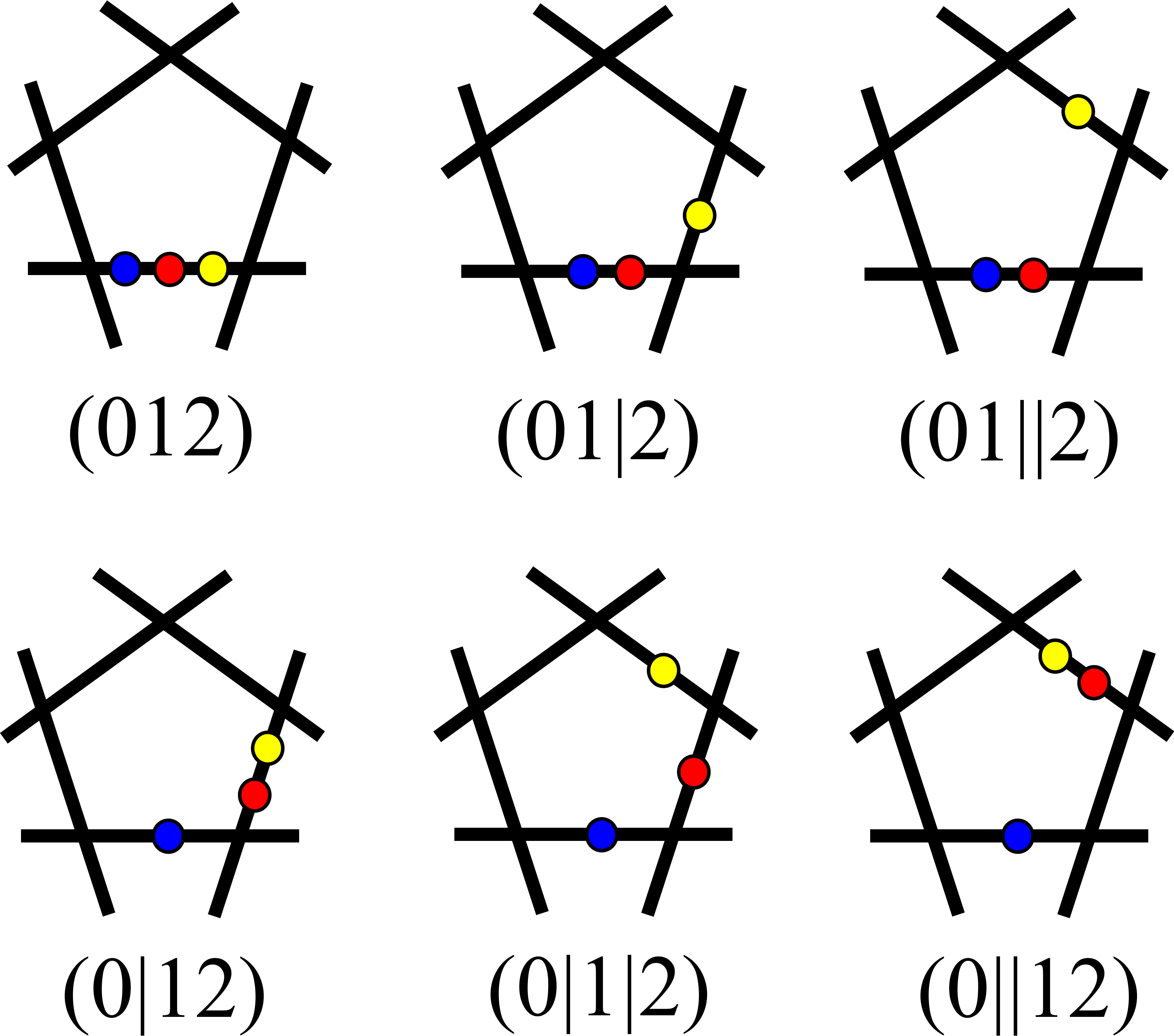}
  \caption{The configurations of fibers for an $SU(5)$ model with two $U(1)$ symmetries. The $I_5$ fiber, realized by the five black lines (corresponding to $\mathbb{P}^1$s in the fiber) gives rise to the $SU(5)$ gauge bosons, and the sections, shown as colored dots corresponding to the zero-section (blue) and the two additional sections (red, yellow), give rise to the additional abelian gauge factors.  }
    \label{fig:I5twoSec}
\end{figure}

The charges that appear in each of these codimension one fibers can be obtained from the charges in \eqref{FCharges} by noting that the charges in an $I_5^{(0|||1)}$ model are simply the negative of those in $I_5^{(0||1)}$. The same statement holds for $I_5^{(0||||1)}$ and $I_5^{(0|1)}$. The two codimension one fibers, $I_5^{(0|||1)}$ and $I_5^{(0||||1)}$, were not considered in the case of a single additional rational section as they are equivalent, under a reordering of the simple roots to $I_5^{(0||1)}$ and $I_5^{(0|1)}$, respectively. In the case of two rational sections it is not always possible to bring both of the sections into one of these forms.

From what was stated above, it is clear that not all of the codimension one fibers in \eqref{Codim12U1sFake} are distinct. For example, the charges in $I_5^{(0|12)}$ are the same as those in $I_5^{(0|1|||2)}$ if the charges under the second $U(1)$ are multiplied by $-1$. The anomaly cancellation conditions (A2.) and (A3.) are invariant under such re-scalings of the $U(1)$ charges therefore these two fibers will give rise to the same set of solutions up to the normalisation of one of the $U(1)$s. In this analysis we will consider the reduced set of codimension one fibers which give rise to distinct $U(1)$ charges given by
 \be 
I_5^{(012)}, \quad I_5^{(01|2)}, \quad I_5^{(01||2)}, \quad I_5^{(0|12)}, \quad I_5^{(0|1|2)}, \quad I_5^{(0||12)}\,.
\label{Codim12U1sReal}
\ee
These configurations are also shown in figure \ref{fig:I5twoSec}.
For these fibers, each additional rational section with the zero-section will generate a $U(1)$ with charges equal to those in \eqref{FCharges}. By taking all possible pairings between these two sets of charges one obtains the charges for a model with two additional $U(1)$ symmetries.


\section{Single $U(1)$ Models}
\label{sec:OneU1}

We begin our analysis by considering $SU(5)$ models with one additional $U(1)$ symmetry, and varying $\mathcal{N}_{\bf 10}$ and $\mathcal{N}_{\obfive}$. In summary: 
a single phenomenologically good model is found for $\mathcal{N}_{\bf 10} =1$ and $\mathcal{N}_{\bf \bar{5}} =4$, denoted I.1.4.a in  table \ref{tab:I.1.4}, where the unwanted operators are not regenerated at the same order as the remaining charged Yukawa couplings. For $\mathcal{N}_{\bf 10} =1$ and $\mathcal{N}_{\bf \bar{5}} =5$  as well as $\mathcal{N}_{\bf 10}>1$ (see appendix \ref{app:Multi10}) solutions are found, which however regenerate some dimension five proton decay operators along-side the charged Yukawas. This in itself is not problematic, as long as the suppression is high enough. However, single $U(1)$ models suffer generically from a poor  flavor structure as generated by a FN-type mechanism. Nevertheless it is interesting to note that there are solutions to the constraints within the F-theoretic $U(1)$ charges, which could be augmented with other mechanisms for generating flavor such as \cite{Marchesano:2015dfa}, to produce a phemomenologically consistent F-theory model. 

Contrary to this, models with two $U(1)$s can satisfy  the constraints from anomalies and couplings, and in addition will generate successful flavor physics via an FN-mechanism as will be discussed in section  \ref{sec:flavor}.

\subsection{$\mathcal{N}_{\bf 10} =1$}

We start the analysis with one {\bf 10} representation. Requiring one top Yukawa coupling implies that not all {\bf 10} charges listed in \eqref{FCharges} can be used. 
The charges, in each codimension one configuration, which have a top Yukawa coupling with one of the possible ${\bf 5}$ charges are 
\be \label{10ChargeswithTY}
\ba
\hbox{$U(1)$ charges of ${\bf 10}$ with ${\bf 10}_{q} {\bf 10}_{q} {\bf 5}_{-2q}$:} &\quad 
\left\{ 
\ba
 I_5^{(01)} \qquad & q_{\bten} \in \left\{- 1, 0, +1 \right\} \cr
I_5^{(0|1)} \qquad &q_{\bten} \in \left\{-7,-2,+3 \right\} \cr
I_5^{(0||1)} \qquad &q_{\bten} \in \left\{-4,+1, + 6 \right\} \,.
\ea
\right.
\cr 
\ea \ee
For the case of one $\bf 10$ representations the solutions to the anomaly equations can be parameterized as follows\footnote{Note that we give the charge of the conjugate of the up-type Higgs, i.e. $q_{H_u}$ is the charge of {\bf 5}, whereas $-q_{H_u}$ is the charge for $\bar{\bf 5}$.}
\be\label{ParaTable}
\begin{array}{c|c|c|c}
{\bf R} &  q({\bf R}) & M & N \cr\hline
\bar{\bf 5}_{H_u} &   -q_{H_u}  &0 & -1 \cr
\bar{\bf 5}_{H_d}  & -q_{H_u} + 5 w_{H_d} &0& 1 \cr
\bar{\bf 5}_{i} &  -q_{H_u} + 5w_{\bar{5}_i} &M_i &N_i  \cr\hline
{\bf 10}  &  q_{\bf 10} &  3 & 0 \cr
\end{array}
\ee
where $i = 1, \ldots, \mathcal{N}_{\bar{\bf 5}}-2$, where $\mathcal{N}_{\bar{\bf 5}}$ is the number of $\bar{\bf 5}$ representations. The integer parameters $w_{H_d}$ and $w_{\bar{5}_i}$ denote the separation between the charges of $H_d$ and $\obfive$ matter from the charge of $H_u$ \footnote{{In this analysis we have multiplied the charges of the fiber type $I_5^{(01)}$ by 5 so that all fiber types can be analysed with the same parameterization.}}. The charges for the $\bf 10$ and  $\obfive$ representations take values in \eqref{10ChargeswithTY} and \eqref{FCharges}, respectively.


\subsubsection{$\mathcal{N}_{\bar{\bf 5}} =3$}

In view of the arguments in (\ref{Conu}) and (\ref{Cond}), 
the minimal number of ${\bf \bar{5}}$ representations is three. However this case always allows the $\mu$-term, which disfavors these models. To see this, parameterize the models as in (\ref{ParaTable}) with one ${\bf\bar{5}}_1$ curve, which has  $M=3$ and $N=0$.
The anomaly condition (A2.) implies
$w_{H_d}  = 0$, which exactly generates the $\mu$-term. 

{\renewcommand{\arraystretch}{1.5}
\begin{sidewaystable}
\centering
\begin{tabular}{|c||c|c|c|c|c|c|}
\hline
 & {\bf I.1.4.a} & I.1.4.b & {I.1.4.c} & I.1.4.d & I.1.4.e &I.1.4.f \\ \hline \hline
 $M$       & 1 & 3 & $2/3$ & 3 & 0 & 3 \\
 $N$       & 2 & $-3$ & $-2$ & $-3$ & 3 & $-3$\\ \hline
 $q_{10}$  & $-1$ & $-1$ & $-7$ & 0 & $-7$ & 6 \\
 $q_{H_u}$ & 2 & 2 & 14 & 0 & 14 & $-12$\\
 $q_{H_d}$ & 2 & 1 & 6 & $-3$ & 1 & $-3$\\
 $q_{\bar{5}_1}$     & $-1$ & 0 & 1 & $-2$ & $-9$ & 2\\
 $q_{\bar{5}_2}$     & 1 & $-1$ & $-9$ & $-1$ & $-4$ & 7\\ \hline
 $Y^b_1$     & 0 & 0 & 0 & $-5$ & $-15$ & 5\\
 $Y^b_2$     & 2 & $-1$ & $-10$ & $-4$ & $-10$ & 10 \\ \hline
 $\mu$     & 4 & 3 & 20 & $-3$ & 15 & $-15$\\ \hline
 C2 & $\{-4,-2\}$ & $\{-3,-4\}$ & $\{-20,-30\}$ & $\{-2,-1\}$  & $\{-30,-25\}$ & $\{20,25\}$ \\
 C3 & $\{1,3\}$ & $\{1,2\}$ & $\{15, 5\}$ & $\{-2,-1\}$ & $\{5,10\}$ & $\{-10,-15\}$\\
 C4 & $\{-3,-1,1\}$ & $\{-3,-2,-1\}$ & $\{-25,-15,-5\}$ & $\{-4,-3,-2\}$ & $\{-25,-10,-15\}$ & $\{10,15,20\}$\\
 C5 & $\{-3,-1\}$ & $\{-2,-1\}$ & $\{-15, -5\}$ & $\{2,1\}$ & $\{-5,-10\}$ & $\{10,5\}$ \\
 C6 & $\{5,7\}$ & $\{4,5\}$ & $\{35, 25\}$ & $\{-5,-4\}$ & $\{20,25\}$ & $\{-25,-20\}$ \\
 C7 & $-1$ & $-2$ & $-15$ & $-3$ & $-20$ & 15\\ \hline
\end{tabular}
\caption{Solutions for 
$\mathcal{N}_{\bf 10} =1$ and $\mathcal{N}_{\obfive}=4$, which do not generate (C1.)$-$(C7.). The charges of the bottom Yukawa couplings are shown in the row $Y^b_i$, where $i = 1,2$ labels the which $\obfive_i$ is involved in the coupling. The charges of the couplings (C2.)$-$(C7.) are shown in the corresponding rows.
The model I.1.4.a is such that the bottom Yukawa coupling does not bring back any of the dangerous couplings and is phenomenologically preferred. The model I.1.4.c brings back dimension five proton decay operators. The remaining models are disfavored as they regenerate dimension four proton decay operators at the same level as the bottom Yukawas.
\label{tab:I.1.4}}
\end{sidewaystable}}


\subsubsection{$\mathcal{N}_{\bar{\bf 5}} =4$}

For four $\obfive$ representations, the anomaly conditions can be solved exactly, and we will find one model, which is phenomenologically viable. 
Consider again the parameterization
\be
\begin{array}{c|c|c|c}
{\bf R} &  q({\bf R}) & M & N \cr\hline
\bar{\bf 5}_{H_u} &  - q_{H_u}  &0 & -1 \cr
\bar{\bf 5}_{H_d}  & -q_{H_u} + 5 w_{H_d} &0& 1 \cr
\bar{\bf 5}_{1} &  -q_{H_u} + 5w_{\bar{5}_1} &M &N  \cr 
\bar{\bf 5}_{2}& -q_{H_u} + 5 w_{\bar{5}_2} &3-M& -N \cr\hline
{\bf 10}  &  -{1\over 2} q_{H_u}&   3&0 
\end{array}
\ee
where $M,N \in \mathbb{Z}^+$. In the analysis of four and more\footnote{This is to avoid repetition of solutions and in all sections that follow each $\obfive$ will have a non-zero net flux restriction.} $\obfive$s we do not allow solutions where $M_i = N_i = 0$ for any of the $\obfive$s. 
The above parameterization already satisfies (A1.), (A4.) and (A5.) by construction. Constraint (A2.) and (A3.) imply
\be
w_{H_d} = N (w_{\bar{5}_2} - w_{\bar{5}_1})\,,\qquad N (w_{\bar{5}_1} - w_{\bar{5}_2}) (w_{\bar{5}_1} + N w_{\bar{5}_1} + w_{\bar{5}_2} - N w_{\bar{5}_2}) = 0 \, ,
\ee
where we exclude cases $N = 0$ as well as $w_{\bar{5}_1} = w_{\bar{5}_2}$ as they imply $ q_{H_u} = -q_{H_d}$. 
If we do not require a bottom Yukawa coupling $q_{H_u}$ is left unconstrained and the charges are given by
\be 
q_{H_d} = -q_{H_u} + \frac{10 w_{\bar{5}_2} N}{1 + N} \,\qquad 
q_{\bar{5}_1} = -q_{H_u} + \frac{5 w_{\bar{5}_2} (N-1)}{1 + N} \,\qquad 
q_{\bar{5}_2} = -q_{H_u} + 5w_{\bar{5}_2}  \,.
\ee
Requiring a bottom Yukawa with $\obfive_1$ gives the additional constraint
\be
q_{H_u} = \frac{6N -2}{N+1} w_{\bar{5}_2} \,.
\ee
This results in the following set of charges
\be
q_{10} = \frac{-3N+1}{N+1}w_{\bar{5}_2} \,,\quad q_{H_d} = \frac{4N+2}{N+1} w_{\bar{5}_2} \,\quad 
q_{\bar{5}_1} = -\frac{N+3}{N+1} w_{\bar{5}_2}\,, \quad q_{\bar{5}_2} = \frac{-N+7}{N+1} w_{\bar{5}_2} \,,
\label{One10Four5bs1U(1)Charges}
\ee
where $w_{\bar{5}_2}$ is unconstrained. In this case  we do not consider the case $M=0$ as we want a bottom Yukawa coupling with $q_{\bar{5}_1}$, which must then contain a down-type quark.

To exemplify our solution process, in this case we summarize all solutions in  table \ref{tab:I.1.4}, which fall within \eqref{FCharges} and \eqref{10ChargeswithTY}. This corresponds to picking a specific value for $w_{\bar{5}_2}$, and $q_{H_u}$ in the case without a leading order bottom Yukawa coupling. The table also displays the charges of the forbidden couplings (C1.)$-$(C7.) as well as the charged Yukawa couplings, $Y^b_i$. The solutions can be summarised as follows:
\begin{itemize}
\item Model I.1.4a is the only phenomenologically viable solution for a single $U(1)$ solving all constraints, without bringing back any of the dangerous operators, when generating the remaining Yukawa couplings. It does  regenerate the $\mu$-term with two singlet insertions. As noted already in general, the flavor physics of 
this model is however quite limited, which is a matter that will be improved upon in the multiple $U(1)$ case.

\item Model I.1.4.c regenerates both dimension five proton decay operators with two and three singlet insertions and all other remaining models regenerate the dimension four proton decay operators (C4.). 
\end{itemize}



\subsubsection{$\mathcal{N}_{\bar{\bf 5}} =5, 6, 7$}

For $\mathcal{N}_{\bar{\bf 5}} > 4$ solving the anomaly cancellation conditions for general charges is difficult, however we provide a method for solving these in general in appendix \ref{app:Mordell}. 
In practice given the finite set of charges, one can simply scan over all possibilities. 
For each ${\bf 10}$ charge in
\eqref{10ChargeswithTY}, one can require the top Yukawa coupling, which fixes the charge of ${\bf 5}_{H_u}$. 
Solving  (A1.)$-$(A5.) {and requiring absence of (C1.)$-$(C7.)}, we find very few solutions, where {\it every} single one regenerates dimension 5 or dimension 4 proton decay operators at the same order as the remaining Yukawa couplings (with exactly the same singlet suppression).
Thus all models are disfavored. 

For $\mathcal{N}_{\obfive} \geq 6$ there are no solutions. The case of six $\bar{\bf 5}$ is maximal for two of the charge sets in \eqref{FCharges}. For these sets the only freedom comes in selecting the charge of the $\bf 10$ representation which will fix $q_{H_u}$. As there are seven possible charges for fundamental matter in the case of $I_5^{(01)}$ we need to consider all possible subsets of six once the charge of the $\bf 10$ has been fixed.  

One can go further and allow for seven distinctly charged $\bar{\bf 5}$ representations in the case of the $I_5^{(01)}$ models. One finds two solutions to the anomaly cancellation conditions for $q_{\bf 10} = \pm 1$, however, these solutions do not forbid (C2.) and are therefore excluded. 

%

\subsection{$\mathcal{N}_{\bf 10} \geq 2$}

The case of multiple ${\bf 10}$ representations for a single $U(1)$ symmetry does not yield any interesting solutions to the constraints. In particular for a single $U(1)$ the flavor physics is very constrained.
The analysis is provided in appendix \ref{app:Multi10} for completeness. In summary we find the following:
\begin{itemize}
\item There are two solutions for $\mathcal{N}_{\bten}=2$ and  $\mathcal{N}_{\obfive} =4$, shown in table \ref{tab:I.2.3}. Both these models regenerate dimension five operators at the same order as the charged Yukawas. In terms of the flavor physics of these models, with only two $\bten$ representations and four $\obfive$s one can not satisfy the mass hierarchies for the up-type and down-type quarks simultaneously.
\item For $\mathcal{N}_{\bten}=3$ there is one model, which has realistic flavor structure for the quark sector. In fact it generates the 
Haba textures (\ref{HabaText}), albeit it does regenerate dimension four proton decay operators at the same order as the Yukawas. 
\item No other solution exists with two or three $\bten$s,  which solved the anomaly cancellation conditions and forbid the dangerous operators.

\end{itemize}
It is clear from the analysis carried out in this section that in order to  construct feasible models, that might give rise to interesting flavor structure, it is necessary to extend to multiple $U(1)$s.



\section{Two $U(1)$ Models with Hypersurface Realization}
\label{sec:TwoU1sGeo}

For two additional $U(1)$ symmetries, the phenomenological properties of the models become much more favorable. 
Allowing models with up to three $\bten$ and eight $\obfive$ representations in the survey, one finds a large number of solutions to the  anomaly cancellation condition with no exotics, which futhermore forbid the unwanted operators. In view of this, it is then useful to focus  on two subclasses of solutions:
\begin{itemize}
\item[1.] Models with charges that have a known geometric realization.
\item[2.] Models, where the $U(1)$ symmetries can be used to construct realistic flavor textures. This is detailed in section \ref{sec:flavor}.
\end{itemize}

We now turn to point 1. and find solutions, which have charges that  are closest to known geometric models.
We will find in this section that there are no solutions, which are within the charges obtained in the literature. However, there are solutions, summarized 
 in tables \ref{tab:II.1.6p1} and \ref{tab:II.1.6p2},  for which we determine new geometric models, that give rise to these charges  in section \ref{sec:GeoRealisation}.

{\renewcommand{\arraystretch}{1.4}
\begin{table}
\centering
\begin{tabular}{|c||c|c|}
\hline
 & {\bf II.1.6.a} & {\bf II.1.6.b} \\ \hline \hline
 $M_1$ & 1  & 1  \\
 $M_2$ & 1  & 1   \\
 $M_3$ & 0  & 0  \\
 $N_1$ & 1  & 1   \\
 $N_2$ & $-1$ & $-1$ \\
 $N_3$ & 1  & 1   \\ \hline
 $q_{10}$ & $(-2, 3)$  & $(-2, 1)$ \\
 $q_{H_u}$ & $(4, -6)$  & $(4, -2)$   \\
 $q_{H_d}$ & $(6, -4)$  & (6, 2) \\
 $q_{\bar{5}_1}$ & $(-4, 1)$  & $(-4, -3)$  \\
 $q_{\bar{5}_2}$ & $(1, -4)$  & $(1, -3)$  \\
 $q_{\bar{5}_3}$ & (1, 6)  & (1, 7)  \\
 $q_{\bar{5}_4}$ & (6, 1)  & (6, 7)   \\ \hline
 $Y^b_1$ & (0,0)  & (0,0)  \\
 $Y^b_2$ & $(5, -5)$  & (5, 0)   \\
 $Y^b_3$ & (5, 5)  & (5, 10)   \\
 $Y^b_4$ & (10, 0)  & (10, 10)  \\ \hline
 $\mu$ & $(10, -10)$  & (10, 0)   \\ \hline
 C2 & $\{(-10, 10),(-5, 5),(-5, 15),(0, 10)\}$ 
    & $\{(-10, 0),(-5,0),(-5, 10),(0, 10)\}$ 
    \\ \hline
 C3 & $\{(0, -5),(5, -10),(5, 0),(10, -5)\}$ 
    & $\{(0, -5),(5, -5), (5, 5),(10, 5) \}$ 
   \\ \hline
\multirow{2}{*}{C4} & $\{(-10, 5),(-5, 0),(-5, 10),(0, 5),(0, -5),$ 
                    & $\{(-10, -5), (-5, -5), (-5, 5), (0, 5),(0, -5),$  \\
  & $(5, 0),(0, 15),(5, 10),(10, 5)\}$ 
  & $(5, 5),(0, 15), (5, 15),(0, 15)\}$ 
  \\ \hline
 C5 & $\{(0, 5), (-5, 10), (-5, 0), (-10, 5)\}$ 
    & $\{(0, 5), (-5, 5), (-5, -5), (-10, -5)\}$ 
    \\ \hline
 C6 & $\{(10, -15), (15, -20), (15, -10), (20, -15)\}$ 
    & $\{(10, -5), (15, -5), (15, 5), (20, 5)\}$
     \\  \hline
 C7 & (0, 5) & (0, 5) \\ \hline
\end{tabular}
\caption{Solutions for $\mathcal{N}_{\bf 10}=1$ and $\mathcal{N}_{\bf \bar{5}}=6$  with 2 $U(1)$s. The charges of the bottom Yukawa couplings are shown in the row $Y^b_i$, where $i = 1,2,3,4$ labels the which $\obfive_i$ is involved in the coupling. The charges of the couplings (C1.)$-$(C7.) are shown in the corresponding rows.
\label{tab:II.1.6p1}}
\end{table}}

{\renewcommand{\arraystretch}{1.4}
\begin{table}
\centering
\begin{tabular}{|c||c|c|}
\hline
 & {\bf II.1.6.c} & {II.1.6.d} \\ \hline \hline
 $M_1$ & 1  & 1/2  \\
 $M_2$ & 1  & 1/2   \\
 $M_3$ & 1  & 0/1  \\
 $N_1$ & 1  & -1   \\
 $N_2$ & $-1$ & $-1$ \\
 $N_3$ & $-1$  & 1   \\ \hline
 $q_{10}$ & (3, 1)  & (3, 1) \\
 $q_{H_u}$ & $(-6, -2)$  & $(-6, -2)$   \\
 $q_{H_d}$ & $(-4, 2)$  & $(-4, 2)$ \\
 $q_{\bar{5}_1}$ & $(1, -3)$  & $(1, -3)$  \\
 $q_{\bar{5}_2}$ & $(-4, -3)$  & $(-4, 7)$  \\
 $q_{\bar{5}_3}$ & (1, 7)  & (1, 7)  \\
 $q_{\bar{5}_4}$ & (6,7)  & (6, -3)   \\ \hline
 $Y^b_1$ & (0,0)  & (0,0)  \\
 $Y^b_2$ & $(-5, 0)$  & $(-5, 10)$   \\
 $Y^b_3$ & (0, 10)  & (0, 10)   \\
 $Y^b_4$ & (5, 10)  & (5,0)  \\ \hline
 $\mu$ & $(-10, 0)$  & $(-10, 0)$   \\ \hline
 C2 & $\{(10, 0), (5, 0), (10, 10), (15, 10)\}$ 
    & $\{(10, 0), (5, 10), (10, 10), (15, 0)\}$ 
    \\ \hline
 C3 & $\{(-5, -5), (-10, -5), (-5, 5), (0, 5)\}$ 
    & $\{(-5, -5), (-10, 5), (-5, 5), (0, -5) \}$ 
   \\ \hline
\multirow{2}{*}{C4} & $\{(5, -5), (0, -5), (5, 5), (10, 5),(-5,-5),$ 
                    & $\{(5, -5), (0, 5), (5, 5), (10, -5),(-5, 15)\}$  \\
  & $(0, 5), (5, 5),(5, 15), (10, 15),(15, 15)\}$ 
  & $\{(0, 15),(5, 15), (10, 5),(15, -5)\}$
  \\ \hline
C5 & $\{(5, 5), (10, 5), (5, -5), (0, -5)\}$ 
    & $\{(5, 5), (10, -5), (5, -5), (0, 5)\}$ 
    \\ \hline
 C6 & $\{(-15, -5), (-20, -5), (-15, 5), (-10, 5)\}$ 
    & $\{(-15, -5), (-20, 5), (-15, 5), (-10, -5)\}$
     \\ \hline
 C7 & (5, 5) & (5, 5) \\ \hline
\end{tabular}
\caption{Solutions for $\mathcal{N}_{\bf 10}=1$ and $\mathcal{N}_{\bf \bar{5}}=6$ with 2 $U(1)$s.  The model II.1.6.d regenerates dimension five proton decay operators with multiple insertions of the singlets regenerating the charged Yukawas.
\label{tab:II.1.6p2}}
\end{table}}

In the following we restrict to the set of $U(1)$ charges which have appeared in explicit realizations of two U(1) models \cite{Borchmann:2013jwa, Cvetic:2013nia, Borchmann:2013hta, Cvetic:2013jta,Lawrie:2014uya}. These charges are given by\footnote{
Note that not all combinations of these charges are realized in geometric models in the literature -- for instance the models we find in tables \ref{tab:II.1.6p1} and \ref{tab:II.1.6p2} are of this type. However in section \ref{sec:GeoRealisation}, we will determine new geometries (based on the generalized cubic model of \cite{CKPT}), which give a concrete realization of these models.}
\be\label{FChargesReduced}
 I_5^{(01)} :\  \left\{ 
 	\ba
 	q_{\bf 10}  \in &\left\{ 0, +1 \right\}  \cr
  	q_{\bf \bar{5} }\in &\left\{- 1, 0, +1 \right\}
	\ea\right.\quad 	
 I_5^{(0|1)} :\  \left\{ 
 	\ba
 	q_{\bf 10}  \in &\left\{-2,+3 \right\}  \cr
  	q_{\bf \bar{5} }\in &\left\{-4,+1, + 6 \right\}
	\ea\right.\quad 
 I_5^{(0||1)} :\  \left\{ 
 	\ba
 	q_{\bf 10}  \in &\left\{-4,+1 \right\}    \cr
  	q_{\bf \bar{5} }\in &\left\{ -3, +2, + 7 \right\}
	\ea\right.		
\ee
Taking this reduced set of charges we look for subsets, which solve the anomaly cancellation conditions, allowing up to $\mathcal{N}_{\bten} = 3$ and $\mathcal{N}_{\obfive}=8$. The set of models, which solve the conditions (A1.)$-$(A5.) are then further filtered down to those which forbid the dangerous couplings (C1.)$-$(C7.) at leading order. These dangerous couplings should also not be regenerated with the same singlet insertion, that regenerates the charged Yukawa couplings.

The phenomenologically good solutions are given in tables \ref{tab:II.1.6p1} and \ref{tab:II.1.6p2}. These models all feature $\mathcal{N}_{\bten}=1$ and $\mathcal{\obfive} =6$ and have a top and bottom Yukawa coupling at leading order. The model II.1.6.d regenerates dimension five proton decay operators with multiple insertions of the singlets, that generate the charged Yukawas.  The remaining three models all give rise to  the $\mu$-term with two singlet insertions. Interesting flavor textures for these models, which have only a single $\bten$, cannot be generated through the $U(1)$ symmetries, however these models have the advantage of having a concrete geometric realization:  none of the geometries in the literature \cite{Borchmann:2013jwa, Cvetic:2013nia, Borchmann:2013hta, Cvetic:2013jta,Lawrie:2014uya} generate this particular combination of charges, however we will determine elliptic fibrations for these models in section \ref{sec:GeoRealisation}.


\section{F-theoretic Froggatt-Nielsen Models with two $U(1)$s}
\label{sec:flavor}


The constructions passing all anomaly and coupling constraints with charges seen in known geometric constructions have not revealed an interesting flavor structure from the $U(1)$s, as we only found solutions with a single $\bten$ curve. We now turn our discussion to the question whether we can find models with two $U(1)$s and three $\bten$ curves with the more general, F-theoretic set of charges in \eqref{FCharges}. This increase in complexity improves the models, which allow for realistic flavor physics. In short, we shall try to identify models that also can lead to a realization of the FN mechanism. Note that we will match the quark Yukawas to several known flavor hierarchies. It certainly would be very interesting to scan through all the possibilities in the solution space, and possibly determine new textures. 
Concretely, we find two classes of flavor models, which appeared in \cite{Haba:1998wf, Babu:2003zz},  for the quark sector that can be realized. This will be the topic of the current section, and the resulting new lepton flavor structure is discussed in section  \ref{sec:leptonmixing}. There are several popular flavor models, that we cannot realize in our class of models, which will be shown in  appendix \ref{app:OtherTextures}. 


\subsection{Models with $\mathcal{N}_{\bf 10} =3$}
\label{sec:Three10s}

We now analyze F-theoretic $U(1)$ models with two $U(1)$s for their potential to solve all the constraints as well as induce realistic flavor hierarchies by a Froggatt-Nielsen type mechanism. 
Each entry in the Yukawa matrix for the up-type quarks, $Y^u$, is given by the couplings,
\be 
Y^u_{ij} \, Q_i \bar{u}_j H_u \,, \quad i, j = 1,2,3 \,.
\ee
The $U(1)$ charges of $\bten$ representations within which these quarks reside will determine the charges of the singlets required to regenerate these couplings and therefore their suppression. If we require that $Y^u$ is rank one at leading order so that only $Y^u_{3,3}$, for the third generation, is uncharged under the additional $U(1)$s and that $Y^u_{1,1}$ and $Y^u_{2,2}$ appear with different suppressions, to match with a large class of known textures, then we are required to consider models with three $\bten$ representations. 

A leading order rank one up-type Yukawa matrix is achieved most easily by having $Q_3$ and $\bar{u}_3$ residing on the same $\bten$ representation, $\bten_3$, with $U(1)$ charges satisfying
\be 
2 q_{10_3} + q_{H_u}  = 0\,.
\ee
In order for the top Yukawa coupling involving $\bten_3$ to only generate a leading order mass for the top quark we require
\be 
 M_{10_3} = 1 \text{ and } N_{10_3} = 0,
\ee
so that only the third generation of left- and right-handed quarks lie within this $\bten$ representation. It is crucial that only the third generation is present on $\bten_3$ otherwise off diagonal terms in the Yukawa matrix will also be regenerated at first order.

Between the remaining two $\bten$ representations, $\bten_1$ and $\bten_2$, one can have the following distribution of the remaining quarks:
\begin{itemize}
\item[T.1:] $M_{10_1} = M_{10_2} = 1$, $N_{10_1} = N_{10_2} = 0$ \\
For these configurations one has
\be 
\bten_A \supset Q_A + \bar{u}_A + \bar{e}_A, \quad A = 1,2,3 \,,
\ee
and the resulting Yukawa matrix is symmetric. These textures could potentially agree with those in \cite{Haba:1998wf,Ross:2007az,Babu:2003zz,Dudas:2009hu}.
\item[T.2:] $M_{10_1} = M_{10_2} = 1$, $N_{10_1} = -1$, $M_{10_2} = 1$ \\
Here, both the remaining right-handed up-type quarks, $\bar{u}_1$ and $\bar{u}_2$, originate from $\bten_1$,
\be \ba
\bten_1 &\supset Q_1 + \bar{u}_1 + \bar{u}_2 \\
\bten_2 &\supset Q_2 + \bar{e}_1 + \bar{e}_2  \,.
\ea 
\label{Three10CaseB}
\ee
The resulting $Y^u$, denoting the singlet insertion which regenerates the top Yukawa coupling between $\bten_A$ and $\bten_B$ as $s_{AB}$, has the following form,
\be 
Y^u \sim \left( 
\begin{array}{ccc}
s_{11} & s_{11} & s_{13} \\
s_{12} & s_{12} & s_{23} \\
s_{13} & s_{13} & 1 
\end{array}
\right) \,,
\ee
where two columns have identical singlet insertions, as the charges for the couplings involving $\bar{u}_A$, $A = 1,2$ are the same. This does not match known textures, where off diagonal terms have a greater suppression compared to their nearest diagonal terms.
\item[T.3:] $M_{10_1} = 2, M_{10_2} = 0$, $N_{10_1} = 0$, $M_{10_2} = 0$ \\
We do not consider this case as $\bten_2$ has no net chirality and therefore this case reduces to two $\bten$ representations.
\end{itemize}
The case T.2 can be shown to not give rise to good flavor textures.
In the following section we focus on case T.1, where each differently charged $\bten$ representation contains a different generation of $Q_A$ and $u_A$, and match to known textures in the literature. 
We show in appendix \ref{app:OtherTextures} that the flavor hierarchies in \cite{Ross:2007az,Dudas:2009hu} cannot be realized within our global F-theoretic charge framework. Note that the  textures in \cite{Dreiner:2003hw}, which do not have a symmetric $Y^u$, cannot be realized.  
The two flavor models that can be realized in our framework are those in Haba \cite{Haba:1998wf} as well as Babu, Enkhbat, and Gogoladze  \cite{Babu:2003zz}, which we now discuss in turn.

\subsection{F-theoretic FN-models (Haba1) and (Haba2)}
\label{subsec:II.3.4}

In this section we determine solutions to our constraints, which furthermore generate the Yukawa textures in Haba \cite{Haba:1998wf} 
\be
Y_{\rm Haba}^u\sim\left(
\begin{array}{c c c}
\epsilon^8 & \epsilon^6 &\epsilon^4\\
\epsilon^6 & \epsilon^4 &\epsilon^2\\
\epsilon^4 & \epsilon^2 & 1\\
\end{array}
\right)\ , \qquad 
 Y_{\rm Haba}^d\sim\left(
\begin{array}{c c c}
\epsilon^4 & \epsilon^4 &\epsilon^4\\
\epsilon^2 & \epsilon^2 &\epsilon^2\\
1 & 1 & 1\\
\end{array}
\right) 
\ee
from a Froggatt-Nielsen type mechanism. 
Let
\be 
M_{10_A} = 1, \quad N_{10_A} = 0, \quad A = 1,2,3 \,.
\ee
The charges of the $\bten$ representations therefore do not contribute to the anomaly cancellation conditions, as $N_{10_A} = 0$. 
We consider $\mathcal{N}_{\obfive} =4$ in the following. 
The sets of $\obfive$ charges, which solve the conditions were determined in appendix \ref{app:Four5bsMultiU1s} and are given in \eqref{II.1.4CaseA}, \eqref{II.1.4CaseB} and \eqref{II.1.4CaseC}. 

In order to match to this texture we need to impose that all $\bar{d}_i$ are from the same $\obfive$ representation, which is achieved by
\be  \ba 
M_{\bar{5}_1} &= 0, \quad N_{\bar{5}_1} = 1,2 \text{ or } 3 \\
M_{\bar{5}_2} &= 3, \quad N_{\bar{5}_2} = - N_{\bar{5}_1} \,.
\ea \ee
The cases $N_{\bar{5}_1} = 1, 3$  give phenomenologically disfavorable models as the solutions either allow the $\mu$-term or regenerate dimension four proton decay with the remaining charged Yukawas. This leaves only $N = 2$, the solutions of which are given in table \ref{tab:II.1.4}. Imposing the presence of a bottom Yukawa coupling of the form
\be 
\bten_3 \obfive_2 H_d \,,
\ee
restricts the set of solutions further. The set of charges with,
\be  \ba 
M_{5_1} &= 0, \quad N_{5_1} = 2\\
M_{5_2} &= 3, \quad N_{5_2} = - 2 \,,
\label{DPMandNs}
\ea \ee
which furthermore allow for a bottom Yukawa coupling are as follows:
\be 
\begin{array}{c||c|c|c|c|c}
   & \bten_3 & \bfive_{H_u} & \obfive_{H_d} & \obfive_1 & \obfive_2 \\ \hline\hline
 q({\bf R})^{1} & -q_{H_u}^{1}/2 & q_{H_u}^{1} & 3q_{H_u}^{1}/7 & -9 q_{H_u}^{1}/14 & q_{H_u}^{1}/14 \\
 q({\bf R})^{2} & - q_{H_u}^{2}/2 & q_{H_u}^{2} & 3q_{H_u}^{2}/7 & -9 q_{H_u}^{2}/14 & q_{H_u}^{2}/14 
\end{array}
\label{II.3.4}
\ee
 where we have imposed the top Yukawa coupling for $\bten_3$.
The charges of $\bten_1$ and $\bten_2$ are given by\footnote{In order to uniformly study all F-theoretic charges we rescaled for convenience the models of type $I_5^{(01)}$ by a factor of $5$. This allows us to study all the models where the unit charges is now set to be $5$ (rather than~1). }
\be
\left(q^{1}_{10_A},q^{1}_{10_A}\right) = \left(-\frac{1}{2} q_{H_u}^{1}, -\frac{1}{2} q_{H_u}^{2}\right) + 5 \left(w^{1}_{10_A}, w^{2}_{10_A}\right) \,,
\label{AdditionalTenCharges}
\ee
where $q^{\alpha}$ denotes the charges under $U(1)_{\alpha}$ and $A= 1,2$. The charges of matter under these two $U(1)$s will therefore be almost identical, the only difference being in the $\bten_1$ and $\bten_2$, which should be chosen so that no dangerous couplings are allowed at leading order. Restricting these general charges to the F-theory charges one finds that there are only two choices for the charge of the Higgs up given by
\be 
(q^1_{H_u}, q^2_{H_u}) = (14, 14) \ \text{ or } \ (0,14)\,.
\ee
The integer separations, parameterized by $w^{\alpha}_{10_A}$, satisfy the constraint,
\be 
w^1_{10_A} \neq w^2_{10_A} , \quad A = 1,2 \,,
\ee
as there were no two $\bten$ models for a single $U(1)$ which were phenomenologically viable. Violating the above constraint for either $\bten_1$ or $\bten_2$ will either bring back dangerous operators or regenerate them with the charged Yukawas. 
This implies the following distribution:
\be
\begin{array}{c||c|cc|c}
  \text{Representation} & \text{Charge} & M & N & \text{Matter} \\ \hline
 {\bf 10}_1      & \left(- \half q^1_{H_u} + 5 w_{10_1}^1, -\half q^2_{H_u} + 5 w_{10_1}^2 \right)   & 1 & 0 & Q_1, \bar{u}_1, \bar{e}_A, \, A = 1,2 \\ 
 {\bf 10}_2    & \left(-\half q^1_{H_u} + 5 w_{10_2}^1, -\half q^2_{H_u} + 5 w_{10_2}^2 \right)    & 1 & 0  & Q_2, \bar{u}_2, \bar{e}_B, \, B \neq A, \,B = 1,2 \\
  {\bf 10}_3    & \left(-\half q^1_{H_u},- \half q^2_{H_u}\right)    & 1 & 0  & Q_3, \bar{u}_3, \bar{e}_3 \\
 \bar{\bf 5}_{H_u} &  (-q^1_{H_u}, -q^2_{H_u})  & 0 & -1 & H_u \\
 \bar{\bf 5}_{H_d} & (\frac{3}{7}q^1_{H_u}, \frac{3}{7}q^2_{H_u})     & 0 & 1  & H_d \\ 
 \bar{\bf 5}_1 & (-\frac{9}{14}q^1_{H_u}, -\frac{9}{14}q^2_{H_u})    & 0 & 2  & L_I, I = 1,2\\
 \bar{\bf 5}_2 & (\frac{1}{14}q^1_{H_u},\frac{1}{14} q^2_{H_u}) & 3 & -2 & L_3, \bar{d}_I, I = 1,2,3 
\end{array}
\ee
The necessary singlet insertions to regenerate the full up and down-type Yukawa matrices can be determined to be
 \be 
Y^u \sim 
\left(
\begin{array}{ccc}
s_1^2 & s_1 s_2 & s_1 \\
s_1 s_2 & s_2^2 & s_2\\
s_1 & s_2 & 1
\end{array}
\right) \,,
\quad 
Y^d \sim 
\left(
\begin{array}{ccc}
s_1 & s_1 & s_1 \\
s_2 & s_2 & s_2\\
1 & 1 & 1
\end{array} 
\right) \,,
\label{DTGenYukawas}
\ee
where $s_i = \frac{\langle S_i \rangle}{M_{GUT}}$. The charges of the singlets, $S_1$ and $S_2$ are given by
\be \ba 
(q^1_{S_1},q^2_{S_1}) &= - 5(w^1_{10_1}, w^2_{10_1}) \\
(q^1_{S_2},q^2_{S_2}) &= -5 (w^1_{10_2}, w^2_{10_2})\,.
\ea \ee
These singlets exactly correspond those which are present in ${\overline{ \bf 10}_3} \bten_A {\bf 1}$  couplings, where $A = 1,2$, as can be seen from their charges.

Choosing $s_1 = \epsilon^2$ and $s_2 = \epsilon^4$ one obtains the Haba texture in \eqref{HabaText}. When the charges of the two singlets are not coprime we have the following relation,
\be 
(q^1_{S_1},q^2_{S_1}) = n (q^1_{S_2},q^2_{S_2})\,,
\ee
for some integer $n$. In \eqref{DTGenYukawas} $s_1$ can be replaced with $s_2^n$, and from this we see that in order to match to the texture in \eqref{HabaText} we must have $n = 2$. In this case we can generate the entire Yukawa matrix by giving a vev to only one singlet. 
 
One choice of $w^{\alpha}_{10_A}$ which avoids all dangerous operators is given by\footnote{The charge of the ${\bf 10}_{1,2}$ are not fixed so far, even when including the constraint of suppressing all dangerous couplings. Here they are chosen to bring them closest to the known geometric models.}
\be \ba
(w^1_{10_1}, w^2_{10_1}) &= (2,0) \\
(w^1_{10_2}, w^2_{10_2}) &= (1,0)\,.
\label{flavorExample}
\ea \ee
Within this setup there are two choices for the  charges for the up-type Higgs:
\be\ba
\hbox{(Haba1)}: & \qquad (q^1_{H_u}, q^2_{H_u}) = (14, 14)\cr
\hbox{(Haba2)}: &  \qquad (q^1_{H_u}, q^2_{H_u}) = (0, 14)\,.
\ea\ee 
The full set of charges for these models are as follows\footnote{Note, that as mentioned earlier we rescaled the charges of the class of models $I_5^{(01)}$ by 5, to allow for a uniform treatment of all models. This means, that instead of the ${\bf 10}$ charges (2,1,0) we write (10, 5, 0).}
\be \label{E11}
\begin{array}{c||c|c|cc|c}
  \text{GUT} & \text{Charges for (Haba1)} &  \text{Charges for (Haba2)} & M & N & \text{MSSM Matter} \\ \hline\hline
 {\bf 10}_1      &  (3,-7)  &  (10,-7) & 1 & 0 & Q_1, \bar{u}_1, \bar{e}_A, \, A = 1,2 \\ 
 {\bf 10}_2    & (-2,-7)  & (5,-7) & 1 & 0  & Q_2, \bar{u}_2, \bar{e}_B, \, B \neq A, \,B = 1,2 \\
  {\bf 10}_3    & (-7,-7)   & (0,-7)& 1 & 0  & Q_3, \bar{u}_3, \bar{e}_3 \\
\bar{\bf 5}_{H_u} & (-14,-14)  & (0,-14)& 0 & -1 & H_u \\
 \bar{\bf 5}_{H_d} &(6,6)   &(0,6)   & 0 & 1  & H_d \\ 
 \bar{\bf 5}_1 &(-9,-9)  &(0,-9)  & 0 & 2  & L_I, I = 1,2\\
 \bar{\bf 5}_2 & (1,1) & (0,1) & 3 & -2 & L_3, \bar{d}_I, I = 1,2,3 
\end{array}
\ee
{\renewcommand{\arraystretch}{1.3}
\begin{table}
\centering
\begin{tabular}{|c||c|c|}
\hline
             & {\bf (Haba1)}                     & {\bf (Haba2)}            \\ \hline \hline
 $M$       & 0                           & 0 \\
 $N$       & 2                           & 2                     \\ \hline
 $M_{10_A}$     & 1                         & 1 \\
 $N_{10_A}$     & 0                         & 0                   \\ \hline 
 $q_{10_1}$  & $(3,-7)      $                  & $(10,-7)$ \\
 $q_{10_2}$  & $(-2,-7)$                      & $(5,-7)$ \\
  $q_{10_3}$  & $(-7,-7)$                      & $(0,-7)$ \\
 $q_{H_u}$   & $(14,14) $                    & $(0,14)$ \\
 $q_{H_d}$   & $(6,6)    $                   & $(0,6)$ \\
 $q_{\bar{5}_1}$       & $(-9,-9)$       & $(0,-9)$ \\
 $q_{\bar{5}_2}$       & $(1,1)$           & $(0,1)$ \\ \hline
 $\mu$       & $(20,20)$                          & $(0,20)$                       \\ \hline
 \multirow{4}{*}{C2}  & $\{(0, -30), (10, -20), (-5, -30), (5, -20),$ & $\{(30, -30), (30, -20), (25, -30), (25, -20),$\\
 & $(-10, -30),(0, -20), (-15, -30), (-5, -20),$ & $(20, -30), (20, -20), (15,-30), (15, -20),$\\
 & $ (-20, -30), (-10, -20),(-25, -30),$   & $ (10, -30), (10, -20), (5, -30), (5, -20),$ \\
 &  $ (-15, -20),(-30, -30), (-20, -20)\}$ & $(0, -30),(0, -20)\} $ \\\hline
 C3          & $\{(5,5),(15,15)\}$                  & $\{(0,5),(0,15)\}$              \\\hline
 \multirow{3}{*}{C4}          & $\{(-15, -25), (-5, -15), (-20, -25),  $ & $\{(10, -25), (10, -15), (5, -25), (5, -15),$ \\
 & $(-10, -15), (-25, -25),(-15, -15), (5, -5),$  & $\{(0, -25), (0, -15), (10,-5), (5, -5),$   \\
 & $ (0, -5),(-5, -5)\}$ & $(0, -5)\}$\\\hline
 \multirow{3}{*}{C5}          & $\{(15, -5), (5, -15), (10, -5), (0, -15), $ & $\{(20, -5), (20, -15), (15, -5), (15, -15),$ \\
& $ (5, -5),(-5, -15), (0, -5),(-10, -15),$    & $(10, -5), (10, -15), (5,-5), (5, -15),$   \\
& $ (-5, -5), (-15, -15)\}$ & $ (0, -5), (0, -15)\}$  \\\hline
 C6          & $\{(25, 25), (35, 35)\}$                 & $\{(0, 25), (0, 35)\}$           \\\hline
 C7          & $\{(-5, -15), (-10, -15), (-15, -15)\}$                & $\{(10, -15), (5, -15), (0, -15)\}$          \\ \hline
\end{tabular}
\caption{F-theoretic FN-models (Haba1) and (Haba2): these models have two $U(1)$s and 
$\mathcal{N}_{\bf 10}= 3$ and $\mathcal{N}_{\obfive} = 4$
and have realistic flavor textures, which for the quark sector match those by Haba in \cite{Haba:1998wf}. \label{tab:E8}
} 
\end{table}}
The models are summarized in table \ref{tab:E8}, including the charges for all the couplings (C1.)$-$(C7.). 
Both models have up- and down-type Yukawas with the same singlet insertion structure,
\be 
\ba
\hbox{(Haba1,2)}: &\qquad 
Y^u\sim 
\left(
\begin{array}{ccc}
\omega_1^4 & \omega_1^3 & \omega_1^2 \\
\omega_1^3 & \omega_1^2 & \omega_1\\
\omega_1^2 & \omega_1 & 1
\end{array}
\right) \,,
\quad 
Y^d \sim 
\left(
\begin{array}{ccc}
\omega_1^2 & \omega_1^2 & \omega_1^2 \\
\omega_1 & \omega_1 & \omega_1\\
1 & 1 & 1
\end{array} 
\right) \,,
\ea\ee
where $\omega_1 = \frac{\langle W_1 \rangle}{M_{GUT}}$, where the charge of the singlet $W_1$ is
\be 
(q_{W_1}^1, q_{W_1}^2) = (-5,0) \,.
\ee
By choosing $\omega_1 = \epsilon^2$ one recovers the Haba flavor texture in \eqref{HabaText}. The lepton Yukawa matrices, from the above sets of charges, have the following singlet structure
\be \ba
\hbox{(Haba1,2)}:& \qquad 
Y^L \sim 
\left(
\begin{array}{ccc}
\omega_1^2 \omega_2& \omega_1 \omega_2 & \omega_2\\
\omega_1^2 \omega_2 & \omega_1 \omega_2 & \omega_2 \\
\omega_1^2 & \omega_1 & 1
\end{array}
\right) \qquad \hbox{with $\bten_A\supset \bar{e}_A$, where $A= 1,2$} \,.
\ea\ee
Here again the second singlet vev is $\omega_2 = \frac{\langle W_2 \rangle}{M_{GUT}}$.
The choice for how the $e_A$ are distributed on the ${\bf 10}_B$ matter loci is made to get the standard hierarchy between first and second generation. To regenerate the entries in these matrices the following charged singlets must gain a vacuum expectation value
\be \ba 
\hbox{(Haba1)}: \qquad &\begin{array}{l} 
(q^1_{W_2}, q^2_{W_2}) = (10,10) \\
\end{array} \\
\hbox{(Haba2)}: \qquad &\begin{array}{l}
(q^1_{W_2}, q^2_{W_2}) = (0,10) \,.
\end{array}
\ea \ee
As one can see from the charges of the (C2.) couplings in table \ref{tab:E8}, regenerating the lepton Yukawas regenerates all the dimension five operators in both models. The dangerous dimension five couplings with coupling constant $\delta^{1}_{112I}$ that are regenerated with certain singlet insertions are shown below for both models, where we take $\omega_2 = O(1)$:
\be \begin{array}{c|c|c|c|c}
\text{Model} & \text{Coupling} & \text{Charge} & \text{Singlet insertions} & \epsilon \text{ suppression}\\ \hline
\hbox{(Haba1)}  & \bten_1 \bten_1 \bten_2 \obfive_1 & (-5,-30) & \omega_1^5 \omega_2^3\ & \leq \epsilon^{10} \\
& \bten_1 \bten_1 \bten_2 \obfive_2 & (5,-20) & \omega_1^5 \omega_2^2 & \leq \epsilon^{10} \\\hline
\hbox{(Haba2)} & \bten_1 \bten_1 \bten_2 \obfive_1 & (25,-30) & \omega_1^5 \omega_2^3 & \leq \epsilon^{10}\\
& \bten_1 \bten_1 \bten_2 \obfive_2 & (25,-20) & \omega_1^5 \omega_2^2 & \leq \epsilon^{10}
\end{array}
 \ee
For both models, in order for dimension five coupling involving $\obfive_2$ to be suppressed within the bound \eqref{dim5bound}, one must have
\be 
\frac{\epsilon^{10}}{M_{GUT}} \approx \frac{10^{-7}}{M_{GUT}} \leq 16 \pi^2 \frac{M_{SUSY}}{M_{GUT}^2} \,,
\ee
where  the Wolfenstein parameter is $\epsilon \approx 0.22$. This translates into the following relation
\be 
M_{SUSY} \geq 10^{-9} M_{GUT} \,,
\label{E8MUSY}
\ee
where  as before $\omega_1 = \epsilon^2$ and $\omega_2 = O(1)$. The latter could be improved upon by considering lepton flavor, where the most constraining factor, for both the models, is the mass ratio between the second and third generation, which is of order $\epsilon^2$. This is discussed in section \ref{sec:leptonmixing}.

Other dimension five operators of type $Q^3L$ are also regenerated with suppressions of $\epsilon^2$ and higher. For example, in this case one also gets the coupling,
\be 
\bten_3 \bten_3 \bten_3 \obfive_2 \supset Q_3 Q_3 Q_3 L_3 \,,
\ee
which can be compared to the bound on $\delta^1_{112I}$ by inserting suppression factors from the CKM between the third generation and the first and second. One finds that this coupling, which has $\epsilon^2$ suppression from the singlets, picks up at least an additional $\epsilon^{10}$ once we take into account the mixing between the quark generations. This coupling does not pose a greater threat than those considered above and the lower bound of $M_{SUSY}$ from this model is unchanged. In section \ref{sec:leptonmixing} we consider the lepton and neutrino flavor physics of these models in more detail.

We extended this analysis to $\mathcal{N}_{\bf \bar{5}} =5$ and $6$, which are all possible choices, however there are no further solutions. Extending the number of ${\bf \bar{5}}$  beyond that results in exotics. Thus the presently analyzed case of $\mathcal{N}_{\bar{\bf 5}}=4$ presents a sort of sweetspot. 

\subsection{F-theoretic FN-models (BaEnGo1)$-$(BaEnGo3)}
\label{sec:BaEnGo}

 Below we consider  solutions which allow for a symmetric up-type Yukawa matrix paired with a down-type Yukawa matrix, which has only two distinct columns. These textures \eqref{BaEnGoText}, as we shall show give rise to a realistic CKM structure. This has appeared in the literature before in \cite{Babu:2003zz}, and will referred to the BaEnGo texture
\be
Y_{\rm BaEnGo}^u \sim\left(\begin{array}{ccc}
\epsilon^8 & \epsilon^6 & \epsilon^4\\
\epsilon^6 & \epsilon^4 & \epsilon^2\\
\epsilon^4 & \epsilon^2 & 1
\end{array}\right) \,,\qquad 
Y_{\rm BaEnGo}^d \sim\left(\begin{array}{ccc}
\epsilon^5 & \epsilon^4 & \epsilon^4\\
\epsilon^3 & \epsilon^2 & \epsilon^2\\
\epsilon & 1 & 1
\end{array}\right)\,.
\ee 
 In this case the structure of singlet insertions is of the form
\be 
Y^u \sim 
\left(
\begin{array}{ccc}
s_1^2 & s_1 s_2 & s_1 \\
s_1 s_2 & s_2^2 & s_2\\
s_1 & s_2 & 1
\end{array}
\right) \,,
\quad 
Y^d \sim 
\left(
\begin{array}{ccc}
s_1 s_3 & s_1 & s_1 \\
s_2 s_3 & s_2 & s_2\\
s_3 & 1 & 1
\end{array} 
\right) \,,
\label{NewTexture1}
\ee
where $s_i = \frac{\langle S_i \rangle}{M_{GUT}}$. The charges of the singlets, $S_1$, $S_2$ and $S_3$ are given by
\be \ba 
(q^1_{S_1},q^2_{S_1}) &= - 5(w^1_{10_1}, w^2_{10_1}) \\
(q^1_{S_2},q^2_{S_2}) &= -5 (w^1_{10_2}, w^2_{10_2}) \\
(q^1_{S_3},q^2_{S_3}) &= -5 (w^1_{\bar{5}_n}-w^1_{\bar{5}_2}, w^2_{\bar{5}_n}-w^2_{\bar{5}_2})\,,
\ea \ee
where one of the $\obfive$s, in this case $\obfive_2$, is the one taken to have a leading order bottom Yukawa coupling and must contain two down-type quarks. One other $\obfive$, in the above, labelled $\obfive_n$, must contain the last down-type quark. Assuming the dominant contribution to the masses comes from the diagonal elements, we choose
\be 
s_1 = \epsilon^4, \quad s_2 = \epsilon^2 \,,
\ee
to satisfy the up-type ratios in \eqref{MassRatios}. Taking $s_3 =1$ we recover a down-type Yukawa matrix in section \ref{subsec:II.3.4}, here we take the third singlet insertion to be
\be 
s_3 = \epsilon  \,.
\ee
Using the formulas for the three mixing angles derived in \cite{Hall:1993ni,Dudas:1995yu} the CKM, neglecting the CP phase, can be calculated to take the form
\be 
V_{CKM} \sim 
\left(
\begin{array}{ccc}
1 & \epsilon^2 & \epsilon^4 \\
\epsilon^2 & 1 & \epsilon^2\\
\epsilon^4 & \epsilon^2 & 1
\end{array} 
\right) \,,
\ee
to leading order in $\epsilon$. The corresponding Yukawas are those shown in  \eqref{BaEnGoText}. 
Below we study the models, which realize these textures with four and five $\obfive$s. These models share the same up-type and down-type Yukawas, which have the structure in \eqref{NewTexture1}, however, they differ on the texture of the lepton Yukawa matrix.

\subsubsection{$\mathcal{N}_{\obfive}=4$}
\label{sec:NewTexturesFour5bs}

One class of such solutions can be obtained by altering the $M,N$s in \eqref{DPMandNs} to
\be  \ba 
M_{5_1} &= 1, \quad N_{5_1} = 2\\
M_{5_2} &= 2, \quad N_{5_2} = - 2 \,,
\ea \ee
which gives rise to models with the same set of possible charges, but the down-type Yukawa matrix now has the structure in \eqref{NewTexture1}. In this distribution of $M,N$s all three generations of leptons reside in $\obfive_1$ which produces lepton Yukawas of the form
\be \ba
\hbox{(BaEnGo1,2)}:& \qquad 
Y^L \sim 
\left(
\begin{array}{ccc}
s_1 s_3 & s_2 s_3 & s_3\\
s_1 s_3 & s_2 s_3 & s_3  \\
s_1 s_3 & s_2 s_3 & s_3
\end{array}
\right)\,,
\ea\ee
where we have chosen the following distribution of right-handed leptons $\bten_A \supset \bar{e}_A$, where $A = 1,2,3$ for both models. These choices ensure that the singlet suppressions generate the correct hierarchy in lepton masses.

Restricting to the charges in \eqref{E11}, where the singlets $s_i$ are not coprime, the up- and down-type Yukawa matrices take the form
\be 
\ba
\hbox{(BaEnGo1,2)}: &\qquad 
Y^u\sim 
\left(
\begin{array}{ccc}
\omega_1^4 & \omega_1^3 & \omega_1^2 \\
\omega_1^3 & \omega_1^2 & \omega_1\\
\omega_1^2 & \omega_1 & 1
\end{array}
\right) \,,
\quad 
Y^d \sim 
\left(
\begin{array}{ccc}
\omega_1^2 \omega_2 & \omega_1^2 & \omega_1^2 \\
\omega_1 \omega_2 & \omega_1 & \omega_1\\
\omega_2 & 1 & 1
\end{array} 
\right) \,,
\ea
\label{YudNewTexture1}
\ee
where $\omega_i = \frac{\langle W_i \rangle}{M_{GUT}}$. The lepton Yukawa matrix for both sets of charges takes the form
\be \ba
Y^L \sim 
\left(
\begin{array}{ccc}
\omega_1^2 \omega_2 & \omega_1 \omega_2 & \omega_2\\
\omega_1^2 \omega_2 & \omega_1 \omega_2 & \omega_2  \\
\omega_1^2 \omega_2 & \omega_1 \omega_2 & \omega_2
\end{array}
\right)\,,
\ea\ee
where the singlets $W_i$ have the following charges
\be \ba 
\hbox{(BaEnGo1)}: \quad  (q_{W_1}^1, q_{W_1}^2) &= (-5,0) \\
(q_{W_2}^1, q_{W_2}^2) &= (10,10) \\
\hbox{(BaEnGo2)}: \quad (q_{W_1}^1, q_{W_1}^2) &= (-5,0) \\
(q_{W_2}^1, q_{W_2}^2) &= (0,10) \,.
\ea \ee
The singlet suppressions, in terms of the Wolfenstein parameter $\epsilon$, are given by,
\be 
w_1 = \epsilon^2, \qquad w_2 = \epsilon\,,
\label{NewTexture1WP}
\ee
to match the suppression of the singlets $s_i$ in the general texture.
{\renewcommand{\arraystretch}{1.3}
\begin{table}
\centering
\begin{tabular}{|c||c|c|}
\hline
             & {\bf (BaEnGo1)}                     & {\bf (BaEnGo2)}           \\ \hline \hline
 $M$       & 1                           & 1 \\
 $N$       & 2                           & 2                     \\ \hline
 $M_{10_A}$     & 1                         & 1 \\
 $N_{10_A}$     & 0                         & 0                   \\ \hline 
 $q_{10_1}$  & $(3,-7)      $                  & $(10,-7)$ \\
 $q_{10_2}$  & $(-2,-7)$                      & $(5,-7)$ \\
  $q_{10_3}$  & $(-7,-7)$                      & $(0,-7)$ \\
 $q_{H_u}$   & $(14,14) $                    & $(0,14)$ \\
 $q_{H_d}$   & $(6,6)    $                   & $(0,6)$ \\
 $q_{\bar{5}_1}$       & $(-9,-9)$       & $(0,-9)$ \\
 $q_{\bar{5}_2}$       & $(1,1)$           & $(0,1)$ \\ \hline
 $\mu$       & $(20,20)$                          & $(0,20)$                       \\ \hline
 \multirow{4}{*}{C2}  & $\{(0, -30), (10, -20), (-5, -30), (5, -20),$ & $\{(30, -30), (30, -20), (25, -30), (25, -20),$\\
 & $(-10, -30),(0, -20), (-15, -30), (-5, -20),$ & $(20, -30), (20, -20), (15,-30), (15, -20),$\\ 
 & $ (-20, -30), (-10, -20),(-25, -30),$   & $ (10, -30), (10, -20), (5, -30), (5, -20),$ \\
 &  $ (-15, -20),(-30, -30), (-20, -20)\}$ & $(0, -30),(0, -20)\} $ \\ \hline
 C3          & $\{(5,5)\}$                  & $\{(0,5)\}$              \\ \hline
 \multirow{3}{*}{C4}          & $\{(-15, -25), (-5, -15), (-20, -25),  $ & $\{(10, -25), (10, -15), (5, -25), (5, -15),$ \\
 & $(-10, -15), (-25, -25),(-15, -15), (5, -5),$  & $\{(0, -25), (0, -15), (10,-5), (5, -5),$   \\
 & $ (0, -5),(-5, -5)\}$ & $(0, -5)\}$\\ \hline
 \multirow{3}{*}{C5}          & $\{(15, -5), (10, -5),(5, -5), (0, -5), $ & $\{(20, -5), (15, -5),(10, -5), (5,-5),$ \\
& $(-5, -5) \}$    & $ (0, -5)\}$   \\ \hline
 C6          & $\{(25, 25)\}$                 & $\{(0, 25)\}$           \\ \hline
 C7          & $\{(-5, -15), (-10, -15), (-15, -15)\}$                & $\{(10, -15), (5, -15), (0, -15)\}$          \\ \hline
\end{tabular}
\caption{F-theoretic FN-models (BaEnGo1) and (BaEnGo2): these models have two $U(1)$s and 
$\mathcal{N}_{\bf 10}= 3$ and $\mathcal{N}_{\obfive} = 4$
and have realistic flavor textures, which for the quark sector match those by BaEnGo in (\ref{BaEnGoText}). 
\label{tab:NewTexture4}
} 
\end{table}

The charges of the unwanted operators are shown in table \ref{tab:NewTexture4}, in this case one regenerates dimension five proton decay operators with the down-type Yukawas. For the couplings involving $\delta^1_{112I}$ one finds the following suppression:
\be \begin{array}{c|c|c|c|c}
\text{Model} & \text{Coupling} & \text{Charge} & \text{Singlet insertions} & \epsilon \text{ suppression}\\ \hline
\hbox{(BaEnGo1)} & \bten_1 \bten_1 \bten_2 \obfive_1 & (-5,-30) & \omega_1^5 \omega_2^3\ & \leq \epsilon^{13} \\
\hbox{(BaEnGo2)} & \bten_1 \bten_1 \bten_2 \obfive_1 & (25,-30) & \omega_1^5 \omega_2^3 & \leq \epsilon^{13}\\
\end{array}
 \ee
The suppression of these dimension five couplings is bounded as,
\be 
\frac{\epsilon^{13}}{M_{GUT}} \approx \frac{10^{-9}}{M_{GUT}} \leq 16 \pi^2 \frac{M_{SUSY}}{M_{GUT}^2}  \,,
\ee
this results in the following bound on the mass of the sparticles participating in the process:
\be
M_{SUSY} \geq 10^{-11} M_{GUT} \,.
\label{NewTexture1MSUSY}
 \ee
As in the earlier case of the Haba textures, the other operators $\delta^1_{3333}$ is further suppressed compared to $\delta^1_{112I}$ and thus not threatening to the consistency of the model. 

\subsubsection{$\mathcal{N}_{\obfive}=5$}
\label{sec:NewTextureFive5bs}

Finally, we discuss a solution, which has a distinct lepton flavor structure, by extending the solution in section \ref{subsec:II.3.4} to five $\obfive$s with 
\be  \ba 
M_{5_1} &= 0, \quad N_{5_1} = 2\\
M_{5_2} &= 2, \quad N_{5_2} = - 2 \\
M_{5_3} &= 1, \quad N_{5_3} = 0\,.
\ea \ee
In this case the charges of three $\bten$s, $\obfive_1$ and $\obfive_2$ are as in \eqref{II.3.4} and \eqref{AdditionalTenCharges}. However the charge of $\obfive_3$ is constrained not by the anomaly cancellation conditions, but by the requirement of suppressing the unwanted operators. The distribution of MSSM matter is
\be
\begin{array}{c||c|cc|c}
  \text{Representation} & \text{Charge} & M & N & \text{Matter} \\ \hline
 {\bf 10}_1      & (q_{10_1}^1,q_{10_1}^2)   & 1 & 0 & Q_1, \bar{u}_1, \bar{e}_3\\ 
 {\bf 10}_2    & (q_{10_2}^1,q_{10_2}^2)     & 1 & 0  & Q_2, \bar{u}_2, \bar{e}_2, \\
  {\bf 10}_3    & (-7,-7) & 1 & 0  & Q_3, \bar{u}_3, \bar{e}_1 \\
\bar{\bf 5}_{H_u} &   (-14,-14)   & 0 & -1 & H_u \\
 \bar{\bf 5}_{H_d} & (6,6) & 0 & 1  & H_d \\ 
 \bar{\bf 5}_1 & (-9,-9) & 0 & 2  & L_I, L_J, I,J = 1,2,3\\
 \bar{\bf 5}_2 & (1,1) & 2 & -2 &  \bar{d}_2, \bar{d}_3 \\
  \bar{\bf 5}_3 & (q_{5_3}^1,q_{5_3}^2) & 1 & 0 & L_K, \bar{d}_1
\end{array}
\ee
where there is a choice in  how the different generations of leptons are distributed, which is unfixed by the anomaly cancellation conditions. The general structure of the lepton Yukawas is given by,
\be 
Y^L \sim 
\left(
\begin{array}{ccc}
s_4 s_1 & s_4 s_2 & s_4 \\
s_4 s_1 & s_4 s_2 & s_4\\
s_3 s_1 & s_3 s_2 & s_3
\end{array} 
\right)\,,
\label{NewTexture5L}
\ee
where the singlets have charges
\be \ba 
(q_{S_3}^1, q_{S_3}^2) &= -5(w^1_{5_3} - w^1_{5_2},w^2_{5_3} - w^2_{5_2}) \\
(q_{S_4}^1, q_{S_4}^2) &= -5(w^1_{5_1} - w^1_{5_2},w^2_{5_1} - w^2_{5_2}) \,.
\ea \ee
The up-type and down-type Yukawa textures are given in \eqref{NewTexture1}.
One choice of charges, which we will denote as model (BaEnGo3), that does not allow unwanted operators at leading order is given by:
\be
\hbox{(BaEnGo3)}:\qquad 
\begin{array}{c||c|cc|c}
  \text{Representation} & \text{Charge} & M & N & \text{Matter} \\ \hline
 {\bf 10}_1      & (-12,13)  & 1 & 0 & Q_1, \bar{u}_1, \bar{e}_3\\ 
 {\bf 10}_2    & (-7,3)    & 1 & 0  & Q_2, \bar{u}_2, \bar{e}_2, \\
  {\bf 10}_3    & (-7,-7) & 1 & 0  & Q_3, \bar{u}_3, \bar{e}_1 \\
\bar{\bf 5}_{H_u} &   (-14,-14)   & 0 & -1 & H_u \\
 \bar{\bf 5}_{H_d} & (6,6) & 0 & 1   & H_d \\ 
 \bar{\bf 5}_1 & (-9,-9) & 0 & 2  & L_1, L_2\\
 \bar{\bf 5}_2 & (1,1) & 2 & -2 &  \bar{d}_2, \bar{d}_3 \\
 \bar{\bf 5}_3 & (-4,-9) & 1 & 0 & L_3, \bar{d}_1
\end{array}
\ee
A scan yields that there are no models, which give a lower bound on $M_{SUSY}$ than that derived in the case for four $\obfive$s. This model has been chosen as it produces the same bound for $M_{SUSY}$ as in \eqref{NewTexture1MSUSY} and does not regenerate any dimension four operators with any number of singlet insertions. 

In this case the charges of the singlets required to regenerate the up-type, $W_1$ and $W_2$, down-type, $W_3$, and lepton Yukawa matrices, $W_4$, 
have charges,
\be \ba
(q_{W_1}^1, q_{W_1}^2) &= (5,-20) \\
(q_{W_2}^1, q_{W_2}^2) &= (0,-10) \\
(q_{W_3}^1, q_{W_3}^2) &= (5,10) \\
(q_{W_4}^1, q_{W_4}^2) &= (10,10) \,,
\ea \ee
where the Yukawa matrices take the general forms \eqref{NewTexture1} and \eqref{NewTexture5L}. The singlet insertions, $\omega_1$ and $\omega_2$, expressed in terms of the Wolfenstein parameter, $\epsilon$, are,
\be 
\omega_1 = \epsilon^4 , \quad \omega_2 = \epsilon^2 , \quad  \omega_3 =\epsilon \,.
\ee

As was the case in the previous textures, regenerating the lepton Yukawas also regenerates dimension five operators. The dangerous couplings with coupling constant $\delta^1_{112I}$ are given below where we have written the singlet insertions which give rise to the lowest suppression,
\be \begin{array}{c|c|c|c|c}
\text{Model} & \text{Coupling} & \text{Charge} & \text{Singlet insertions} & \epsilon \text{ suppression}\\ \hline
\hbox{(BaEnGo3)} & \bten_1 \bten_1 \bten_2 \obfive_1 & (-40,20) & \omega_1^2 \omega_2 \omega_4^3 & \epsilon^{13} \\
& \bten_1 \bten_1 \bten_2 \obfive_3 & (-35,20) & \omega_1 \omega_2^3 \omega_4^3 + \omega_1^2 \omega_2 \omega_4^2 \omega_3 & \leq \epsilon^{13} 
\end{array}
 \ee
In the coupling involving $\obfive_3$ we have taken $\omega_4 =\epsilon$, which is consistent with the lepton mass hierarchies, in the estimation of the suppression. These suppression levels are the same as those derived in the previous section and give rise to the bound on $M_{SUSY}$ in \eqref{NewTexture1MSUSY}.
In the next section we examine model (BaEnGo1-3), as well as the  models (Haba1) and (Haba2) of section \ref{subsec:II.3.4} under the constraints of lepton and neutrino flavor.

Finally, we should note, that extending the current analysis to more ${\bf \bar{5}}$ matter we find, that for $\mathcal{N}_{\obfive} = 6$ there no solutions with suitable flavor structure. It would be interesting to extend this to  $\mathcal{N}_{\obfive} = 7$ (which is the largest for this type of models without introducing exotics), however increasing the number of $\obfive$s usually brings back proton decay operators. 


\section{Lepton and Neutrino Flavor}
\label{sec:leptonmixing}

Let us now turn to the lepton and neutrino flavor properties of the F-theoretic FN-models of the last section. Unlike the quark sector and the lepton masses \eqref{MassRatios}, the neutrino sector is far less experimentally constrained. Nevertheless let us state the respective experimental bounds on the masses\footnote{These are best-fit values and $3\sigma$ allowed ranges for neutrino masses with a normal hierarchy ($m_1<m_2<m_3$). The sum of neutrino masses and the other values can be found in the Neutrino mass, mixing and oscillations chapter of \cite{pdg}.}
\be
\ba
\Delta m_{12}^2 \left[10^{-5}{\rm eV}^2\right] &= 7.54^{+0.64}_{-0.56}\cr
\Delta m_{23}^2 \left[10^{-3}{\rm eV}^2\right] &= 2.43^{+0.18}_{-0.20}\cr
\sum m_{\nu_j} &< 0.66\ {\rm eV} 
\ea
\ee
and mixing angles
\be
\theta_{12}= 0.59^{+0.05}_{-0.06}\ ,\quad  \theta_{23} =  0.72^{+0.19}_{-0.06}\ ,\quad  \theta_{13} = 0.15^{+0.02}_{-0.02} \,.
\label{eq:neutrinoconstraints}
\ee
The absolute masses for the neutrinos are not known and various different hierarchies could be accommodated within these constraints. Furthermore the mixing angles are not hierarchical. Nevertheless, we show that our F-theoretic FN-models from above can accommodate the mixing angles. 

The neutrino masses can arise from a so-called standard type I seesaw mechanism, for which we introduce three right-handed neutrinos that are $SU(5)$ singlets but are charged under the additional $U(1)$s. The couplings needed are
\be
(Y_\nu)_{IJ} \bfive_{L_I} \bfive_{H_u} {\bf 1}_{\nu_R^J},\quad  M_{IJ} {\bf 1}_{\nu_R^I}{\bf 1}_{\nu_R^J} \,,
\ee
where $M_{IJ}$ is generated by singlets with a vev. Below the mass scale of the right-handed neutrinos $\Lambda$ this leads to an effective neutrino mass via the Weinberg operator
\be
\frac{1}{\Lambda}L_IL_J H_u H_u \,.
\ee
Again this operator can be forbidden by the additional $U(1)$ symmetries, but regenerated by appropriate singlet insertions. Note that the charges of the right-handed neutrinos do not enter the effective Weinberg operator and are not relevant for the discussion of neutrino mixing.
For the flavor models in section \ref{sec:flavor}, the three distinct phenomenlogical scenarios are studied in turn in the following.

\subsection{Models (Haba1) and (Haba2)}

Including the structure of the neutrino Yukawa matrix {arising from the} Weinberg operators, the models from section \ref{subsec:II.3.4} have the following Yukawa structures
\be \ba
Y^u_{\rm Haba} &\sim\left(\begin{array}{ccc}
\epsilon^8 & \epsilon^6 & \epsilon^4\\
\epsilon^6 & \epsilon^4 & \epsilon^2\\
\epsilon^4 & \epsilon^2 & 1
\end{array}\right) &
 Y^d_{\rm Haba} &\sim\left(\begin{array}{ccc}
\epsilon^4 & \epsilon^4 & \epsilon^4\\
\epsilon^2 & \epsilon^2 & \epsilon^2\\
1 & 1 & 1
\end{array}\right)\cr
Y^L &\sim\left(\begin{array}{ccc}
\epsilon^{4+a} & \epsilon^{2+a} & \epsilon^a\\
\epsilon^{4+a} & \epsilon^{2+a} & \epsilon^a\\
\epsilon^4 & \epsilon^2 & 1
\end{array}\right) &
M_\nu &\sim 
\left(\begin{array}{ccc}
\kappa_1 & \kappa_1 & \kappa_2 \\
\kappa_1 & \kappa_1 & \kappa_2\\
\kappa_2 & \kappa_2 & \kappa_1 \kappa_2 + \kappa_1^3 
\end{array}\right) \,,
\label{eq:pheno1}
\ea\ee
where the charges of singlets regenerating entries in the neutrino Yukawa matrices are given by
\be \ba
\hbox{(Haba1)}:\qquad & q_{K_1} = (-10,-10), \ q_{K_2} = (-20,-20) \\
\hbox{(Haba2)}:\qquad & q_{K_1} = (0,-10), \ q_{K_2} = (0,-20) \,.
\ea \ee

Note also that despite the fact that the quark mixing are those in Haba \cite{Haba:1998wf} (and BaEnGo \cite{Babu:2003zz} in section \ref{sec:BaEnGoLep})  the lepton and neutrino textures are distinct from the models in the literature. 

For each choice of hierarchical singlet scalings, we scan over the $O(1)$ coefficients in front of each coupling and identify experimentally viable masses and mixings using the {\tt Mathematica} package {\tt Mixing Parameter Tools} \cite{Antusch:2005gp}. For each of the three different scalings we find consistent mixing angles with suitable choices for the $O(1)$ coefficients and mass hierarchies  that are consistent with \eqref{MassRatios}. We allow the $O(1)$ coefficients, $z$, for the Yukawa matrices to be within the range
\be 
0.8 < |z| < 1.2 \,,
\ee
where in the case of the lepton Yukawa matrices $z$ is complex.

For models (Haba1), (Haba2) with $a = 0.05, \kappa_1 = 0.1$ and $\kappa_2 = 0.3$ one choice of $O(1)$ coefficients for the lepton and neutrino Yukawa matrices which give consistent mixing angles is given by
\be   \ba
Y^L &\sim \left(\begin{array}{ccc}
1.00 \epsilon^{4.05} & -(0.68 + 0.97  i)\epsilon^{2.05} & (0.30 - 0.78i)\epsilon^{0.05}\\
(-0.89 + 0.54 i)\epsilon^{4.05} & -(0.14 + 1.13i)\epsilon^{2.05} & -(0.43 + 1.10 i)\epsilon^{0.05}\\
(0.64 + 0.63 i)\epsilon^4 & -(0.12 + 0.97 i)\epsilon^2 & -0.81 - 0.10i
\end{array}\right) \\
M_{\nu} &\sim \left(\begin{array}{ccc}
1.17 \kappa_1 & 1.13  \kappa_1 & 0.92  \kappa_2 \\
1.13  \kappa_1 & 0.92 \kappa_1 & 0.83 \kappa_2\\
0.92 \kappa_2 & 0.83 \kappa_2 & 0.96( \kappa_1 \kappa_2 + \kappa_1^3)
\end{array}\right) \,.
\ea \ee
This choice for $a$ means that $s_2$ in these models is a $O(1)$ number as was assumed in the calculation of the bound for $M_{SUSY}$ in section \ref{subsec:II.3.4}. This set of matrices gives the following mass ratios and mixing angles
\be \ba 
\theta_{12} = 0.60, \quad \theta_{13} &= 0.20, \quad \theta_{23} =  0.72 \\
m_{\tau} : m_{\mu} : m_e &= 1: 0.88 \epsilon^2 : 0.63 \epsilon^4 \,,
\ea \ee
which are consistent with the constraints in \eqref{MassRatios} and \eqref{eq:neutrinoconstraints}. 

More interestingly, one can take $a = 1$ and still find $O(1)$ coefficients which give rise to good mixing angles and lepton mass hierarchies. This choice for $a$ improves the bound on $M_{SUSY}$ in \eqref{E8MUSY} to
\be 
M_{SUSY} \geq 10^{-10} M_{GUT} \,,
\ee
as now $s_2 = 0.22$. In this case the other singlets take values $\kappa_1 = 0.7, \kappa_2 = 0.7$ and the Yukawa matrices are given by
\be   \ba
Y^L &\sim \left(\begin{array}{ccc}
1.00 \epsilon^{5} & (-0.79 + 0.27 i)\epsilon^{3} & (0.72 - 0.61 i)\epsilon\\
-(0.70 + 0.87 i)\epsilon^{5} & (-0.86 + 0.57 i)\epsilon^{3} & (0.99 + 0.01i)\epsilon\\
(0.98 - 0.30 i)\epsilon^4 & (0.32 - 1.08 i)\epsilon^2 & 0.34 - 0.80 i
\end{array}\right) \\
M_{\nu} &\sim \left(\begin{array}{ccc}
0.90\kappa_1 & 0.98  \kappa_1 & 1.07  \kappa_2 \\
0.98  \kappa_1 & 1.16\kappa_1 & 0.89 \kappa_2\\
1.07\kappa_2 & 0.89 \kappa_2 & 1.00 ( \kappa_1 \kappa_2 + \kappa_1^3)
\end{array}\right) \,.
\ea \ee
The mixing angles and mass hierarchies are in very good agreement with those in \eqref{eq:neutrinoconstraints} and \eqref{MassRatios}
\be \ba 
\theta_{12} = 0.56, \quad \theta_{13} &= 0.14, \quad \theta_{23} =  0.71 \\
m_{\tau} : m_{\mu} : m_e &= 1: 0.68 \epsilon^2 : 1.02 \epsilon^5 \,.
\ea \ee

\subsection{Models (BaEnGo1)$-$(BaEnGo3)}
\label{sec:BaEnGoLep}

For the matter distributions in the F-theoretic FN-models (BaEnGo1) and (BaEnGo2) of section \ref{sec:BaEnGo} we find that the leptons and neutrinos are different from the models in \cite{Babu:2003zz}, and are given by
\be \ba
Y_{\rm BaEnGo}^u &\sim\left(\begin{array}{ccc}
\epsilon^8 & \epsilon^6 & \epsilon^4\\
\epsilon^6 & \epsilon^4 & \epsilon^2\\
\epsilon^4 & \epsilon^2 & 1
\end{array}\right) &
Y_{\rm BaEnGo}^d &\sim\left(\begin{array}{ccc}
\epsilon^5 & \epsilon^4 & \epsilon^4\\
\epsilon^3 & \epsilon^2 & \epsilon^2\\
\epsilon & 1 & 1
\end{array}\right)\cr
Y^L &\sim\left(\begin{array}{ccc}
\epsilon^5 & \epsilon^3 & \epsilon^1\\
\epsilon^5 & \epsilon^3 & \epsilon^1\\
\epsilon^5 & \epsilon^3 & \epsilon^1
\end{array}\right) &
M_\nu &\sim \left(\begin{array}{ccc}
\kappa_1 & \kappa_1 & \kappa_1\\
\kappa_1 & \kappa_1 & \kappa_1\\
\kappa_1 & \kappa_1 & \kappa_1
\end{array}\right) \,,
\label{eq:pheno2}
\ea\ee
where the singlets have charges
\be 
\hbox{(BaEnGo1)}:\  q_{K_2} = (-10,-10), \quad \hbox{(BaEnGo2)}:  q_{K_2} = (0,-10) \,.
\ee
Likewise for model (BaEnGo3) we get
\be \ba
Y_{\rm BaEnGo}^u &\sim\left(\begin{array}{ccc}
\epsilon^8 & \epsilon^6 & \epsilon^4\\
\epsilon^6 & \epsilon^4 & \epsilon^2\\
\epsilon^4 & \epsilon^2 & 1
\end{array}\right) &
Y_{\rm BaEnGo}^d &\sim\left(\begin{array}{ccc}
\epsilon^5 & \epsilon^4 & \epsilon^4\\
\epsilon^3 & \epsilon^2 & \epsilon^2\\
\epsilon & 1 & 1
\end{array}\right)\cr 
Y^L &{\sim\left(\begin{array}{ccc}
\epsilon^{4+c} & \epsilon^{2+c} & \epsilon^c\\
\epsilon^{4+c} & \epsilon^{2+c} & \epsilon^c\\
\epsilon^5 & \epsilon^3 & \epsilon
\end{array}\right)} &
M_\nu &\sim \left(\begin{array}{ccl}
\kappa_1 & \kappa_1 & \kappa_1 \kappa_2 \\
\kappa_1 & \kappa_1  & \kappa_1 \kappa_2\\
\kappa_1 \kappa_2  & \kappa_1 \kappa_2 &  \kappa_1 \kappa_2^2
\end{array}\right) \,,
\label{eq:pheno3}
\ea\ee
where
\be
q_{K_1} = (-10,-10)\,, \quad  q_{K_2} = (-5,0)\,.
\ee
{For (BaEnGo1) and (BaEnGo2) the following $O(1)$ coefficients in the lepton and neutrino Yukawa matrices
\be   \ba
Y^L &\sim \left(\begin{array}{ccc}
(1.09 - 0.04i) \epsilon^{5} & (-1.11 + 0.42i)\epsilon^{3} & (-0.13 - 0.91i)\epsilon \\
(0.19 + 1.05i)\epsilon^{5} & (-0.88 + 0.31i)\epsilon^{3} & (1.04 - 0.52i)\epsilon\\
(-0.21 + 0.93 i)\epsilon^5 & (-0.24 + 1.00i)\epsilon^3 & 0.97 + 0.10i \epsilon
\end{array}\right) \\
M_{\nu} &\sim \left(\begin{array}{ccc}
0.81 \kappa_1 & 1.10  \kappa_1 & 1.03  \kappa_1 \\
1.10  \kappa_1 & 1.11 \kappa_1 & 1.05 \kappa_1\\
1.03 \kappa_1 & 1.05 \kappa_1 & 1.01\kappa_1
\end{array}\right) 
\ea \ee
with $\kappa_1 = 0.22$ result in PMNS mixing angles and lepton mass hierarchies, which are consistent with the phenomenological constraints \eqref{MassRatios} and \eqref{eq:neutrinoconstraints}
\be \ba 
\theta_{12} = 0.60, \quad \theta_{13} &= 0.18, \quad \theta_{23} =  0.69 \\
m_{\tau} : m_{\mu} : m_e &= 1: 0.92\epsilon^2 : 0.59 \epsilon^5 \,.
\ea \ee
This model fits precisely the anarchy models in \cite{Haba:2000be}.

Finally, consider FN-model  (BaEnGo3), where in addition to the quark Yukawa matrices in (\ref{eq:pheno3}) one finds the following set of Lepton and neutrino Yukawa matrices for $c = 1, \kappa_1 = 0.2$ and $\kappa_2 = 0.4$
\be   \ba
Y^L &\sim \left(\begin{array}{ccc}
(-0.92+0.08i) \epsilon^{5} &(1.06+0.36i)\epsilon^{3} & (0.69-0.57i)\epsilon \\
(0.89-0.33i)\epsilon^{5} & (1.00+0.10i)\epsilon^{3} &(0.30-0.77i)\epsilon\\
(0.38+0.82i)\epsilon^5 & (1.02+0.19i)\epsilon^3 & (0.53+0.61i)\epsilon
\end{array}\right) \\
M_{\nu} &\sim \left(\begin{array}{ccc}
0.82\kappa_1 &  0.88\kappa_1 & 0.85\kappa_1 \kappa_2 \\
 0.88 \kappa_1 & 0.94\kappa_1 &  1.10\kappa_1\kappa_2\\
 0.85\kappa_1\kappa_2 &  1.10\kappa_1\kappa_2 & 0.93\kappa_1 \kappa_2^2
\end{array}\right) \,.
\ea \ee
The resulting mixing angles and lepton mass ratios are 
\be \ba 
\theta_{12} = 0.61, \quad \theta_{13} &= 0.16, \quad \theta_{23} = 0.71 \\
m_{\tau} : m_{\mu} : m_e &= 1: 0.92\epsilon^2 :0.97 \epsilon^4 \,,
\ea \ee
which again are phenomenologically sound. 
}

\section{Geometric Realization}
\label{sec:GeoRealisation}
In this section we discuss how some of the phenomenologically viable models can be realized geometrically. For the case of the two $U(1)$ models, some of the solutions in section \ref{sec:TwoU1sGeo} can be realized in terms of a general cubic in $\mathbb{P}^2$. For the F-theoretic FN-models, we have not determined a 
geometric construction, however we provide the necessary fiber types, that realize the charge patterns underlying these flavor models. 

\subsection{Single $U(1)$ Models}

For one $U(1)$ there is exactly one model that is consistent, denoted by I.1.4.a in table \ref{tab:I.1.4}. All other models bring back in one way or another the dimension four or five proton decay operators. In addition the single $U(1)$ models have very limited scope with respect to flavor. Nevertheless to geometrically engineer the solution I.1.4.a one has to consider the codimension one fiber type $I_5^{(01)}$. As one can see however, the charges in the model are wider separated than in known constructions. We will focus our attention on the phenomenologically more interesting multiple $U(1)$ models.


\subsection{Two $U(1)$ Models}
\label{sec:TwoU1Geo}

In section \ref{sec:TwoU1sGeo}, the charge spectrum of the four models, with two $U(1)$ symmetries, which solved the anomaly cancellation conditions and forbid dangerous proton decay operators were detailed. In this section we show how three of these models can be constructed by considering elliptic fibrations with two additional rational sections, described by enhancing the singularity type of the general cubic in $\mathbb{P}^2$ \cite{CKPT}. 

The elliptically fibered Calabi-Yau four-fold, as a hypersurface in an ambient five-fold, is given by the following cubic equation
\be 
w(s_1 w^2 + s_2 w x + s_3 x^2 + s_5 wy + s_6 xy + s_8 y^2) + \prod_{i=1}^3 (a_i x+ b_i y) =0
\label{GenP111} 
\ee
where $[w:x:y]$ are projective coordinates in $\mathbb{P}^2$. This fibration has three rational sections given by
\be 
\sigma_0: [0 : -b_1 : a_1], \qquad \sigma_1: [0 : -b_2 : a_2], \qquad \sigma_2: [0 : -b_3 : a_3] \,,
\ee
By expanding the coefficients above, which we denote generally as $c_i$, along a coordinate in the base, $z$, as
\be 
c_i = \sum_{j = 1}^{\infty} c_{i,j} z^j \,,
\ee
singularities can be tuned along the locus $z = 0$. The coefficients $s_{i,j}, a_{i,j}$ and $b_{i,j}$ are sections of the following holomorphic line bundles over the base shown in table \ref{tab:sections}, where $K_B$ is the pullback of the canonical class of the base, $B$, and $S_G$ is the class of $z$.
\begin{table}
$$\begin{array}{c|c} 
\text{Section} & \text{Line bundle} \\ \hline
s_{1,j} & \mathcal{O}(-6K_B - 2[a_1] - 2[a_2] - 2[a_3] -3[s_8]-jS_G) \\
s_{2,j} & \mathcal{O}(-4K_B - [a_1] - [a_2] - [a_3] -2[s_8]-jS_G) \\
s_{3,j} & \mathcal{O}(-2K_B - [s_8]-jS_G) \\
s_{5,j} & \mathcal{O}(-3K_B - [a_1] - [a_2] - [a_3] -[s_8]-jS_G) \\
s_{6,j} & \mathcal{O}(-K_B -jS_G) \\
s_{8,j} & \mathcal{O}([s_8]-jS_G) \\
a_{1,j} & \mathcal{O}([a_1]-jS_G) \\
a_{2,j} & \mathcal{O}([a_2]-jS_G) \\
a_{3,j} & \mathcal{O}([a_3]-jS_G) \\
b_{1,j} & \mathcal{O}(K_B + [a_1] + [s_8]-jS_G) \\
b_{2,j} & \mathcal{O}(K_B + [a_2] + [s_8]-jS_G) \\
b_{3,j} & \mathcal{O}(K_B + [a_3] + [s_8]-jS_G)
\end{array}$$
\caption{Classes of the sections for the elliptic fibration realised in terms of a general cubic in $\mathbb{P}^2$.\label{tab:sections}}
\end{table}

As we are interested in $SU(5)$ GUTs we will only consider models which realize $I_5$ singularities. To determine this, we apply Tate's algorithm to the general cubic. 
Resolving the $I_5$ singularities introduces four exceptional curves $F_i$ into the fiber. The fibration of each $F_i$ over the singular locus $z$ gives a divisor $D_{F_i}$.   With each rational section, in addition to the zero-section $\sigma_0$, we can define the Shioda map, $S(\sigma_i)$ such that
\be 
S(\sigma_i)\cdot_Y F_j = 0 \,.
\ee
Here $\cdot_Y$ denotes that the intersection is taken in the fourfold $Y$.
The Shioda map constructs from each rational section a divisor which corresponds to the generator of the $U(1)$ symmetry. The $U(1)$ charges of matter are found by intersecting $S(\sigma_i)$ with the matter curves obtained from the splitting of the $F_j$ in codimension two. The resolutions and intersections carried out in this paper are computed using the {\tt Mathematica} package {\tt Smooth} \cite{Smooth}.

Here, we label our models as in \cite{Lawrie:2014uya},  where the vanishing orders, $n_{c_i}$, are given in the order
\be 
(n_{s_1},n_{s_2},n_{s_3},n_{s_5},n_{s_6},n_{s_8},n_{a_1},n_{b_1},n_{a_2},n_{b_2},n_{a_3},n_{b_3})\,.
\ee
Furthermore it will be necessary to consider so-called non-canonical models, where the enhancement of the discriminant to $O(z^5)$, occurs not by simply specifying the vanishing order of the coefficients, but by subtle cancellations between the coefficients, which are non-trivially related see e.g. \cite{Katz:2011qp, Kuntzler:2014ila,Lawrie:2014uya}. In the models we consider here the enhancement to $I_5$ requires solving 
\be 
AB - CD = 0 \,,
\ee 
in terms of the coefficients of the hypersurface equation. This has to be  solved over the coordinate ring of the base of the elliptic fibration, which is a unique factorization domain. Applying the standard Tate's algorithm in this context \cite{Katz:2011qp, Kuntzler:2014ila,Lawrie:2014uya} the enhancement is obtained as a so-called non-canonical solution in terms of sections $\xi_i$
\be 
A = \xi_1 \xi_2, \quad
B = \xi_3 \xi_4, \quad
C = \xi_1 \xi_3, \quad
D = \xi_2 \xi_4 \,.
\ee
In addition to specifying the vanishing order, the labelling of the non-canonical models also includes the specialisation of the coefficients in terms of $\xi_i$, which is given underneath the vanishing orders.

The models which realize the solutions in section \ref{sec:TwoU1sGeo} are given in table \ref{tab:GeoModels}. For each model the fiber type, vanishing orders and non-canonical specialisation is given along with the charges of the {\bf 10} and $\bar{\bf 5}$ matter in the model. The equations for the matter loci refered to in the table are given below:
\be
\xi_2 \xi_3 \xi_4^2 s_{3,0} + \xi_1 \xi_4 (\xi_3^2 a_{3, 1} + \xi_2^2 a_{1,0} b_{2, 0} s_{5, 0} - \xi_2 \xi_3 s_{6,1}) - \xi_1^2 (\xi_3^2 b_{3, 1} + \xi_2^2 b_{1, 0} b_{2, 0} s_{5, 0} - \xi_2 \xi_3 s_{8, 1}) =0
\label{Model1Last5}
\ee
\be 
a_{1, 0}^2 b_{2, 0} b_{3, 0} s_{5, 0} + b_{1, 0} s_{6, 0}^2 - a_{1, 0} s_{6, 0} s_{8, 0} =0
\label{Model2Last5}
\ee
\be 
\xi_4^2 (\xi_3^2 s_{1,1} - \xi_2 \xi_3 s_{2,1} + \xi_2^2 s_{3, 0}) - \xi_1 \xi_4 (\xi_2^2 a_{1, 0} a_{3, 0} b_{2, 0} + \xi_3^2 s_{5,1} - \xi_2 \xi_3 s_{6,   1}) + \xi_1^2 \xi_3 (-\xi_2 a_{1, 0} b_{2, 0} b_{3,0} + \xi_3 s_{8, 0}) =0
\label{Model3Last5}
\ee
These models provide new charge configurations that have thus far not been obtained in the literature. 

Each of the models in table \ref{tab:GeoModels} have additional charged matter, which is not present in the corresponding solutions given in section \ref{sec:TwoU1sGeo}, which can be forbidden in the base.  
As an aside: As noted in the table, the charges for the non-canonical model $(3,2,1,1,0,0,0,0,1,0,0,0)$ under the first $U(1)$ is reversed to those in solution II.1.6.a. 
This can be further seen by the fact that the fiber type of this model is not one considered in the analysis in section \ref{sec:TwoU1sGeo}, as was noted earlier. This is justified as the charges in an $I_5^{(i|j|k)}$ model are the same as those in $I_5^{(i|jk)}$ except with the sign of one of the $U(1)$s reversed. As the anomaly cancellation conditions are unaffected by global rescalings of the $U(1)$ charges, the model in the table  solves the anomaly cancellation condition as in solution II.1.6.a.

\begin{sidewaystable}
\centering
{\footnotesize \begin{tabular}{|c|c|c|c|c|}
\hline
Solution & Fiber & Vanishing Orders & Matter Locus & Matter \\ \hline
\multirow{9}{*}{II.1.6.a} & \multirow{9}{*}{$I_{5,nc}^{(1|0|2)}$} & \multirow{9}{*}{(3,2,1,1,0,0,0,0,1,0,0,0)} &  $\xi_1$ & $\bten_{2,-2} \oplus \overline{\bten}_{-2,2}$ \\
\multirow{9}{*}{(Sign of charges under}& & \multirow{9}{*}{$[-,-,-,-, \xi_1 \xi_3, \xi_3 \xi_4, -,-,-,-,\xi_1 \xi_2, \xi_2 \xi_4]$} & $\xi_3$ & $\bten_{2,3} \oplus \overline{\bten}_{-2,-3}$  \\
\multirow{9}{*}{first $U(1)$ is reversed)}& & & $\xi_2$ & $\bfive_{6,-1}\oplus\obfive_{-6,1}$ \\
& & & $a_{1,0}$ & $\bfive_{-4,-6}\oplus\obfive_{4,6}$ \\
& & & $b_{2,0}$ & $\bfive_{1,-6}\oplus\obfive_{-1,6}$ \\
& & & $\xi_4 a_{1, 0} - \xi_1 b_{1, 0}$ & $\bfive_{6,4}\oplus\obfive_{-6,-4}$ \\
& & & $\xi_1 \xi_3 a_{2, 0} - b_{2, 0} s_{3, 0}$ & $\bfive_{1,4}\oplus\obfive_{-1,-4}$ \\
& & & $\xi_1^2 \xi_3^2 s_{1, 0} - \xi_1 \xi_3 s_{2, 0} s_{5,0} + s_{3, 0} s_{5, 0}^2$ & $\bfive_{1,-1}\oplus\obfive_{-1,1}$  \\
& & & \eqref{Model1Last5} & $\bfive_{-4,-1}\oplus\obfive_{4,1}$ \\ \hline
\multirow{8}{*}{II.1.6.b} & \multirow{8}{*}{$I_5^{(0|2|1)}$} & \multirow{8}{*}{(3,2,1,1,0,1,0,1,1,0,0,0)} &  $s_{6,0}$ & $\bten_{-2,1} \oplus \overline{\bten}_{2,-1}$ \\
& & & $a_{1,0}$ & $\bfive_{-6,-7}\oplus\obfive_{6,7}$ \\
& & & $a_{3,0}$ & $\bfive_{4,-2}\oplus\obfive_{-4,2}$ \\
& & & $b_{2,0}$ & $\bfive_{-1,-7}\oplus\obfive_{1,7}$ \\
& & & $b_{3,0}$ & $\bfive_{-6,-2}\oplus\obfive_{6,2}$ \\
& & & $b_{2, 0} s_{3, 0} - a_{2, 0} s_{6, 0}$ & $\bfive_{-1,3}\oplus\obfive_{1,-3}$ \\
& & & $s_{3, 0} s_{5, 0}^2 - s_{6, 0}s_{2, 0} s_{5, 0} + s_{1, 0} s_{6, 0}^2$ & $\bfive_{-1,-2}\oplus\obfive_{1,2}$  \\
& & & \eqref{Model2Last5} & $\bfive_{4,3}\oplus\obfive_{-4,-3}$ \\ \hline
\multirow{9}{*}{II.1.6.c} & \multirow{9}{*}{$I_{5,nc}^{(0|2|1)}$} & \multirow{9}{*}{(2,1,1,1,0,1,0,1,1,0,0,0)} &  $\xi_3$ & $\bten_{-2,1} \oplus \overline{\bten}_{2,-1}$ \\
& & \multirow{9}{*}{$[\xi_1\xi_2,\xi_1\xi_3,-,\xi_2\xi_4,\xi_3\xi_4,-,-,-,-,-,-,-]$} & $\xi_4$ & $\bten_{3,1} \oplus \overline{\bten}_{-3,-1}$  \\
& & & $a_{1,0}$ & $\bfive_{-6,-7}\oplus\obfive_{6,7}$ \\
& & & $a_{3,0}$ & $\bfive_{4,-2}\oplus\obfive_{-4,2}$ \\
& & & $b_{2,0}$ & $\bfive_{-1,-7}\oplus\obfive_{1,7}$ \\
& & & $b_{3,0}$ & $\bfive_{-6,-2}\oplus\obfive_{6,2}$ \\
& & & $\xi_3 \xi_4^2 a_{2, 0} - b_{2, 0}\xi_4 s_{3, 0} +\xi_1 a_{1, 0} a_{3, 0} b_{2, 0}^2)$ & $\bfive_{-1,3}\oplus\obfive_{1,-3}$ \\
& & & $\xi_3^2 \xi_4 b_{1, 0} + \xi_2 a_{1, 0}^2 b_{2, 0} b_{3,0} - \xi_3 a_{1, 0} s_{8, 0}$ & $\bfive_{4,3}\oplus\obfive_{-4,-3}$  \\
& & & \eqref{Model3Last5} & $\bfive_{-1,-2}\oplus\obfive_{1,2}$ \\ \hline
\end{tabular}
\caption{Geometric realizations of the models with two $U(1)$s  corresponding to the solutions II.1.6.a, II.1.6.b and II.1.6.c. 
The charges under the first $U(1)$ in solution II.1.6.a are reversed with respect to those in the model. 
\label{tab:GeoModels} }
}
\end{sidewaystable}

\subsection{Fibers for Models (Haba1) and (Haba2)}
\label{sec:FibsFN}
The F-theoretic FN-models in section \ref{subsec:II.3.4} have particularly nice phenomenology in addition to satisfying all anomaly constraints and absence of dangerous couplings. The charges for those models are within the classification of the F-theory charges \cite{Lawrie:2015hia}, however so far no concrete geometric realization is known. To guide the construction of these geometries, we now provide the possible fiber types necessary for these models in the following for the models in table \ref{tab:E8}. 

The models are based on $I_5^{(02|1)}$, where the two additional sections $\sigma_1$ and $\sigma_2$ generate the two extra $U(1)$ symmetries. For simplicity we discuss the model (Haba2) in table \ref{tab:E8} -- for model (Haba1) the only change is that the two extra sections have the same charges, for the ${\bf \bar{5}}$ matter loci, and thus have the same configurations. The fibers in codimension two, including the configuration of the sections is shown in figure \ref{fig:FibersForFlavor}. We shall refrain from providing the details of this result and refer the reader to \cite{Lawrie:2015hia}, where a comprehensive discussion of these fibers was obtained. 

The main difficulty in constructing this class of models is that the charges are separated, e.g.  the ${\bf\bar{5}}$ charges have a range from $q_2= -14$ to $6$, i.e. $q^{max}_{\bf \bar{5}}-q^{min}_{\bf \bar{5}} = 20$,
which is in current algebraic constructions not observed. Generically the charge differences are $q^{max}_{\bf \bar{5}}-q^{min}_{\bf \bar{5}} = 10$, with the only example, known to us, with this difference given by $15$ is a toric construction obtained in \cite{Braun:2013yti}. It would be very interesting to systematically search for models with wider separation of charges. 
One complication is of course, that the codimension two fibers will have to be more and more wrapped, i.e. there will be components in the codimension two fibers that are contained within the section, as shown in figure \ref{fig:FibersForFlavor}.


\begin{figure}
  \centering
  \includegraphics[width= 15cm]{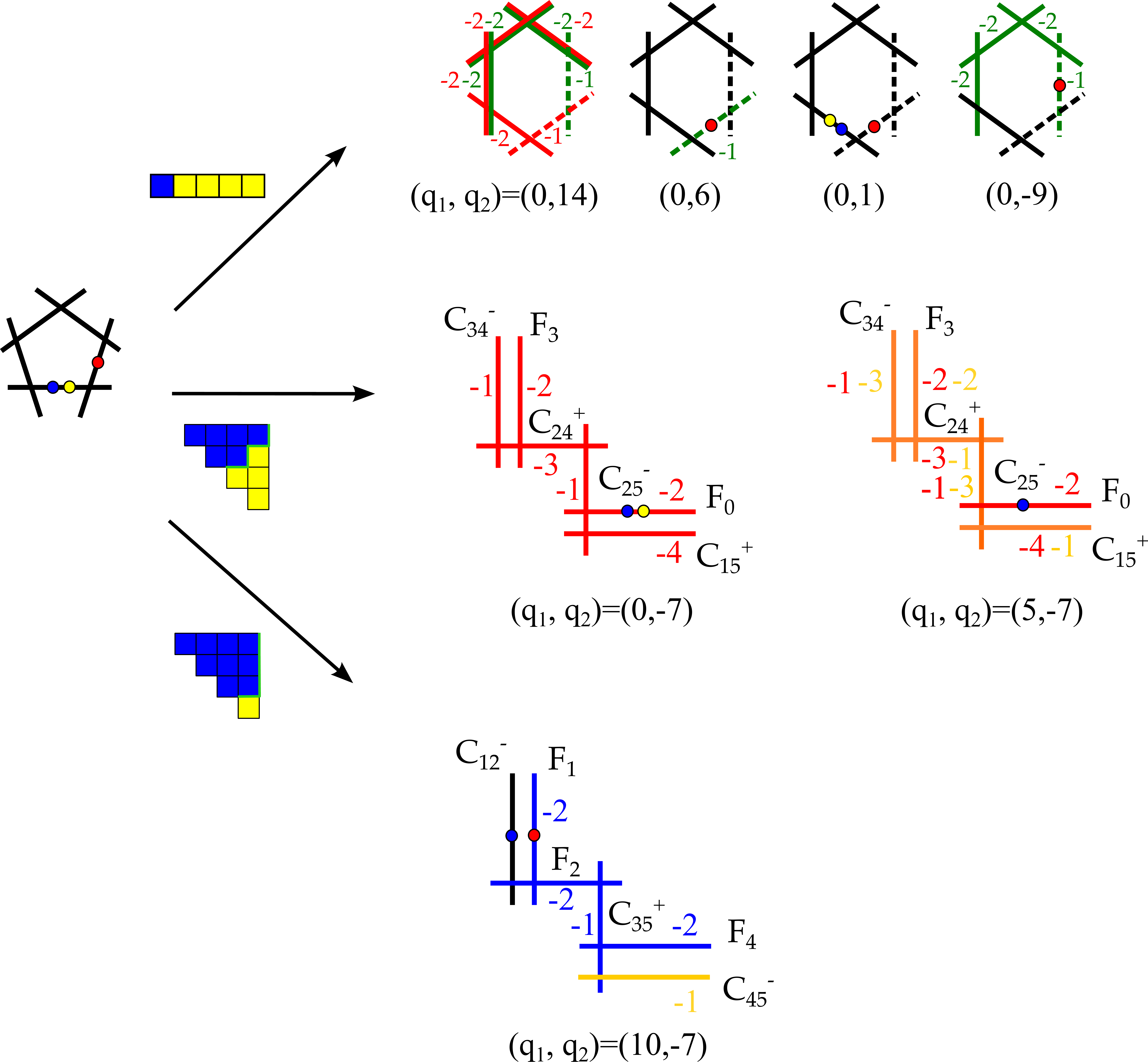}
  \caption{Fibers for the F-theoretic FN-model (Haba2) of table \ref{tab:E8}. The codimension one type $I_5^{(02|1)}$ is as in figure \ref{fig:I5twoSec} (up to permutation of the two extra sections), with the zero-section shown in blue. 
  The nomenclature is as in \cite{Lawrie:2015hia}: The codimension two fibers realizing the ${\bf \bar{5}}$ matter ($I_6$) as well as ${\bf 10}$ matter ($I_1^*$ fibers) are shown together with their charges. The coloring correspond to the wrapping of the fibers, and the labels along the wrapped components correspond to the degrees of the normal bundle, which in turn determine the charges. 
For the ${\bf \bar{5}}$ matter, the blue and yellow sections have to have the same configurations, as the charge is zero. These are shown in terms of green coloring.
  The blue/yellow colored representation graphs (box graphs) indicate the phases of the respective resolution type, see \cite{Hayashi:2014kca}. }
    \label{fig:FibersForFlavor}
\end{figure}
\section{Conclusions}

We have shown that there are viable models in the class of F-theory charge configurations from the classification in \cite{Lawrie:2015hia}, which satisfy all consistency requirements (A1.)$-$(A5.) and (C1.)$-$(C7.), and have realistic flavor physics, however these are very scarce. 

We considered one or two $U(1)$ symmetries, although our analysis can be easily extended to three or more  $U(1)$s. 
For single $U(1)$ models there is one solution, which does not regenerate any of the dangerous couplings at the same order as the Yukawa couplings, however single $U(1)$ models have very limited scope with regards to FN-type modeling. 
For two $U(1)$s we studied two sets of solutions: one which solves all the constraints and has an explicit geometric construction -- albeit coming short on the flavor physics. The second class of solutions have realistic flavor textures generated by an FN-mechanism, which we studied in both quark and lepton sectors, however their geometric construction is unknown -- these were denoted by (Haba1)-(Haba2) and (BaEnGo1)-(BaEnGo3), according to the Yukawa textures for the quarks. We provided the required fiber types for (Haba1) and (Haba2) and hope that our result gives a guidance to the geometric efforts to construct more elaborate F-theory compactifications.
It would be very exciting to find a geometric construction of these models summarized in table \ref{tab:E8}. This includes the construction of the  elliptic fibration with two rational sections as well as the $G$-flux, in particular also the hypercharge flux, that induces the necessary matter distributions as detailed by the $M$ and $N$ values. 
{Furthermore, combining our general insights from the structure of the elliptic fiber with recent advances on the understanding of the base of the fibration of four-folds \cite{Morrison:2014lca, Halverson:2015jua} would lead to a very powerful way to constrain the set of phenomenologically viable F-theory vacua. }


\subsection*{Acknowledgements}

We thank  Herbi Dreiner, Denis Klevers, Craig Lawrie, Eran Palti, J\"urgen Rohrwild, Graham Ross, Jim Talbert for discussions and the authors of \cite{CKPT} for sharing the generalization of the two $U(1)$ models to general cubics, which allowed us to determine the geometric models in section \ref{sec:TwoU1Geo} using Tate's algorithm. The work of SK is supported by the ERC Starting Grant ``Supersymmetry Breaking in String Theory". The work of SSN and JW is supported in part by STFC grant ST/J002798/1.

\appendix

\section{Multiple {\bf 10} curves for single $U(1)$ Models}
\label{app:Multi10}

In this appendix we provide details on multiple ${\bf 10}$ representations for single $U(1)$ models, completing the analysis in section \ref{sec:OneU1}. 
We find only models with $\mathcal{N}_{\bten} = 2$ and $\mathcal{N}_{\obfive} = 4$ solve the anomaly cancellation conditions and forbid the unwanted operators at leading order. These models regenerate dimension five proton decay operators with the remaining charged Yukawas, which if sufficiently suppressed, could still leave these models phenomenologically viable. However their flavor physics is highly constrained and does not yield phenomenologically interesting textures.

\subsection{$\mathcal{N}_{\bf 10} =2$}

For the case of multiple $\bf 10$ representations with one $U(1)$ symmetries 
it is possible to have top Yukawa couplings of the form,
\be
{\bf 10}_{q_1} {\bf 10}_{q_2} {\bf 5}_{-q_1 - q_2} \,,
\label{DiffTenTY}
\ee
where the two $\bf 10$ representations do not have the same charge under the $U(1)$. This means we can make use of the full set of charges in \eqref{FCharges} and, in particular, we do not require one of the $\bf 10$ representations to have a U(1) charge within the set given in \eqref{10ChargeswithTY}.

In this case the general parameterization will be of the form,
\be
\begin{array}{c|c|c|c}
{\bf R} &  q({\bf R}) & M & N \cr\hline
\bar{\bf 5}_{H_u} &   -q_{H_u}  &0 & -1 \cr
\bar{\bf 5}_{H_d}  & q_{H_d} &0& 1 \cr
\bar{\bf 5}_{i} &  q_{\bar{5}_i} &M_i &N_i  \cr\hline
{\bf 10}_1  &  q_{10_1} &  M_{10} & N_{10} \cr
{\bf 10}_2  &  q_{10_2} &  3- M_{10} & - N_{10}
\end{array}
\ee
where $i = 1, \ldots, \mathcal{N}_{\bar{\bf 5}}$, the latter being the number of $\bar{\bf 5}$ representations.


\subsubsection{$\mathcal{N}_{\bar{\bf 5}} =3$}
Here the anomaly cancellation conditions can be solved for general charges.   There are two possible parameterizations, which differ in the structure of the top Yukawa coupling.
\be
\begin{array}{cccc}
\begin{array}{c|c|c|c}
{\bf R} &q({\bf R}) & M & N \cr\hline
\bar{\bf 5}_{H_u} &  - q_{H_u} &0 & -1 \cr
\bar{\bf 5}_{H_d}  &  -q_{H_u} +  5  w_{H_d}   &0& 1 \cr
\bar{\bf 5}_{1} &  -q_{H_u} + 5 w_{\bar{5}_1}  & 3 & 0   \cr
\hline
{\bf 10}_1  &  -{1\over 2} q_{H_u}&   M_{10} & N_{10} \cr
{\bf 10}_2  &  -{1\over 2} q_{H_u} + 5 w_{10} &   3 - M_{10} & -N_{10}
\end{array}
&&&
\begin{array}{c|c|c|c}
{\bf R} &q({\bf R}) & M & N \cr\hline
\bar{\bf 5}_{H_u} & -  q_{H_u} &0 & -1 \cr
\bar{\bf 5}_{H_d}  &  -q_{H_u} +  5  w_{H_d}   &0& 1 \cr
\bar{\bf 5}_{1} &  -q_{H_u} + 5 w_{\bar{5}_1}  & 3 & 0   \cr
\hline
{\bf 10}_1  &  q_{10} &   M_{10} & N_{10} \cr
{\bf 10}_2  &  -q_{10}-q_{H_u} &   3 - M_{10} & -N_{10}
\end{array} \\
\text{(a)} &&& \text{(b)}
\end{array}
\ee
In the above parameterization, $N_{10} = 0,\pm 1$, this is to ensure the absence of exotics. However setting $N_{10} = 0$ only gives solutions where the $\mu$-term is allowed at leading order, therefore we neglect this case and focus on $N_{10} = \pm 1$. For parameterization (a) the top Yukawa coupling is of the standard form,
\be 
{\bf 10}_{q_1} {\bf 10}_{q_1} H_u \,,
\ee
where the two $\bf 10$s have the same charge under the $U(1)$. In (b) the top Yukawa coupling is of the form given in \eqref{DiffTenTY}. Below we outline the solution for (a) but a very similar analysis can be done for (b).



\begin{sidewaystable}
\centering
{\footnotesize
\begin{tabular}{|c||c|c|}
\hline
            & {I.2.3.a}      & {I.2.3.b} \\ \hline \hline
 $M_{10}$      & 1/2          & 1/2  \\
 $N_{10}$      & 1            & 1  \\ \hline
 $q_{10_1}$ & $-1$           & 0  \\
 $q_{10_2}$ & 0            & $-1$  \\
 $q_{H_u}$  & 2           & $1$ \\
 $q_{H_d}$  & $-1$         & $-2$  \\
 $q_{\bar{5}_1}$      & $w_{\bar{5}_1} -2$  & $w_{\bar{5}_1}-1$ \\ \hline
 $Y^t_{ab}$ & $\{\{0,1\},\{1,2\}\}$  & $\{\{1,0\},\{0,-1\}\}$ \\ 
 $Y^b_a$    & $\{w_{\bar{5}_1}-4, w_{\bar{5}_1}-3\}$ & $\{w_{\bar{5}_1}-3,w_{\bar{5}_1}-4\}$ \\ \hline
 $\mu$      & 1            & $-1$           \\ \hline
 \multirow{2}{*}{C2}         &$\{w_{\bar{5}_1}-5,w_{\bar{5}_1}-4,$     & $\{w_{\bar{5}_1}-1,w_{\bar{5}_1}-2$   \\
		& $w_{\bar{5}_1}-3,w_{\bar{5}_1}-2\}$ & $w_{\bar{5}_1}-3,w_{\bar{5}_1}-4\}$  \\ \hline
 C3         &$w_{\bar{5}_1}$        & $w_{\bar{5}_1}$ \\ \hline
 \multirow{2}{*}{C4}         & $2w_{\bar{5}_1}-5$   & $2w_{\bar{5}_1}-2$ \\
           & $2w_{\bar{5}_1}-4$   & $2w_{\bar{5}_1}-3$ \\ \hline
 \multirow{2}{*}{C5}         & $\{-w_{\bar{5}_1},1-w_{\bar{5}_1},$  & $\{1-w_{\bar{5}_1},-w_{\bar{5}_1},$ \\
 & $2-w_{\bar{5}_1}\}$  & $-1-w_{\bar{5}_1}\}$ \\ \hline
 C6         & $w_{\bar{5}_1}+1$   & $w_{\bar{5}_1}-1$ \\ \hline
 C7         & $\{-4,-3\}$ & $\{-3,-4\}$\\ \hline
\end{tabular}
\quad \quad 
\begin{tabular}{|c||c|c|}
\hline
             & {I.2.4.a}                         & {I.2.4.b}                     \\ \hline \hline
 $M_1$       & 0                           & 0/1                      \\
 $N_1$       & 3                           & 2                     \\ \hline
 $M_{10}$     & 1/2                         & 1/2                    \\
 $N_{10}$     & $-1$                          & 0                   \\ \hline 
 $q_{10_1}$  & $-3 $                        & $-3 $                    \\
 $q_{10_2}$  & $-1$                          & $-1$                      \\
 $q_{H_u}$   & 2                          & 2                    \\
 $q_{H_d}$   & 2                          & 2                      \\
 $q_{\bar{5}_1}$       & $-3 $                        & $-1$                      \\
 $q_{\bar{5}_2}$       & $-1 $                         & 1                       \\ \hline
 $Y^t_{a,b}$ & $\{\{-4,-2\},\{-2,0\}\}$ & $\{\{-4,-2\},\{-2,0\}\}$ \\ 
 $Y^b_{a,j}$ & $\{\{-4,-2\},\{-2,0\}\}$ & $\{\{-2,0\},\{0,2\}\}$ \\ \hline
 $\mu$       & 4                          & 20                       \\ \hline
 C2          & $\{-12,-10, -8,-6,-4\}$   & $\{-10,-8,-6,-4,-2\}$ \\
 C3          & $\{-1,1\}$                  & $\{1,3\}$              \\
 C4          & $\{-9,-7,-5,-3\}$       & $\{-5,-3,-1,1\}$     \\
 C5          & $\{-3,-5,1,-1\}$          & $\{-5,-7,-1,-3\}$       \\
 C6          & $\{3,5\}$                 & $\{5,7\}$           \\
 C7          & $\{-3,-1\}$                & $\{-3,-1\}$          \\ \hline
\end{tabular}
\caption{Solutions for one $U(1)$, 
$\mathcal{N}_{\bf 10} = 2$  and $\mathcal{N}_{\obfive} = 3$ (LHS) and $4$ (RHS). The charges of the Yukawa couplings are labelled by $Y^t_{ab}$ and $Y^b_{aj}$, where $a = 1,2$ ($j = 1,2$) specifies the ${\bf 10}$ ($\bar{\bf 5}$). All models regenerate either dimension four or five proton decay operators, although in the latter case the operators are less problematic as will be discussed in the main text. 
\label{tab:I.2.3} }
}
\end{sidewaystable}
The anomaly condition (A2.) imposes $w_{H_d} = w_{10} N_{10}$, which upon  imposing  (A3.) yields
\be \label{quadcon210s}
w_{10} N_{10} (q_{H_u} + 5 w_{10} (N_{10}-3)) = 0 \,.
\ee
As we require the two $\bten$s to be different charged and $w_{H_d} \neq 0$ to avoid the $\mu$-term thus
$w_{10} , N_{10} \neq 0$,
the only allowed solution to \eqref{quadcon210s} is given by 
$q_{H_u} =- 5 w_{10}(N_{10}-3)$.
The charges, which satisfy the anomaly conditions are 
\be
\begin{array}{c|c|c|c}
{\bf R} &  q({\bf R}) & M & N \cr\hline
\bar{\bf 5}_{H_u} &  5w_{10}(N_{10}-3) &0 & -1 \cr
\bar{\bf 5}_{H_d}  &  5 w_{10} (2N_{10}-3)   &0& 1 \cr
\bar{\bf 5}_{1} & 5 (w_{\bar{5}_1} + w_{10} (N_{10}-3))  & 3 & 0   \cr
\hline
{\bf 10}_1  &  \frac{5}{2} w_{10} (N_{10} - 3) &   M_{10} & N_{10} \cr
{\bf 10}_2  & \frac{5}{2} w_{10} (N_{10}-1) &   3 - M_{10} & -N_{10}
\end{array}
\ee 
Imposing a bottom Yukawa coupling with ${\bf 10}_1$  gives the additional constraint,
\be 
w_{\bar{5}_1} = \frac{w_{10}}{2}(15 - 7N_{10}) \,.
\ee
There are solutions to the above set of charges, which satisfy the F-theory charge pattern, which we summarize in  table \ref{tab:I.2.3}. Here we have not imposed the presence of the bottom Yukawa coupling explicitly as these solutions correspond to one particular choice for $w_{\bar{5}_1}$. Restricting to the F-theory charge range, \eqref{10ChargeswithTY} and \eqref{FCharges}, we are constained to take $w_{10}\pm 1$, and, without loss of generality, we take $w_{10} = 1$ as the two choices differ by an overall factor of $-1$ in normalisation of the $U(1)$ charges. Likewise we have taken $q_{H_u} = 5$ in case (b). All the possible choices for $w_{\bar{5}_1}$, which are within the F-theory charge range, are given by, 
\be  
\ba
\hbox{$w_{\bar{5}_1}$ for} &\quad 
\left\{ 
\ba
\text{I.2.3.a} &\in \left\{-1,2,3,4,5 \right\} \cr
\text{I.2.3.b} &\in \left\{-2,1,2,3,4 \right\} \,.
\ea
\right.
\cr
\ea
\ee
As one can see from the general solutions, all but one of these allow either the dimension five proton decay operators (C2.) or (C6.) and are therefore excluded. The case which forbids the unwanted operators at leading order is given by $w_{\bar{5}_1} = -2$ in I.2.3.b however in this model dimension four proton decay operators are regenerated with bottom Yukawa couplings and therefore is also not a viable model. 


\subsubsection{$\mathcal{N}_{\bar{\bf 5}} =4$}

For $N_{\bar{\bf 5}} \geq 4$, the strategy for finding solutions to the anomaly conditions is to take all possible sets of $\bf 10$ and $\bar{\bf 5}$ charges, selected from \eqref{FCharges} and find those, which can solve (A1.)$-$(A5.) for allowed $M, N$s. The two solutions shown in table \ref{tab:I.2.3} solve the anomaly cancellation conditions, forbid operators (C1.)$-$(C7.) and do not regenerate dimension four proton decay operators with the charged Yukawas. They do, however, regenerate dimension five proton decay operators, which, if sufficiently suppressed, could still give viable models.

The matter in the MSSM can be allocated to the $U(1)$ charged $\bf 10$ and $\bar{\bf 5}$ representations in model I.2.4.a as follows:
\be
\begin{array}{c||c|cc|c}
  \text{Representation} & \text{Charge} & M & N & \text{Matter} \\ \hline
 {\bf 10}_1      & -3    & 1 & -1 & Q_1, \bar{u}_1, \bar{u}_2, \\ 
 {\bf 10}_2    & -1     & 2 & 1  & Q_2, Q_3, \bar{u}_3, \bar{e}_A,A = 1,2,3\\
 \bar{\bf 5}_{H_u} & -2    & 0 & -1 & H_u \\
 \bar{\bf 5}_{H_d} & 2     & 0 & 1  & H_d \\ 
 \bar{\bf 5}_1 & -3   & 0 & 3  & L_I, I = 1,2,3\\
 \bar{\bf 5}_2 & -1     & 3 & -3 & \bar{d}_I, I = 1,2,3 
\end{array}
\label{SolI.2.4.a}
\ee
In this spectrum the following couplings are allowed by the additional $U(1)$ symmetry
\be \ba 
Y^t_{22} {\bf 10}_2 {\bf 10}_2 {\bf 5}_{H_u} & \supset Q_3 \bar{u}_3 H_u \\
Y^b_{24} {\bf 10}_2 \bar{\bf 5}_{H_d} \bar{\bf 5}_2 & \supset Q_3 \bar{d}_3 H_d + Q_2 \bar{d}_2 H_d \,.
\ea \ee
In order to regenerate the remaining Yukawa couplings one needs the singlet of charge 2  to acquire a vev, which however, also regenerates all dimension five operator, with various suppressions. This model may still be viable from the point of view of proton decay, with sufficient suppresion, however, the flavor physics
based on an FN-type model is not very realistic, and we therefore discard these solutions. 

For model I.2.4.b the spectrum is given by
\be
\begin{array}{c||c|cc|c}
  \text{Representation} & \text{Charge} & M & N & \text{Matter} \\ \hline
 {\bf 10}_1    & -3   & 2 & 0  & Q_1, Q_2, \bar{u}_1, \bar{u}_2, \bar{e}_1, \bar{e}_2 \\ 
 {\bf 10}_2    & -1     & 1 & 0  & Q_3, \bar{u}_3, \bar{e}_3\\
 \bar{\bf 5}_{H_u} & -2    & 0 & -1 & H_u \\
 \bar{\bf 5}_{H_d} & 2     & 0 & 1  & H_d \\ 
 \bar{\bf 5}_1 & -1    & 1 & 2  & \bar{d}_3, L_I, I = 1,2,3\\
 \bar{\bf 5}_2 & 1    & 2 & -2 & \bar{d}_1, \bar{d}_2 
\end{array}
\label{SolI.2.4.b}
\ee
For this model there were two sets of $M,N$s which solved the anomaly cancellation conditions, the values displayed in \eqref{SolI.2.4.b} are the ones compatible with having rank one top and bottom Yukawa matrices at tree level.
In order to regenerate the top Yukawa coupling involving the two differently charged $\bten$s a singlet of charge 2 is required, and the same remarks as for I.2.4.a apply. 


\subsubsection{$\mathcal{N}_{\bar{\bf 5}} \geq 5$}

All but one of the models, for $N_{\obfive} = 5$,regenerate the dimension four operator (C4.) with the missing Yukawa couplings. 
The remaining model however is inconsistent with the hierarchy of Yukawa couplings. 
 For the cases of six and seven $\bar{\bf 5}$ representations there are no solutions, which both solve the anomaly cancellation conditions and forbid the unwanted operators, in agreement with what was found for a single $\bf 10$ representation. 



\subsection{$\mathcal{N}_{\bf 10} =3$}
This case is maximal for the number of $\bten$ representations and has the greatest potential for generating a Yukawa texture with good quark mass ratios. However, by increasing the number of $\bten$s one increases the chance of generating forbidden couplings, in particular the operator (C5.) becomes unavoidable in most models. There is only one solution to the anomaly cancellation conditions which forbids the unwanted couplings at leading order.
This solution, I.3.4.a
\be
\begin{array}{c||c|cc}
  \text{Representation} & \text{Charge} & M & N  \\ \hline
 {\bf 10}_1    & -3    & 1 & 0  \\ 
 {\bf 10}_2    & -2    & 1 & 0 \\
 {\bf 10}_3    & -1     & 1 & 0 \\
 \obfive_{H_u}  & -2    & 0 & -1  \\
 \bar{\bf 5}_{H_d} & 1      & 0 & 1   \\ 
 \bar{\bf 5}_1 & -1    & 0 & 3  \\
 \bar{\bf 5}_2 & 0      & 3 & -3 
\end{array}
\label{SolI.3.4.a}
\ee
A full rank Yukawa matrix can be generated by giving a vev to the singlet of charge $1$. 
This model interestingly generates the Haba textures (\ref{HabaText}), however 
one also regenerates  dimension four proton decay operators with a singlet insertion of the singlet, which is phenomenologically inacceptable. 

In conclusion we see that for a single $U(1)$ the solution space is very limited -- even disregarding flavor problems -- and for  solutions to the anomalies and constraints on couplings, 
generically the Yukawas bring back the unwanted couplings at subleading order. 


\section{General Solution for $\mathcal{N}_{\bf 10} =1$ and $\mathcal{N}_{\bar{\bf 5}} =4$ with multiple $U(1)$s}
\label{app:Four5bsMultiU1s}

\subsection{Two $U(1)$s}
In this appendix the general solution for the case of one $\bten$ and four $\obfive$s is derived for two $U(1)$s. This class of solutions, which give rise to good phenomenological models, is given in table \ref{tab:II.1.4}. The extension of the solutions for the case of two $U(1)$s to multiple $U(1)s$ is also discussed.
Consider a model with two abelian factors, parameterized as
\be
\begin{array}{c|c|c|c}
{\bf R} &  q({\bf R})^{\alpha} & M & N \cr\hline
\bar{\bf 5}_{H_u} &  - q_{H_u}^{\alpha}  &0 & -1 \cr
\bar{\bf 5}_{H_d}  & -q_{H_u}^{\alpha} + 5 w_{H_d}^{\alpha} &0& 1 \cr
\bar{\bf 5}_{1} &  -q_{H_u}^{\alpha} + 5w_{\bar{5}_1}^{\alpha} &M &N  \cr 
\bar{\bf 5}_{2}& -q_{H_u}^{\alpha} + 5 w_{\bar{5}_2}^{\alpha} &3-M& -N \cr\hline
{\bf 10}  &  q_{\bf 10} = -{1\over 2} q_{H_u}^{\alpha}&   3&0 
\end{array}
\label{Four5bsMultiU1s}
\ee
where $q_i^{\alpha}$ denotes the charges under $U(1)_{\alpha}$, $\alpha = 1,2$. Without loss of generality, we take $N\geq 0$.
The linear anomaly (A2.) for each abelian factor is of the form
\be
w_{H_d}^{\alpha} + N(w_{\bar{5}_1}^{\alpha} - w_{\bar{5}_2}^{\alpha})=0\,,
\label{II.1.4A2}
\ee
which can be solved for $w_{H_d}^{\alpha}$. Inserting this equation into the quadratic set of anomalies (A3.), we have,
\be \ba 
N(w_{\bar{5}_1}^1 - w_{\bar{5}_2}^1) (w_{\bar{5}_1}^1 + w_{\bar{5}_2}^1 + (w_{\bar{5}_1}^1 - w_{\bar{5}_2}^1) N) &= 0\\
N(w_{\bar{5}_1}^2 - w_{\bar{5}_2}^2) (w_{\bar{5}_1}^2 + w_{\bar{5}_2}^2 + (w_{\bar{5}_1}^2 - w_{\bar{5}_2}^2) N) &= 0\\
N (w_{\bar{5}_1}^1 w_{\bar{5}_1}^2 - w_{\bar{5}_2}^1 w_{\bar{5}_2}^2  + (w_{\bar{5}_1}^1  - w_{\bar{5}_2}^1) (w_{\bar{5}_1}^2 - w_{\bar{5}_2}^2) N) &= 0\,.
\ea 
\label{II.1.4A3}
\ee
Setting $N= 0$ solves all the anomaly conditions simultaneously but from \eqref{II.1.4A2} we see that this results in the presence of the $\mu$-term at tree-level, which is unfavorable. We therefore neglect this class of solutions. The first two quadratic anomalies can be solved in three distinct ways:
\begin{itemize}
\item[a)] $w_{\bar{5}_1}^1 = w_{\bar{5}_2}^1, w_{\bar{5}_1}^2 = w_{\bar{5}_2}^2$
\item[b)] $w_{\bar{5}_1}^1 = \frac{(N-1)}{N+1} w_{\bar{5}_2}^1, w_{\bar{5}_1}^2 = \frac{(N-1)}{N+1}w_{\bar{5}_2}^2$
\item[c)] $w_{\bar{5}_1}^1 = w_{\bar{5}_2}^1, w_{\bar{5}_1}^2 = \frac{(N-1)}{N+1}w_{\bar{5}_2}^2$
\end{itemize}
The sets of charges from these three possibilities are given below.
\begin{itemize}
\item[a)] Upon the insertion of $w_{\bar{5}_1}^1 = w_{\bar{5}_2}^1, w_{\bar{5}_1}^2 = w_{\bar{5}_2}^2$ into the third anomaly condition in \eqref{II.1.4A3} the mixed quadratic anomaly is automatically solved. The $U(1)$ charges in this case are
\be 
{\renewcommand{\arraystretch}{1.2}
\begin{array}{c||c|c|c|c|c}
   & \bten & \bfive_{H_u} & \obfive_{H_d} & \obfive_1 & \obfive_2 \\ \hline\hline
 q^1({\bf R}) & -\frac{1}{2} q_{H_u}^1 & q_{H_u}^1 & -q_{H_u}^1 & -q_{H_u}^1 + 5 w_{\bar{5}_2}^1 & -q_{H_u}^1 + 5 w_{\bar{5}_2}^1 \\
 q^2({\bf R}) & -\frac{1}{2} q_{H_u}^2 & q_{H_u}^2 & -q_{H_u}^2 & -q_{H_u}^2 + 5 w_{\bar{5}_2}^2& -q_{H_u}^2 + 5 w_{\bar{5}_2}^2
\end{array}}
\label{II.1.4CaseA}
\ee 
This pair of $U(1)$s always gives rise to the $\mu$-term at leading order and therefore does not give phenomenologically favorable models.
\item[b)] Here the solutions for $w_{\bar{5}_1}^1$ and $w_{\bar{5}_1}^2$ have the same form as in the single $U(1)$ case. The mixed anomaly in \eqref{II.1.4A3} is automatically solved and the charges for each $U(1)$ are 
\be 
{\renewcommand{\arraystretch}{1.2}
\begin{array}{c||c|c|c|c|c}
   & \bten & \bfive_{H_u} & \obfive_{H_d} & \obfive_1 & \obfive_2 \\ \hline\hline
 q^1({\bf R}) & -\frac{1}{2} q_{H_u}^1 & q_{H_u}^1 & -q_{H_u}^1 + \frac{10 N}{1 + N}w_{\bar{5}_2}^1 & -q_{H_u}^1 + \frac{5 (N-1)}{1+N} w_{\bar{5}_2}^1 & -q_{H_u}^1  + 5w_{\bar{5}_2}^1 \\
 q^2({\bf R}) & -\frac{1}{2} q_{H_u}^2 & q_{H_u}^2 & -q_{H_u}^2 + \frac{10 N}{1 + N}w_{\bar{5}_2}^2 & -q_{H_u}^2 + \frac{5(N-1)}{1+N} w_{\bar{5}_2}^2  & -q_{H_u}^2  + 5w_{\bar{5}_2}^2
\end{array}}
\label{II.1.4CaseB}
\ee
\item[c)] The mixed anomaly in \eqref{II.1.4A3} is not automatically solved, but instead it reduces to
\be 
\frac{w_{\bar{5}_2}^1 w_{\bar{5}_2}^2 N}{1 + N} = 0 \,.
\label{II.1.4CaseC}
\ee
The charges for the two different solutions to \eqref{II.1.4CaseC} are:
\begin{itemize}
\item[i)] $w_{\bar{5}_2}^1 = 0$
\be 
{\renewcommand{\arraystretch}{1.2}
\begin{array}{c||c|c|c|c|c}
   & \bten & \bfive_{H_u} & \obfive_{H_d} & \obfive_1 & \obfive_2 \\ \hline\hline
 q^1({\bf R}) & -\frac{1}{2} q_{H_u}^1 & q_{H_u}^1 & -q_{H_u}^1 & -q_{H_u}^1  & -q_{H_u}^1  \\
 q^2({\bf R}) & -\frac{1}{2} q_{H_u}^2 & q_{H_u}^2 & -q_{H_u}^2 + \frac{10 N}{1 + N}w_{\bar{5}_2}^2 & -q_{H_u}^2 + \frac{5(N-1)}{1+N}w_{\bar{5}_2}^2 & -q_{H_u}^2  + 5w_{\bar{5}_2}^2
\end{array}}
\label{II.1.4CaseCi}
 \ee
\item[ii)] $w_{\bar{5}_2}^2 = 0$
\be 
{\renewcommand{\arraystretch}{1.2}
\begin{array}{c||c|c|c|c|c}
   & \bten & \bfive_{H_u} & \obfive_{H_d} & \obfive_1 & \obfive_2 \\ \hline\hline
 q^1({\bf R}) & -\frac{1}{2} q_{H_u}^1 & q_{H_u}^1 & -q_{H_u}^1  & -q_{H_u}^1 + 5 w_{\bar{5}_2}^1 & -q_{H_u}^1 + 5 w_{\bar{5}_2}^1 \\
 q^2({\bf R}) & -\frac{1}{2} q_{H_u}^2 & q_{H_u}^2 & -q_{H_u}^2  & -q_{H_u}^2 & -q_{H_u}^2 
\end{array}}
\label{II.1.4CaseCii}
\ee
This set of charges also does not forbid the $\mu$-term at leading order and therefore is disfavored.
\end{itemize}
\end{itemize}

{\renewcommand{\arraystretch}{1.4}
\begin{table}
\centering
\begin{tabular}{|c||c|c|}
\hline
                 & {\bf II.1.4.a}                             & {\bf II.1.4.b} \\ \hline \hline
 $M$             & 0/1                                    & 0/1  \\
 $N$             & 2                                      & 2  \\ \hline
 $q_{10_1}$      & $(-\half q^1_{H_u}, -\half q^2_{H_u})$ & $(-\half q^1_{H_u}, -\half q^2_{H_u})$   \\
 $q_{H_u}$       & $(q^1_{H_u}, q^2_{H_u})$               & $(q^1_{H_u}, q^2_{H_u})$\\
 $q_{H_d}$       & $(-q^1_{H_u} + \frac{20}{3}w^1_{5_2}, -q^2_{H_u}+\frac{20}{3}w^2_{5_2})$ & $(-q^1_{H_u}, -q^2_{H_u}+\frac{20}{3}w^2_{5_2})$ \\
$q_{\bar{5}_1}$  & $(-q^1_{H_u} +\frac{5}{3}w^1_{5_2}, -q^2_{H_u}+\frac{5}{3}w^2_{5_2})$    & $(-q^1_{H_u} , -q^2_{H_u}+\frac{5}{3}w^2_{5_2})$ \\
$q_{\bar{5}_2}$  & $(-q^1_{H_u} +5w^1_{5_2}, -q^2_{H_u}+5w^2_{5_2})$                        & $(-q^1_{H_u}, -q^2_{H_u}+5w^2_{5_2})$  \\ \hline
 $Y^b_1$           & $(-\frac{5}{2} q^1_{H_u} + \frac{25}{3}w^1_{5_2}, -\frac{5}{2} q^2_{H_u} + \frac{25}{3}w^2_{5_2})$                                  & $(-\frac{5}{2} q^1_{H_u} , -\frac{5}{2} q^2_{H_u} + \frac{25}{3}w^2_{5_2})$\\
 $Y^b_2$           & $(-\frac{5}{2} q^1_{H_u} + \frac{35}{3}w^1_{5_2}, -\frac{5}{2} q^2_{H_u} + \frac{35}{3}w^2_{5_2})$                                  & $(-\frac{5}{2} q^1_{H_u} , -\frac{5}{2} q^2_{H_u} + \frac{35}{3}w^2_{5_2})$ \\ \hline
\end{tabular}
\caption{Solution for $\mathcal{N}_{\bf \bar{5}}=4$, $\mathcal{N}_{\bf 10} =1$ for two $U(1)$s. 
\label{tab:II.1.4}}
\end{table}}

Excluding the cases where the tree-level $\mu$-term is not forbidden by the additional $U(1)$ symmetries we are left with only case b) and ci). In both cases setting $N = 1$ results in the separation between the charges of $\obfive_1$ and $\obfive_{H_u}$ becoming zero, this produces a leading order coupling of the form (C5.)
\be
\bten_1 \bten_1 \bfive_1 \,,
\ee
which is forbidden. Similarly, $N = 3$ can be excluded as these cases always regenerate dimension four proton decay operators with the remaining charged Yukawa couplings. This can be seen from the charge relation
\be 
q^{\alpha}_{H_d} + q^{\alpha}_{\bar{5}_1} = 2 q^{\alpha}_{\bar{5}_2} \,, \quad \alpha = 1,2\, ,
\ee
which is true only when $N =3$. This relation implies that the charge of the dimension four proton operator coupling $\bten_1$ and $\obfive_2$ will be the same as the bottom Yukawa couplings for $\obfive_1$.

The charges for case (b) and (ci) for $N = 2$ are given in table \ref{tab:II.1.4}. If the charges are to remain within the F-theory charge set then $w^{\alpha}_{5_2} = \pm 3$, $\alpha = 1,2$ and the charges of $H_u$ are restricted to,
\be  
\ba
\hbox{$q^{\alpha}_{H_u}$} \quad \hbox{ for} &\quad 
\left\{ 
\ba
 I_5^{(01)} &\in \left\{-2,+2 \right\} \cr
I_5^{(0|1)} &\in \left\{-14,+ 6 \right\} \cr
I_5^{(0||1)} &\in \left\{- 8, + 12 \right\}  \,.
\ea
\right.
\cr
\ea
\ee
Each distinct pair of charges $(q^1_{H_u}, q^2_{H_u})$ gives a phenomenologically viable model, which forbids the unwanted operators at leading order.

Imposing the presence of a bottom Yukawa coupling further constrains the sets of possible $U(1)$ charges. For II.1.4.a the requirement of a bottom Yukawa coupling with either $\obfive_1$ or $ \obfive_2$ gives solutions where all matter is charged the same under both $U(1)$s. In model II.1.4.b requiring a bottom Yukawa coupling forces all matter to be completely uncharged under one of the two $U(1)$s.  Thus in both cases, the solutions reduce to the single $U(1)$ models I.1.4.a and I.1.4.c given in table \ref{tab:I.1.4}. Extending to two additional $U(1)$ symmetries results in no new models, if one requires the presence of a bottom Yukawa coupling.

\subsection{Extension to Multiple $U(1)$s}
The pairs of matter charges for two $U(1)$s, determined above, can be combined to give models charged under multiple $U(1)$s. Every pair of $U(1)$s must solve the anomaly cancellation conditions in one of the cases a), b), ci) or cii). From examining the charges in each case one can rule out certain combinations of the four different pairs of $U(1)$ charges. One obtains four types of models with multiple $U(1)$s: 
\begin{itemize}
\item[] Type A: Charges from case a) and case cii) are combined in one model
\item[] Type B: Charges from case a) are combined in one model
\item[] Type C: Charges from case b) and case ci) are combined in one model
\item[] Type D: Charges from case b) are combined in one model
\end{itemize}
Models of type A and B are phenomenologically disfavored as the $\mu$-term is always present at leading order. This can be seen from the charges in \eqref{II.1.4CaseA} and \eqref{II.1.4CaseCii}. All models of type C and D can be obtained by combining the charges which arise in II1.4.a and II.1.4.b in table \ref{tab:II.1.4}, however none of these allow for a leading order bottom Yukawa coupling. This implies that all multiple $U(1)$ models in this case, with F-theory charges and a bottom Yukawa coupling at leading order, are trivial extensions of the single $U(1)$ solutions I.1.4.a and I.1.4.c in table \ref{tab:I.1.4}.


\section{General solutions to Anomaly Equations}
\label{app:Mordell}

Solving the anomaly constraints  in generality for multiple matter curves can be quite difficult. 
Here we provide some systematic approach how to do so.  The quadratic anomaly (A3.) is a diophantine equation in terms of the 
$U(1)$ charges and integer multiplicities $M$ and $N$, and we will use some methods from Mordell's work in \cite{Mordell} to find general solutions. 
Note that for the case of the restricted F-theory charge range (as we can simply scan through all the possibilities), these methods are not necessary, however it provides an elegant approach to finding closed forms of the solutions. 

{We would like to stress that this approach can be used to classify all possible solutions allowed after imposing the constraints (A1.)-(A5.) and (C1.)-(C7.). This approach allows to classify all phenomenologically allowed solutions and can be used to survey all field-theoretically allowed FN models. It is similar to the approach taken in~\cite{Ibanez:1991pr,Dreiner:2005rd} where anomaly free, flavor universal gauge symmetry extensions to the MSSM were classified.}

\subsection{Mordell's solution for  Diophantine equations}
\label{sec:GeneralSol}

Consider one $U(1)$ with $\mathcal{N}_{\bf 10}=1$ and  $\mathcal{N}_{\bar{\bf 5}} =n$. We will now solve the system of anomaly constraints using a method of Mordell.
First let us set up the equations: the matter spectrum in this case  takes the following form, where the top Yukawa coupling is already imposed 
\be
\begin{array}{l|c|c|c}
{\bf R} &  q({\bf R}) & M & N \cr\hline
\bar{\bf 5}_{H_u} & -  q_{H_u}  &0 & -1 \cr
\bar{\bf 5}_{H_d}  & q_{H_d} &0& 1 \cr
\bar{\bf 5}_{i\not=n} &  q_i &M_i &N_i  \cr 
\bar{\bf 5}_{n}& q_n &3- \sum_{i=1}^{n-1}M_i & -\sum_{i=1}^{n-1} N_i \cr\hline
{\bf 10}  &  q_{\bf 10} = -{1\over 2} q_{H_u}&   3&0 
\end{array}
\ee
The constraints on the integers $M_i$ and $N_i$ is 
\be
0 \leq  M_i \leq  3\,, \qquad 
0 \leq  M_i + N_i \leq  3 \,,\qquad 
\sum_{i=1}^{n-1} {M_i}\leq  3 \,,\qquad  (N_1, \ldots, N_n) \not= (0, \ldots, 0) \,.
\ee
Imposing the anomaly constraint
\be
\hbox{(A2.)} \quad \Rightarrow \quad q_{H_d} =  - q_{H_u} +  \sum_{i=1}^{n-1} N_i (q_n-q_i)  \,.
\ee
This automatically implies that the charge of the $\mu$-term is
\be
q_{\mu} = q_{H_u} + q_{H_d} =   \sum_{i=1}^{n-1} N_i (q_n-q_i) \,.
\ee
Next, impose the bottom Yukawa coupling, without loss of generality, for $\bar{\bf 5}_1$
\be\ba
\hbox{(Y2.)}:\quad  q_1 + q_{H_d} + q_{\bf 10} =0 \quad \Rightarrow \quad &
q_{H_u} = {2\over 3} \left(q_1+ \sum_{i=1}^{n-1} N_i (q_n-q_i)\right)\cr
& 
q_{H_d} =  {1\over 3} \left(-2q_1+ \sum_{i=1}^{n-1} N_i (q_n-q_i)\right) \,.
\ea
\ee
Finally, we impose the anomaly (A3.), which results for a single $U(1)$ in a quadratic constraint
\be\label{A3Gen}
q_{\mu}\, (q_{H_d} -q_{H_u}  )+ \sum_{i=1}^{n-1}N_i (q_i^2 - q_n^2) =0 \,,
\ee
which after inserting the solution of the charges for the Higgs doublets takes the form of a Diophantine equation
\be\label{Diophanti}
\sum_{i, j=1}^{n-1} a_{ij} q_i q_j =0 \,,
\ee
where the integers $a_{ij}$ depend on the  mutiplicities $N_i$. 
From the form (\ref{A3Gen}) it is clear that each term in the anomaly is proportional to the difference of two charges, so that one inital seed solution is
\be
q_i =q_0 \qquad i=1, \ldots, n \,.
\ee
Starting from this solution, we can generate all solutions to this with the method from Mordell \cite{Mordell}. 


The theorem in Mordell \cite{Mordell} states, that if a non-zero integer solution to  
\be\label{Mordellf}
a q_1^2 + b q_2^2 + c q_3^2 + 2 f q_2 q_3  + 2 g q_1 q_3 +  2 h q_1 q_2 =0 
\ee
exists, then the general solution with all $q_i$ coprime, i.e. $(q_1, q_2, q_3)=1$, is given by expressions
\be
q_i = a_i p^2 + b_i p q + c_i q^2 \,, \qquad (p,q)=1  \,,\quad p,q\in\mathbb{Z}\,,
\ee
with  $a_i, b_i, c_i\in \mathbb{Z}$ constants. In fact a constructive method is given: consider an initial seed solution 
$(q_1^0, q_2^0, q_3^0)$. Then let 
\be\label{MordellShift}
q_1 =  r q_1^0  +  p \,,\qquad 
q_2 = r q_2^0 + q \,,\qquad 
q_3 = rq_3^0 \,.
\ee
Inserting this back into (\ref{Mordellf}) results in a linear equation for $r$, which can be solved and thus one determines the expressions for $q_i$ from (\ref{MordellShift}). 

This method can be applied more generally for $n>2$. The ans\"atze are 
\be\ba
q_i  &= q_i^0r + p_i \,,\ \hbox{for} \  i= 1, \ldots, n-1 \cr
q_n & = q_n^0 r  \,.
\ea\ee
Again, the resulting equation (\ref{Diophanti}) becomes lines in $r$, and can be solved in each case to yield the charges $q_i$ for all $i$. 
In general this leaves $n-1$ charges unfixed by the constraints imposed thus far. 
For each case we will now consider in the following the charges of the unwanted couplings (C1.)$-$(C7.), in order to determine the phenomenological soundness of the models.


\subsection{General Solutions for $\mathcal{N}_{\bar{\bf 5}} =5$}

To exemplify the method in the last section, consider the case of
three matter $\bar{\bf 5}$ representations, in addition to the two Higgs ones, which will be parameterized as
\be
\begin{array}{c|c|c|c}
{\bf R} &  q({\bf R}) & M & N \cr\hline
\bar{\bf 5}_{H_u} &  - q_{H_u}  &0 & -1 \cr
\bar{\bf 5}_{H_d}  & q_{H_d} &0& 1 \cr
\bar{\bf 5}_{1} &  q_1 &M_1 &N_1  \cr 
\bar{\bf 5}_{2}& q_2 &M_2 &  N_2 \cr
\bar{\bf 5}_{3}& q_3 &3-M_1-M_2& -N_1 - N_2 \cr\hline
{\bf 10}  &  q_{\bf 10} = -{1\over 2} q_{H_u}&   3&0 
\end{array}
\ee
Note that for fewer, the equations always factor and can be solved easily. The first non-trivial case is $n=5$.
The constraints on the integers $M_i$ and $N_i$ is 
\be
0 \leq  M_i \leq  3\,, \qquad 
0 \leq  M_i + N_i \leq  3 \,,\qquad 
M_1 + M_2 \leq  3 \,,\qquad  (N_1, N_2) \not= (0, 0) \,.
\ee
There are 90 solutions, however only 40 will be eventually of interest and distinct from earlier cases with fewer, distinctly charged matter. 

Again, we first solve the anomaly constraint (A2.) which yields
\be
\hbox{(A2.)}  \quad \Rightarrow \quad  q_{H_d} = q_3 (N_1+N_2)-N_1 q_1-N_2 q_2-q_{H_u} \,.
\ee
Furthermore, without loss of generality, we impose the bottom Yukawa coupling for the $\bar{\bf 5}_{1}$ matter, i.e. 
\be
\hbox{(Y2.)} \quad \Rightarrow \quad q_1 + q_{H_d} + q_{\bf 10} =0 \quad \Rightarrow \quad
q_{H_u} = -{2\over 3} (N_1 q_1 - N_1 q_3 + N_2 q_2 - N_2 q_3 - q_1)\,,
\ee
where $q_{H_d}$ from the anomaly was used. 
Furthermore as we impose the bottom Yukawa for $\bar{\bf 5}_{1}$, we require $M_1\not=0$. 
Note that the $\mu$-term has charge
\be
q_{\mu} = q_{H_u} + q_{H_d}= \left(N_1+N_2\right) q_3-N_1 q_1-N_2 q_2  \not= 0 \,.
\ee
This in particular implies
\be\label{N1N20}
(N_1, N_2) \not= (0,0) \,.
\ee
The anomaly (A2.) constraint now reads 
\be\ba\label{55barsA3}
(A3.)\quad \Rightarrow \quad 
& \left(7-N_1\right) N_1 q_1^2+\left(3-N_2\right) N_2 q_2^2-\left(N_1+N_2\right) \left(N_1+N_2+3\right) q_3^2 \cr
&+2 \left(N_1-2\right) \left(N_1+N_2\right) q_1 q_3-2 \left(N_1-2\right) N_2 q_1 q_2
+2 N_2 \left(N_1+N_2\right) q_2 q_3 =0 \,,
\ea\ee
which is a homogeneous quadratic equation in $q_i$ with integer coefficients. We are searching for rational solutions, although by rescaling, we can consider integer solutions. Such Diophantine equations are for instance discussed in \cite{Mordell}, which gives a systematic construction of its solution, starting with a seed solution.

Applying this to the anomaly constraint  (\ref{55barsA3}) with the seed solutions 
\be
q_1^0 = 4 N_1 N_2-3 N_1+9 N_2 \,,\quad 
q_2^0 = 4 N_1 N_2+17 N_1-11 N_2\,,\quad 
q_3^0 = 4 N_1 N_2-3 N_1-11 N_2 \,,
\ee
which is non-trivial as $N_i$ cannot both vanish. {We now need to choose these integers so that the seed solution satisfies $(q_1^0, q_2^0, q_3^0) =1$.
Examples of these are 
\be
(N_1, N_2, N_3) = (-1, -2, 3),\  (1, -2, 1),\  (-3, 2, 1),\  (-2, 1, 1), \ (2, -1, -1) \,.
\ee
}The resulting charges from the Mordell argument are 
\be
\ba
q_1 &=p-\frac{\left(9 N_2+N_1 \left(4 N_2-3\right)\right) \left(\left(N_1-7\right) N_1 p^2+2 \left(N_1-2\right) N_2 p q+\left(N_2-3\right) N_2
   q^2\right)}{10 \left(N_1 \left(N_1 \left(4 N_2+3\right)-25 N_2\right) p+N_2 \left(3 N_2+N_1 \left(4 N_2-9\right)\right) q\right)}
   \cr
q_2 &=q-\frac{\left(N_1 \left(4 N_2+17\right)-11 N_2\right) \left(\left(N_1-7\right) N_1 p^2+2 \left(N_1-2\right) N_2 p q+\left(N_2-3\right) N_2
   q^2\right)}{10 \left(N_1 \left(N_1 \left(4 N_2+3\right)-25 N_2\right) p+N_2 \left(3 N_2+N_1 \left(4 N_2-9\right)\right) q\right)}\cr
q_3 &= -\frac{\left(N_1 \left(4 N_2-3\right)-11 N_2\right) \left(\left(N_1-7\right) N_1 p^2+2 \left(N_1-2\right) N_2 p q+\left(N_2-3\right) N_2
   q^2\right)}{10 \left(N_1 \left(N_1 \left(4 N_2+3\right)-25 N_2\right) p+N_2 \left(3 N_2+N_1 \left(4 N_2-9\right)\right) q\right)}\,.
\ea
\ee
Here, $p, q\in \mathbb{Z}$ and coprime. The $\mu$-term is 
\be
q_{\mu} = -\frac{3 \left(N_1+N_2\right) \left(N_1 p-N_2 q\right){}^2}{N_1 \left(N_1 \left(4 N_2+3\right)-25 N_2\right) p+N_2 \left(3 N_2+N_1 \left(4
   N_2-9\right)\right) q} \not= 0 \,.
\ee
The remaining bottom Yukawa couplings have charge
\be\ba
q(\lambda^b_2) &= -\frac{\left(N_1+N_2\right) \left(N_1 p-N_2 q\right) \left(\left(2 N_1-11\right) p+\left(2 N_2-3\right) q\right)}{N_1 \left(N_1 \left(4
   N_2+3\right)-25 N_2\right) p+N_2 \left(3 N_2+N_1 \left(4 N_2-9\right)\right) q} \cr
q(\lambda^b_3)&= -\frac{\left(N_1 p-N_2 q\right) \left(-11 N_2 p+N_1 \left(2 N_2+3\right) p+2 \left(N_2-3\right) N_2 q\right)}{N_1 \left(N_1 \left(4 N_2+3\right)-25
   N_2\right) p+N_2 \left(3 N_2+N_1 \left(4 N_2-9\right)\right) q} \,.
\ea\ee
Similarly one can solve for more ${\bf \bar{5}}$ curves using this Mordell approach. In the main text we will constrain ourselves to the F-theoretic charges, which comprise a finite set, and thus do not necessarily need to use this method. 


\section{Search for Other Known Textures}
\label{app:OtherTextures}

In section \ref{sec:flavor} we saw that the case of four $\obfive$ representations produced Yukawa textures matching \eqref{HabaText} and \eqref{BaEnGoText}. Extending the analysis to five and six $\obfive$s we find that there are no solutions to the anomaly cancellation conditions, which produce the same Yukawa hierarchies. Here we consider whether other known flavor models can be realized within our F-theory framework. We find  no fits to other known flavor textures. 

\subsection{Symmetric Textures}
Consider first the Yukawa hierarchies in \cite{Ross:2007az} given by
\be 
Y^u \sim 
\left(
\begin{array}{ccc}
\epsilon^4 & \epsilon^3 & \epsilon^3  \\
\epsilon^3  & \epsilon^2 & \epsilon^2 \\
\epsilon^3  & \epsilon^2 & 1
\end{array}
\right) \,,
\quad 
Y^d \sim 
\left(
\begin{array}{ccc}
\epsilon^4 & \epsilon^3 & \epsilon^3 \\
\epsilon^3 & \epsilon^2 & \epsilon^2\\
\epsilon^3 & \epsilon^2 & 1
\end{array} 
\right) \,.
\label{RSflavorTexture}
\ee
In this section we will show it is not possible to match to this texture in our framework, due to the form of the down-type Yukawa matrix.
To see this, note that we need at least five $\obfive$s so that each down-type quark resides within a differently charged $\obfive$. Consider the parameterization \be
\begin{array}{c|c|c|c|c}
{\bf R} &  q^1({\bf R}) & q^2({\bf R}) & M & N \cr\hline
\bar{\bf 5}_{H_u} &   -q^1_{H_u}  & -q^2_{H_u} & 0 & -1  \cr
\bar{\bf 5}_{H_d}  & \frac{3}{2} q^1_{H_u} - 5 w^1_{5_3} & \frac{3}{2} q^2_{H_u} - 5 w^2_{5_3} & 0 & 1 \cr
\bar{\bf 5}_{1} &  -q^1_{H_u} + 5w^1_{5_1} & -q^2_{H_u} + 5w^2_{5_1} & 1 & N_1 \cr 
\bar{\bf 5}_{2} &-q^1_{H_u} + 5w^1_{5_2} & -q^2_{H_u} + 5w^2_{5_2} & 1 & N_2\cr
\bar{\bf 5}_{3} &-q^1_{H_u} + 5w^1_{5_3} & -q^2_{H_u} + 5w^2_{5_3}  & 1 & -N_1 -N_2\cr\hline
{\bten_1}  & -{1\over 2} q^1_{H_u} + 5 w^1_{10_1} & -{1\over 2} q^2_{H_u} + 5 w^2_{10_1} & 1 & 0\cr
{\bten_2}  & -{1\over 2} q^1_{H_u} + 5 w^1_{10_2} & -{1\over 2} q^2_{H_u} + 5 w^2_{10_2} & 1 0\cr
{\bten_3}  &  -{1\over 2} q^1_{H_u} & -{1\over 2} q^2_{H_u}  & 1 & 0
\end{array}
\ee
where the charge of $H_d$ has been chosen to allow an order one bottom Yukawa coupling (which we choose to be ${\bf \bar{5}}_3$) at leading order. This set of charges gives rise to the following Yukawa textures written in terms of the singlet insertions required to regenerate each entry
\be 
Y^u \sim 
\left(
\begin{array}{ccc}
s_1^2 & s_1s_2& s_1 \\
s_1 s_2  & s_2^2 & s_2 \\
s_1  & s_2 & 1
\end{array}
\right) \,,
\quad 
Y^d \sim 
\left(
\begin{array}{ccc}
s_4 s_1 & s_1 s_3 & s_1 \\
s_4 s_2 & s_2 s_3 & s_2 \\
s_4 & s_3 & 1
\end{array} 
\right) \,.
\label{MatchRS}
\ee
where $s_i = \frac{\langle S_i \rangle}{M_{GUT}}$ and the charges of the singlets  $S_i$ are
\be \ba 
(q_{S_1}^1, q_{S_1}^2) &= -5(w_{10_1}^1, w_{10_1}^2) \\
(q_{S_2}^1, q_{S_2}^2) &= -5(w_{10_2}^1, w_{10_2}^2) \\
(q_{S_3}^1, q_{S_3}^2) &= -5(w_{\bar{5}_2}^1-w_{\bar{5}_3}^1, w_{\bar{5}_2}^2-w_{\bar{5}_3}^2) \\
(q_{S_4}^1, q_{S_4}^2) &= -5(w_{\bar{5}_1}^1-w_{\bar{5}_3}^1, w_{\bar{5}_1}^2-w_{\bar{5}_3}^2) \,.
\ea \ee
From the structure of the singlet insertions in the Yukawa matrices shown above one can see that it is not possible to match to the $\epsilon$ suppressions shown in \eqref{RSflavorTexture}. The problem lies in the texture of the down-type matrix in \eqref{MatchRS}, if the singlet insertions in (2,3) and (3,2) are chosen to have $\epsilon^2$ suppression then the (2,2) entry is automatically of order $\epsilon^4$. This is in disagreement with \eqref{RSflavorTexture} therefore it is not possible achieve the texture in \cite{Ross:2007az}. 

\subsection{$E_8$-model Textures}

Consider the Yukawa hierarchies discussed in \cite{Dudas:2009hu}\footnote{The down-type Yukawa matrix has been transposed to match the convention defined in \eqref{DownTypeYukawa}.}, which was discussed in the context of local models in F-theory in the context of models obtained by higgsing $E_8$,
\be 
Y^u \sim 
\left(
\begin{array}{ccc}
\epsilon^6 & \epsilon^5 & \epsilon^3  \\
\epsilon^5  & \epsilon^4 & \epsilon^2 \\
\epsilon^3  & \epsilon^2 & 1
\end{array}
\right) \,,
\quad 
Y^d \sim 
\left(
\begin{array}{ccc}
\epsilon^6 & \epsilon^5 & \epsilon^3 \\
\epsilon^4 & \epsilon^3 & \epsilon \\
\epsilon^3 & \epsilon^2 & 1
\end{array} 
\right) \,.
\label{DPflavorTexture}
\ee
One finds that it is not possible to match to this set of textures either. It is not surprising that the local analysis in  \cite{Dudas:2009hu} is not consistent with the analysis here, as it relied on local $U(1)$ charges and does not consider the quadratic anomaly (A3.). 
To see that the global F-theory charges do not allow for these texture in \eqref{DPflavorTexture}, note that each down-type quark must originate from a differently charged $\obfive$ representation which requires
\be  \ba 
M_{5_i} &= 1,  \\
N_{5_i}  &= -1,0,1,2 , \quad i = 1,2,3 \,,
\ea \ee
where the restriction on $N_{5_i}$ stems from imposing the absence of exotics.
For general $\obfive$ charges there are three distinct cases to consider
\be
(N_{5_1} , N_{5_2} , N_{5_3}) = \{(0,0,0), (1,-1,0), (2,-1,-1) \} \,.
\ee
The first case is excluded as the cancellation of the linear anomaly (A2.) requires the presence of the $\mu$-term at leading order which is unfavorable. We shall see in the following that we find no phenomenologically good models for the second and third cases either. For the second case, the anomaly cancellation conditions can be solved exactly for the following parameterization,
\be
\begin{array}{c|c|c|c|c}
{\bf R} &  q^1({\bf R}) & q^2({\bf R}) & M & N \cr\hline
\bar{\bf 5}_{H_u} &   -q^1_{H_u}  & -q^2_{H_u}  &0 & -1 \cr
\bar{\bf 5}_{H_d}  & -q^1_{H_u} + 5 w^1_{H_d} & -q^2_{H_u} + 5 w^2_{H_d} &0& 1 \cr
\bar{\bf 5}_{1} &  -q^1_{H_u} + 5w^1_{5_1} & -q^2_{H_u} + 5w^2_{5_1} &1 &1  \cr 
\bar{\bf 5}_{2} &-q^1_{H_u} + 5w^1_{5_2} & -q^2_{H_u} + 5w^2_{5_2} & 1 & -1 \cr
\bar{\bf 5}_{3} &-q^1_{H_u} + 5w^1_{5_3} & -q^2_{H_u} + 5w^2_{5_3} & 1 & 0 \cr\hline
{\bten_1}  & -{1\over 2} q^1_{H_u} + 5 w^1_{10_1} & -{1\over 2} q^2_{H_u} + 5 w^2_{10_1} & 1 &0  \cr
{\bten_2}  & -{1\over 2} q^1_{H_u} + 5 w^1_{10_2} & -{1\over 2} q^2_{H_u} + 5 w^2_{10_2} & 1 &0  \cr
{\bten_3}  &  -{1\over 2} q^1_{H_u} & -{1\over 2} q^2_{H_u} &  1&0 
\end{array}
\label{flavorFive5bs}
\ee
The third generation quarks are taken to reside within $\bten_3$, the charge of which has been fixed to allow for a leading order top Yukawa coupling. Inserting this set of charges and $M,N$s into the linear anomaly we obtain, 
\be 
w^{\alpha}_{H_d} + w^{\alpha}_{5_1} - w^{\alpha}_{5_2} = 0\,,
\label{LinearSol}
\ee
where $\alpha = 1,2$. Solving for $w_{H_d}^{\alpha}$ and inserting into the quadratic anomaly (A3.) we obtain,
\be \ba 
 w^{\alpha}_{5_1}( w^{\alpha}_{5_1} - w^{\alpha}_{5_2}) &=0, \\
 2 w^1_{5_1} w^2_{5_1} - w^{2}_{5_1} w^{1}_{5_2} + w^{1}_{5_1} w^{2}_{5_2} &=0  \,.
\ea \ee
This set of three equations has two distinct solutions however neither of them lead to phenomenologically good models
\begin{itemize}
\item $ w^1_{5_1} = w^2_{5_1} =0$ \\
Substituting this into the charges in \eqref{flavorFive5bs} we observe that,
\be 
(q^1_{H_u}, q^2_{H_u}) = (q^1_{5_1}, q^2_{5_1})\,,
\ee
which means that the unwanted operator (C5.) is present at leading order through the coupling
$\bten_3 \bten_3 \obfive_1$.
This set of solutions is therefore not viable.
\item $ w^1_{5_1} = w^1_{5_2}$ and $ w^2_{5_1} = w^2_{5_2}$ \\
Substituting this solution into \eqref{LinearSol} one observes that $w^{\alpha}_{H_d} =0$, which results in a leading order $\mu$-term which is unfavorable. 
\end{itemize}
To find solutions for the last case, given by the choice,
$N_{5_1} = 2$, $N_{5_2} = N_{5_3} = -1$
we scan through the possible charges of $\bten$ and $\obfive$ matter under two $U(1)$s for the six codimension one fibers in \eqref{Codim12U1sReal}. We find no sets of charges which solve the anomaly cancellation conditions for this set of $N_{5_i}$s. Therefore, in order to obtain a model that is consistent with the flavor texture in \eqref{DPflavorTexture}, anomaly cancellation and absence of dangerous operators one must go to greater than five $\obfive$ representations. However, on extending this analysis to six $\obfive$ representations, there are again no solutions matching to flavor texture in \eqref{DPflavorTexture}. Possibly, by including more $U(1)$s these other textures become accessible in this class of models as well. We leave this for future investigations.

\newpage
\bibliographystyle{JHEP}

\providecommand{\href}[2]{#2}\begingroup\raggedright\endgroup


\end{document}